\documentclass[twoside,fleqn]{ActaStyle}
\pdfoutput=1
\usepackage{times,cite}
\usepackage{amssymb}
\usepackage{subfigure}
\usepackage{rotating}
\usepackage{booktabs}

\def\authorheading{J. Wagner-Kuhr}
\def\shorttitle{Top Quark Production}

\begin{document}
\pagerange{1}{68} 

\title{Experimental Studies of Top Quark Production}
\author{Jeannine Wagner-Kuhr\email{jeannine.wagnerkuhr@lmu.de}$^*$} 
       {$^*$Ludwig-Maximilians-Universit\"at M\"unchen,\\ Am Coulombwall 1, D-85748 Garching, Germany}

\abstract{
In this review article three promising aspects of top quark production are discussed:
the charge asymmetry in top quark pair production, the search for resonant top quark pair
production, and electroweak single top quark production. 
First, an overview of the theoretical predictions of top quark pair
and single top quark production is given.
Then, for each topic the general analysis strategy and improvements are exemplarily explained using selected analyses 
and are put into the context of the global status at the beginning of LHC Run II and progress 
in this field. The example analyses discussed in more detail in this article use data from the LHC experiment CMS
and for the charge asymmetry studies also data
from the Tevatron experiment CDF have been used.
}


\newpage
\tableofcontents



\section{Introduction}
\label{sec:Introduction}


The top quark is the heaviest known elementary particle and after 17 years of intensive search it was discovered in 1995 at Fermilab's proton anti-proton collider Tevatron by the experiments CDF and D0~\cite{Abe:1995hr,Abachi:1995iq}.\\ 

In the sixties of the last century hundreds of new strongly interacting particles were discovered. Only when 
M.~Gell-Mann (Nobel Prize in 1969) and independently G. Zweig introduced the quark model~\cite{GellMann:1964nj} in the year 1964, this particle zoo could be explained. In this model the particles discovered at that time are not fundamental but consist of either a quark and an anti-quark (meson) or of three quarks (baryons). At the time this model was developed three quark flavors ($u$, $d$, and $s$) were known. To explain the experimentally observed very small branching ratio of $K^0\rightarrow \mu^+\mu^-$, S.~L.~Glashow, J. Iliopolus and L. Maiani (Nobel Prize to S.~L.~Glashow in 1979) predicted in the year 1970 a further quark, the charm quark ($c$)~\cite{Glashow:1970gm}. In the year 1972 M.~Kobayashi and T.~Maskawa (Nobel Prize in 2008) predicted a third quark generation ($t$, $b$)~\cite{Kobayashi:1973fv} to explain the occurrence of the CP-violating decay $K^0_L\rightarrow\pi^+\pi^-$~\cite{Christenson:1964fg}. With the simultaneous discovery of the fourth quark~\cite{Aubert:1974js,Augustin:1974xw}, the $c$-quark, in 1974 by S. Ting et al. and B.~Richter et al. (Nobel Prize to S.~Ting and B.~Richter in 1976), the second quark generation ($c$, $s$) has been established. First indications for a new heavy lepton~\cite{Perl:1975bf}, the tauon, have been observed in 1975 and the discovery of the tauon~\cite{Perl:1976rz,Perl:1977se} was established in 1977 (Nobel Prize to M.~Perl in 1995), 
indicating a third family of leptons. Also in 1977, L.~Lederman et al. (Nobel Prize to L.~Lederman in 1988) discovered the fifth quark, 
the $b$-quark~\cite{Herb:1977ek}. This indicated the existence of a third quark generation and a frenetic search for the weak isospin partner of the $b$-quark, the top quark, began. Measurements of the charge~\cite{Berger:1978dm,Darden:1978dk,Bienlein:1978bg} and of the 3rd component of the weak isospin of the $b$-quark~\cite{Bartel:1984rg} in 1978 at the Doris storage ring and 1984 at the JADE experiment, respectively, allowed to predict the charge and the weak isospin of the top quark.\\

First searches for a top quark were conducted at $e^+e^-$ colliders and the lower mass limit on the top quark mass could be increased in the years between 1979 and 1990 from $m_t>23.3\,\mbox{GeV/c}^2$~\cite{Adeva:1984vq,Behrend:1984ub,Bartel:1983ym,Althoff:1984us} measured at Petra (DESY, centre-of-mass energy $\sqrt{s}=12-46.8\,\mbox{GeV}$), to $m_t>30.2\,\mbox{GeV/c}^2$~\cite{Abe:1989qx,Adachi:1987cg,Sagawa:1987sb} at Tristan (KEK, $\sqrt{s}=50-61.4,\mbox{GeV}$) and finally to $m_t>45.8\,\mbox{GeV/c}^2$~\cite{Decamp:1989fk,Akrawy:1989rh,Abreu:1990za,Abrams:1989hs} at LEP I (CERN, $\sqrt{s}=M_Z$) and SLC (SLAC, $\sqrt{s}=M_Z$). Parallel to the searches at the $e^+e^-$-colliders the top-quark has been searched for at the first proton anti-proton collider $Sp\bar{p}S$ (CERN, $\sqrt{s}=546-630\,\mbox{GeV}$). Due to $m_p/m_e\approx 2000$ much higher centre-of-mass energies are feasible at hadron colliders. Because protons are not fundamental particles the collision events are more complicated (large number of final state particles, unknown initial partonic centre-of-mass energy) at hadron colliders than at $e^+e^-$-colliders. At the $Sp\bar{p}S$, the experiments were sensitive to the strong interaction process $p\bar{p}\rightarrow t\bar{t}$ and the electroweak process $p\bar{p}\rightarrow W\,X\rightarrow t\bar{b}\,X$ process. In 1984 the UA1 collaboration claimed first indication for a top quark with $m_t\approx 40\,\mbox{GeV/c}^2$~\cite{Arnison:1984iw}, but with larger data statistics and with a better understanding of background processes it turned out to be a misinterpretation and in 1990 the lower limit on the top quark mass could be increased to $m_t>60\,\mbox{GeV}/c^2$~\cite{Albajar:1990cq} (UA1) and $m_t>69\,\mbox{GeV}/c^2$~\cite{Akesson:1989us} (UA2).\\

A first indirect indication of the top quark mass came from the discovery of the $B^0-\bar{B}^0$-oscillation 
by the Argus collaboration in 1987~\cite{Albrecht:1987dr}. Since the oscillation rate is in good approximation 
proportional to $m_t^2$, the large observed oscillation rate was a first hint for a heavy top quark. 
Starting in 1989 precision measurements on the $Z$-resonance at the colliders SLC and LEP-I provided better 
and better indirect predictions of the top quark mass~\cite{Schaile:1993wh,Quast:1999sh,ALEPH:2005ab}. 
Radiative electroweak loop corrections to $e^+e^-\rightarrow Z/ \gamma\rightarrow f\bar{f}$ ($f$: fermion) 
containing top quarks have due to the heaviness of the top quark a sizable effect on measurable quantities 
like the $Z$-boson mass, the width, and the couplings. A top mass dependent fit to all electroweak precision 
measurements was performed, which allowed the prediction of the top mass within the standard model (SM).\\

In 1988 data-taking of the CDF detector at the new proton anti-proton collider Tevatron at a centre-of-mass energy of $\sqrt{s}=1.8\,\mbox{TeV}$ started. At such energies, the top quark pair production $p\bar{p}\rightarrow t\bar{t}$ via the strong interaction dominates. Already in 1990 the lower limit on the top quark mass of $m_t>77\,\mbox{GeV/c}^2$~\cite{Abe:1989nc,Abe:1990jw} set by the CDF collaboration was better than the limit obtained at the $Sp\bar{p}S$ collider. In 1992 the D0 experiment was commissioned and in 1994 the CDF collaboration saw a first indication of the top quark~\cite{Abe:1994xt,Abe:1994st}. One year later, the top quark was finally observed by both Tevatron experiments~\cite{Abe:1995hr,Abachi:1995iq} and the observed top quark mass of around $176\pm 8\pm 10\,\mbox{GeV/c}^2$ (CDF) matched amazingly well the top quark mass of $m_t=178\pm 11^{+18}_{-19}\,\mbox{GeV/c}^2$~\cite{LEPEW:1994aa} predicted by the global fit to electroweak precision data.\\ 

The top quark mass is with $m_t=173.34\pm 0.76\;\mbox{GeV/c}^2$~\cite{ATLAS:2014wva} about a 
factor 40 heavier than the second heaviest quark, the bottom quark. It is the only quark with a 
mass of the order of the electroweak symmetry breaking scale and a predicted Yukawa coupling of 
one to the Higgs boson, discovered by the ATLAS and CMS collaborations in 
2012~\cite{Aad:2012tfa, Chatrchyan:2012ufa} (Nobel Prize to Francois Englert and Peter W. Higgs 
in 2013 for the theoretical prediction of the Higgs mechanism). 
As such, the top quark plays a special role in many electroweak symmetry breaking theories 
beyond the SM (BSM) and additional or modified top quark production mechanisms are predicted in 
many BSM theories. 
A further unique feature of the top quark arising due to its huge mass is that it decays before 
it can hadronise. Hence, the top quark gives the unique opportunity to investigate the properties 
of a quasi-free quark. For example, the top quark spin is accessible via the angular distribution 
of its decay products, while this information is lost for all other quarks during the hadronisation process.
Being produced at very small distances ($1/m_t$), the top quark is an excellent perturbative object, as
the characteristic strong coupling constant at its production is $\alpha_s \approx 0.1$ meaning that 
perturbative series converges quickly. So the top quark is a good instrument for testing quantum 
chromodynamics (QCD).\\


With the discovery of the top quark a new realm of particle physics opened, namely the
exploration of the properties of the top quark. For Tevatron's second data taking period, 
Run II, from 2002 up to 2011, the centre-of-mass energy was increased to $\sqrt{s}=1.96\,\mbox{GeV}$ 
and the CDF and D0 detectors were improved substantially. One important topic of Tevatron's Run II 
physics program was the detailed survey of the top quark properties: the production, the intrinsic
properties like charge and mass and the decay. In March 2010 the proton-proton collider LHC at CERN
in Geneva started data taking at an initial centre-of-mass energy of $7\,\mbox{TeV}$. In 2011 and 2012
the centre-of-mass energy was increased to $8\,\mbox{TeV}$ and in Run II started in spring 2015 the 
centre-of-mass energy was further increased to $13\,\mbox{TeV}$.  
While the Tevatron was the only place where top quarks could be investigated for about 15 years, 
the LHC accelerator is a top quark factory allowing much more precise top quark investigations.\\

In this review article three promising aspects of top quark production are discussed:
\begin{itemize}
\item Charge asymmetry in top quark pair production
\item Search for resonant top quark pair production
\item Electroweak single top quark production
\end{itemize}
As measurements of the charge asymmetry in top quark pair production by both Tevatron experiments, CDF and D0, resulted in a 
larger value than expected~\cite{Aaltonen:2008hc,Abazov:2007ab,Aaltonen:2011kc,Abazov:2011rq}, those measurements were one of 
the rare cases where a deviation compared to the SM was observed.
This deviation triggered a large activity on the theory side, beyond SM explanations as well as the improvement of the existing SM calculations, 
but showed also the importance to study the charge asymmetry at the LHC.\\
While the Tevatron experiments were only sensitive to the sub-TeV regime in the invariant mass of top-quark pairs, the TeV 
regime of the invariant top-quark pair spectrum became accessible for the first time with the LHC, making searches for heavy resonances
decaying into top quark pairs extremely interesting. Furthermore, moving from the sub-TeV to the TeV regime leads to
a new challenge, namely the handling of the altered ("boosted") event topology.\\
Electroweak single top quark production was discovered in 2009 by both Tevatron 
ex\-peri\-ments \cite{Aaltonen:2009jj,Abazov:2009ii} as "needle in the haystack" and sophisticated 
multivariate
analyses and the combination of those were essential. At the LHC not only the single top quark production cross section is increased but also
the lagest background at the Tevatron, $W$-boson production in association with jets, is substantially more moderate at the LHC. All this,
makes the LHC a unique place to study electroweak single top quark production in great detail.\\

The review article is structured as follows. After a short overview of theoretical predictions of top quark pair and 
single top quark production in section 2, the three aspects of top production mentioned above are described: the charge asymmetry in section 3, 
the search for resonant top quark pair production in section 4 and electroweak single top quark production in section 5. A summary is given
in the last section.\\
For each topic the general analysis strategy and improvements are exemplarily explained using selected analyses and are put 
into the context of the global status at the beginning of LHC Run II and progress in this field.
The example analyses discussed in more detail in this article use data from the LHC experiment CMS and for the charge asymmetry studies 
also data from the Tevatron experiment CDF have been used. A detailed review of the D0 and CDF-II detectors is given in 
Ref.~\cite{Abazov:2005pn,Blair:1996kx,Acosta:2004yw}, while detailed information about the ATLAS and CMS detectors can be found in 
Ref.~\cite{Aad:2008zzm,Chatrchyan:2008aa}.

\section{Top Quark Production and Event Signature}
\label{sec:topproduction}

In this section, first a brief overview of theoretical aspects of top quark production at 
hadron colliders relevant for this review article is given. Then the expected event 
signatures in the detector are described for the different standard model (SM)
top quark production mechanisms.\\ 

The only place where top quarks have been produced are the Tevatron proton anti-proton collider 
at Fermilab close to Chicago and the proton-proton collider LHC at CERN close to Geneva. 
At the Tevatron and the LHC colliders top quarks are dominantly pair produced via the strong 
interaction. Beside the production in pairs top quarks can be singly produced in charged current 
electroweak interaction. In 1995 the top quark pair production was discovered at Fermilab's proton 
anti-proton collider Tevatron by both experiments CDF and D0~\cite{Abe:1995hr, Abachi:1995iq} and 
in spring 2009 single top quark production has been observed by both Tevatron 
experiments~\cite{Aaltonen:2009jj, Abazov:2009ii}.

\subsection{Top Quark Pair Production}
\label{sec:ttbar}

\begin{figure}[htb]
\subfigure[]{
\includegraphics[width=0.23\textwidth]{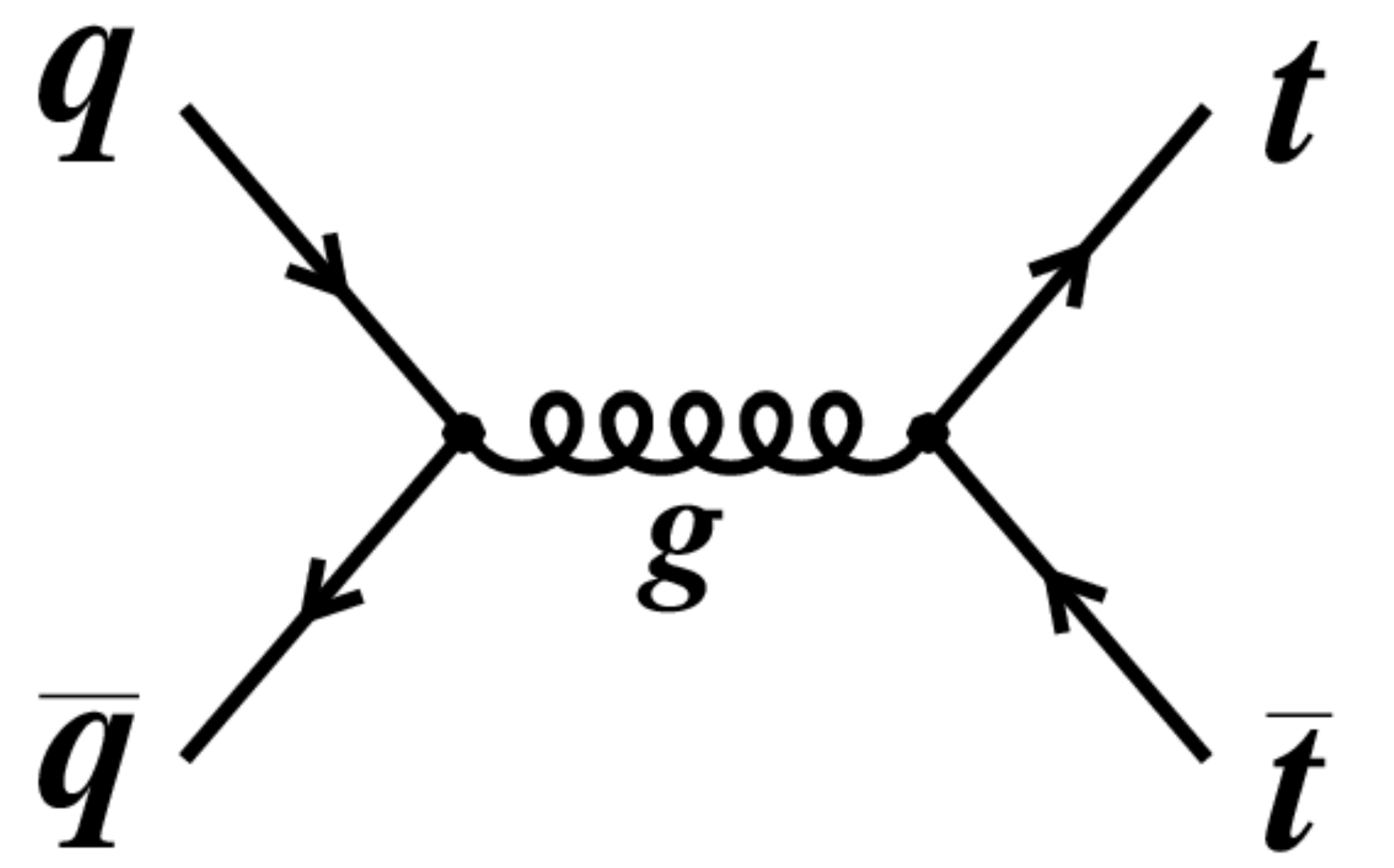}
}
\subfigure[]{
\includegraphics[width=0.23\textwidth]{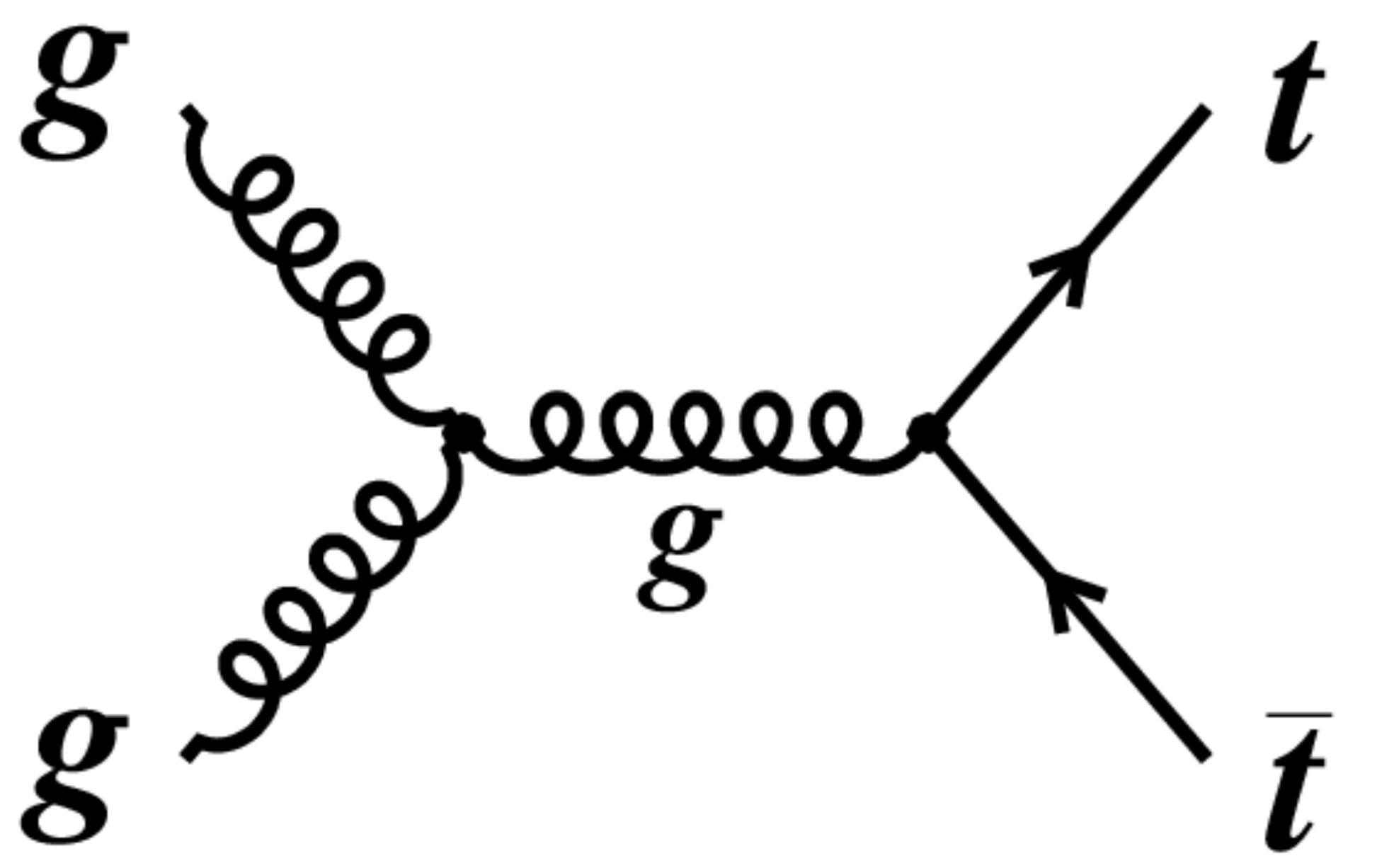}
}
\subfigure[]{
\includegraphics[width=0.23\textwidth]{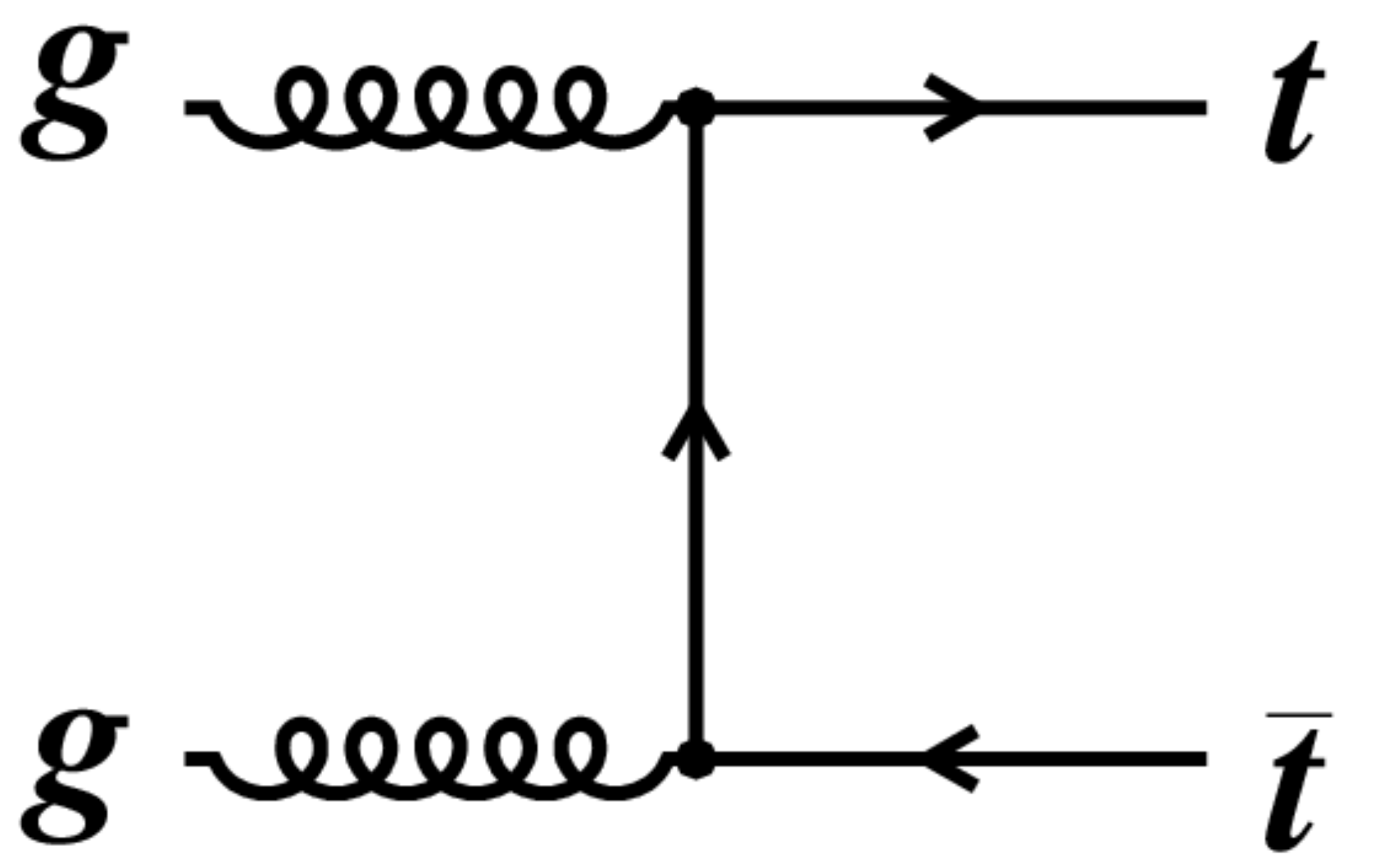}
}
\subfigure[]{
\includegraphics[width=0.23\textwidth]{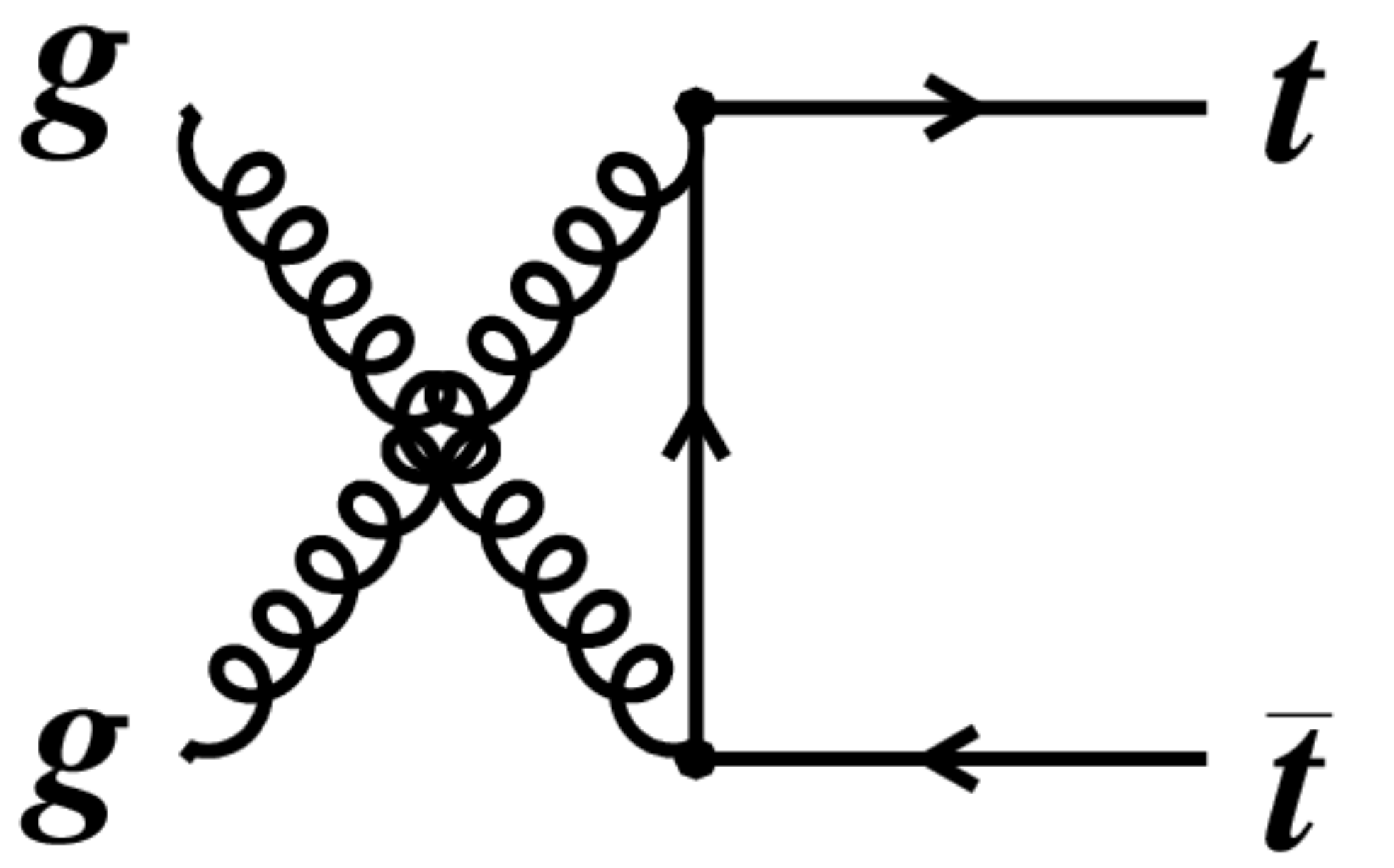}
}
\caption{Feynman diagrams of top quark pair production at leading-order. (a) Quark anti-quark annihilation ($q\bar{q}\rightarrow t\bar{t}$), (b)-(d) gluon fusion ($gg \rightarrow t\bar{t}$) processes.}
\label{fig:ttbarLO}
\end{figure}

In leading order (LO) top quark pairs are produced via quark anti-quark annihilation or via gluon fusion (see Figure~\ref{fig:ttbarLO}) 
and are of the order $O(\alpha_s^2)$ in the strong coupling constant. In higher orders also quark gluon processes exist.\\

At the $p\bar{p}$ collider Tevatron the quark anti-quark annihilation is with about $85\%$~\cite{Cacciari:2003fi} the 
dominant production mode, while at the $pp$ collider LHC about 90\% is from gluon fusion processes at 
$\sqrt{s}=$14 TeV~\cite{Moch:2008qy} ($\approx 80\%$ at $\sqrt{s}=$7 TeV~\cite{Agashe:2014kda}). 
Increasing the centre-of-mass energy lower and lower momentum fractions of the initial partons are accessible and with 
that the contribution of gluon fusion processes becomes larger and larger. At the LHC centre-of-mass energies of 7 TeV and 
8 TeV the difference in the total $t\bar{t}$ production cross section for $pp$ and $p\bar{p}$ collisions is only of the 
order of 3\% and 2\%~\cite{ttbarsigmaplot}, respectively.\\ 

Full  next-to-leading order (NLO) quantum chromodynamics (QCD) corrections~\cite{Nason:1987xz}, which are of the 
order $O(\alpha_s^3)$, to the inclusive LO top quark pair ($t\bar{t}$) cross section and with that an inclusive 
$t\bar{t}$ cross section computation at NLO accuracy appeared in 1988. These NLO QCD corrections turned out to be 
significant~\cite{Nason:1987xz}: they increase the LO cross section by about 40\% for top quark masses above 100 GeV 
and $\sqrt{s}=1.8\;\mbox{TeV}$, and they are large in particular close to the production threshold, and larger for 
gluon fusion processes than for the quark anti-quark annihilation process. The next step were semi-inclusive $t\bar{t}$ 
cross sections~\cite{Beenakker:1988bq} and finally fully differential NLO computations~\cite{Mangano:1991jk}.\\ 

The NLO result has been improved by considering soft gluon corrections, which are dominant near threshold, by resumming 
first the leading-logarithmic (LL)~\cite{Laenen:1991af}, then also the next-to-leading logarithmic 
(NLL)~\cite{Catani:1996dj,Berger:1996ad,Kidonakis:1996zd,Bonciani:1998vc,Kidonakis:2001nj} and recently also the 
next-to-next-to-leading logarithmic 
(NNLL)~\cite{Ahrens:2011mw,Beneke:2011mq,Cacciari:2011hy,Beneke:2009ye,Langenfeld:2009wd,Ahrens:2011px,Kidonakis:2011tg} soft gluon 
corrections to all orders of perturbation theory. According to Ref.~\cite{Beneke:2011ys} the resummation effects (NNLL accuracy) 
amount to about 8\% of the fixed-order NLO result at the Tevatron and to about 3\% at the LHC (with $\sqrt{s}=7\,\mbox{TeV}$) and 
lead to a significant  reduction of the theoretical uncertainty. Cross section calculations at ``mixed order'', where the NNLL 
resummation to all orders in perturbation theory is matched to the fixed-order NLO 
calculation~\cite{Ahrens:2011mw,Beneke:2011mq,Cacciari:2011hy}, as well as approximate fixed order NNLO 
calculations~\cite{Beneke:2009ye,Langenfeld:2009wd,Ahrens:2011px,Kidonakis:2011tg} obtained from re-expanding the NNLL 
resummation are available. As pointed out in~\cite{Beneke:2011ys} the total $t\bar{t}$ cross section results at NNLL or 
approximated NNLO level obtained by several groups differ and a comparison of different results would thus give an estimate 
of the ambiguities inherent to resummation (about 10\% of inclusive cross section~\cite{Beneke:2011ys}). According 
to~\cite{Beneke:2011ys} differences in the result occur due to differences in the resummation formalism adopted: whether 
the total cross section is resummed, or differential distributions in different kinematic limits, which are then integrated 
to obtain the inclusive cross section, are resummed, how terms attributed to the exchange of virtual (Coulomb) gluons are 
treated beyond NLO, how constant terms at $o(\alpha^4_s)$ are treated, and whether sets of power-suppressed contributions 
in $\beta $, the velocity of the two top quarks, are included.\\

Furthermore, the top quark pair production has been studied based on NLO cross section formulae in the non-relativistic QCD 
(NRQCD) framework~\cite{Kiyo:2008bv,Hagiwara:2008df}. According to~\cite{Kiyo:2008bv} the large width of the top quark in 
combination with the large contribution from gluon fusion into a (loose bound) colour singlet $t\bar{t}$ system leads at the 
LHC (gluon fusion is here the dominant production mechanism) to a sizable cross section for masses of the $t\bar{t}$ system 
significantly below the nominal threshold. So the $t\bar{t}$ cross section enhancement in the region up to 
$m_{t\bar{t}}=350\,\mbox{GeV/c}^2$ is about a factor three with a significant shift of the threshold~\cite{Kiyo:2008bv}. 
The increase in the total cross section compared to the NLO prediction is only at the 1\% level.\\

Calculations of NLO QCD differential cross sections for the production of $t\bar{t}$ in association with one 
($t\bar{t}$+jet)~\cite{Dittmaier:2007wz,Dittmaier:2008uj,Melnikov:2010iu} and two 
($t\bar{t}$+2 jets)~\cite{Bevilacqua:2010ve,Bevilacqua:2011aa} extra jets became available in 2007 and 2010, respectively.
In 2015 a technique (MEPS@NLO) to merge the NLO QCD calculations for $t\bar{t}$+0,1,2 jets and
parton showers to form a single inclusive sample became available~\cite{Hoeche:2014qda}. 
This approach represents the state of the art in Monte Carlo simulations of $t\bar{t}$ 
production and here all $t\bar{t}$+jets final states with 0,1,2 jets are described at NLO QCD accuracy.\\ 

Great effort has been put into the calculation of the fixed order NNLO $(O(\alpha_s^4))$ $t\bar{t}$ cross section and in 
2013 full results for the total cross section~\cite{Czakon:2013goa, Baernreuther:2012ws,Czakon:2012zr,Czakon:2012pz}
and very recently also differential cross sections at NNLO QCD accuracy~\cite{Czakon:2015owf,Czakon:2016ckf,Czakon:2016dgf} became available. 
The NNLO corrections increase the $t\bar{t}$ cross section at fixed order NLO by about 5\%~\cite{Czakon:2013xaa}. Matching the soft 
gluon resummation at NNLL accuracy to the fixed order NNLO calculation, referred to as NNLO+NNLL, the most accurate calculation as 
of today is obtained~\cite{Czakon:2013goa}, containing for the threshold not only terms of NNLO accuracy but of all orders 
perturbation theory. The resummation shifts the fixed order NNLO cross section up by about (2-3)\%~\cite{Czakon:2013xaa} and 
the theoretical uncertainty of the $t\bar{t}$ cross section at NNLO+NNLL due to unknown higher order corrections (scale uncertainty) 
is about (2-3)\%~\cite{Czakon:2013goa}.
Recently, also approximate NNNLO computations for the $t\bar{t}$ cross section appeared~\cite{Kidonakis:2014isa,Muselli:2015kba}, 
by re-expanding the NNLL resummation. These calculations contain only terms up to the fixed order NNNLO. Table~\ref{tab:ttbarsigmaNNLO} 
summarises the NNLO+NNLL $t\bar{t}$ cross section predictions at the Tevatron and at the LHC taken from~\cite{Czakon:2013goa}.\\

\begin{table}[t]
\centering
\begin{tabular}{l|cc}
\toprule
         & \multicolumn{2}{c}{$\sigma_{t\bar{t}}$ [pb]}\\
Collider & NNLO  & NNLO+NNLL\\ \midrule
Tevatron, $\sqrt{s}=1.96\,\mbox{TeV}$ & $7.009^{+0.259}_{-0.374}\,^{+0.169}_{-0.121}$ & $7.164^{+0.111}_{-0.200}\,^{+0.169}_{-0.122}$  \\
\midrule
LHC, $\sqrt{s}=7\,\mbox{TeV}$ & $167.0^{+6.7}_{-10.7}\,^{+4.6}_{-4.7}$ & $172.0^{+4.4}_{-5.8}\,^{+4.7}_{-4.8}$\\[0.1cm]
LHC, $\sqrt{s}=8\,\mbox{TeV}$ & $239.1^{+9.2}_{-14.8}\,^{+6.1}_{-6.2}$ & $245.8^{+6.2}_{-8.4}\,^{+6.2}_{-6.4}$ \\[0.1cm]
LHC, $\sqrt{s}=14\,\mbox{TeV}$ & $933.0^{+31.8}_{-51.0}\,^{+16.1}_{-17.6}$ & $953.6^{+22.7}_{-33.9}\,^{+16.2}_{-17.8}$\\
\bottomrule
\end{tabular}
\caption{Top quark pair cross section prediction at pure fixed order NNLO and at NNLO+NNLL accuracy at 
the Tevatron and the LHC~\cite{Czakon:2013goa}. All numbers are computed using a top quark mass of $m_t=173.3\,\mbox{GeV/c}^{2}$ and 
the MSTW2008nnlo68cl pdf set~\cite{Martin:2009iq}. The first uncertainty indicates the uncertainty due to an 
independent restricted variation of the scales (between $m_t/2<\mu<2m_t$, with the constraint that the ratio of the two 
scales can never be larger than two) and is defined through the envelope of the resulting cross sections~\cite{Czakon:2013tha}. 
The second uncertainty is due to parton distribution functions (PDF) and is assessed using the MSTW2008 NNLO PDF sets at 68\% C.L.}
\label{tab:ttbarsigmaNNLO}
\end{table}
Besides QCD corrections also electroweak corrections have been computed~\cite{Beenakker:1993yr,Kuhn:2005it,Bernreuther:2005is,Kuhn:2006vh,Bernreuther:2006vg,Hollik:2007sw,Kuhn:2013zoa}. Lowest order electroweak contributions of $O(\alpha^2)$ to the Drell-Yan annihilation via the exchange of a photon or a $Z$-boson are totally negligible since the top quark mass is substantially larger than $M_Z/2$. Due to the different colour flow in the lowest order QCD and lowest order weak diagrams, there is no interference of these processes and corrections of the order $O(\alpha\alpha_s)$ to the $t\bar{t}$ cross section are absent. Hence, first electroweak contributions occur at the order of $O(\alpha\alpha_s^2)$ and are small for the inclusive $t\bar{t}$ cross section, but could become sizable at large invariant top quark masses $m_{t\bar{t}}$, e.g. $\approx 5\%$ at $m_{t\bar{t}}=1.5\,\mbox{TeV}$, or would be sensitive to the Yukawa coupling close to the $t\bar{t}$ production threshold~\cite{Kuhn:2013zoa}.\\

To get calculations that preserve the full spin information of the top quark pair and that are hence able to reproduce the kinematic of the decay particles correctly, the spin degrees of freedom of the $t$ and $\bar{t}$ have to be taken into account as well as the decay of the top quark. Treating top quarks as truly unstable particles all QCD corrections can be decomposed into factorisable and non-factorisable corrections. As pointed out for example in~\cite{Melnikov:2009dn}, the latter implies a cross-talk between production and decays of top quarks, and vanishes in the limit of $\Gamma_t/m_t\rightarrow 0$, where $\Gamma_t$ is the top quark width. In the narrow-width approximation (NWA), top quarks are considered to be produced on-shell. Here, the non-factorisable QCD corrections are ignored and a full description of $t\bar{t}$ production and decay including all the spin correlations is achieved by computing NLO QCD corrections to both production and decay of a polarised $t\bar{t}$ pair~\cite{Bernreuther:2001bx,Bernreuther:2001rq,Bernreuther:2001jb,Czarnecki:1990pe,Brandenburg:2002xr,Bernreuther:2004jv,Bernreuther:2004wz,Melnikov:2009dn}. Later, also electroweak corrections have been included in such calculations~\cite{Bernreuther:2005is,Bernreuther:2006vg,Bernreuther:2008md,Bernreuther:2010ny}. In 2012 NLO QCD corrections to both production and decay in the narrow-width approximation became available for $t\bar{t}$ production in association with one jet~\cite{Melnikov:2011qx}.\\ 

Calculations for the more general $WWb\bar{b}$ production, where the NWA approximation of the top quark decay was dropped and where off-shell contributions and non-factorisable corrections were included became available, recently. First, the full $WWbb$ calculation appeared in the massless $b$-quark approximation (5Flavour, 5F scheme)~\cite{Denner:2010jp,Bevilacqua:2010qb,Denner:2012yc}, later also massive $b$-quarks have been considered~\cite{Frederix:2013gra,Cascioli:2013wga}. In~\cite{AlcarazMaestre:2012vp} a comparison between the $t\bar{t}$ production including the decay in NWA and the full $WWb\bar{b}$ (5F scheme) calculation has been performed. It was shown that the deviation in the inclusive cross section due to the neglected finite top quark width is no larger than 1\%, but that effects can be larger in differential distributions.

\subsection{Single Top Quark Production}
\label{sec:singletop}

\begin{figure}[htb]
\subfigure[]{
\includegraphics[width=0.23\textwidth]{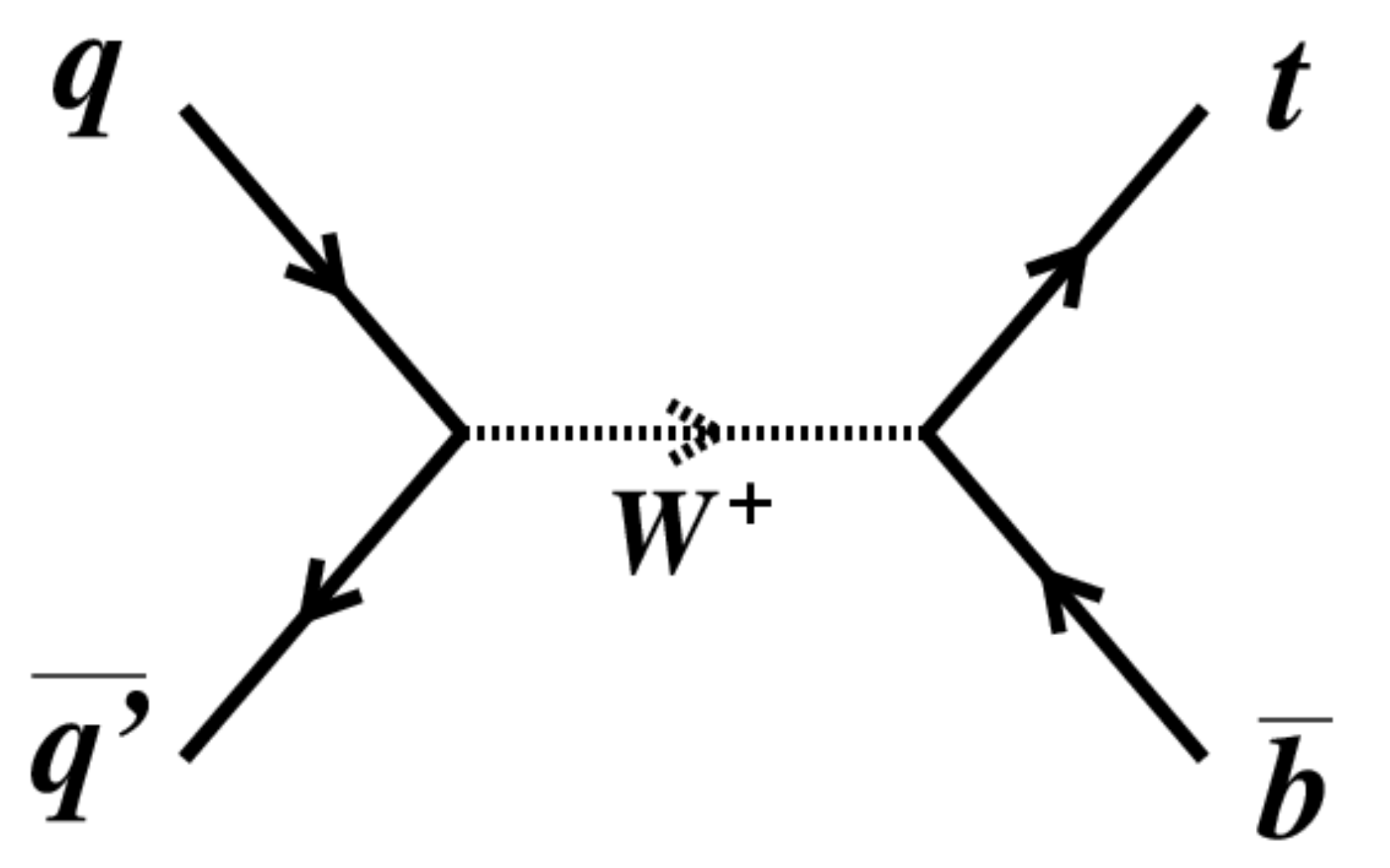}
}
\subfigure[]{
\includegraphics[width=0.23\textwidth]{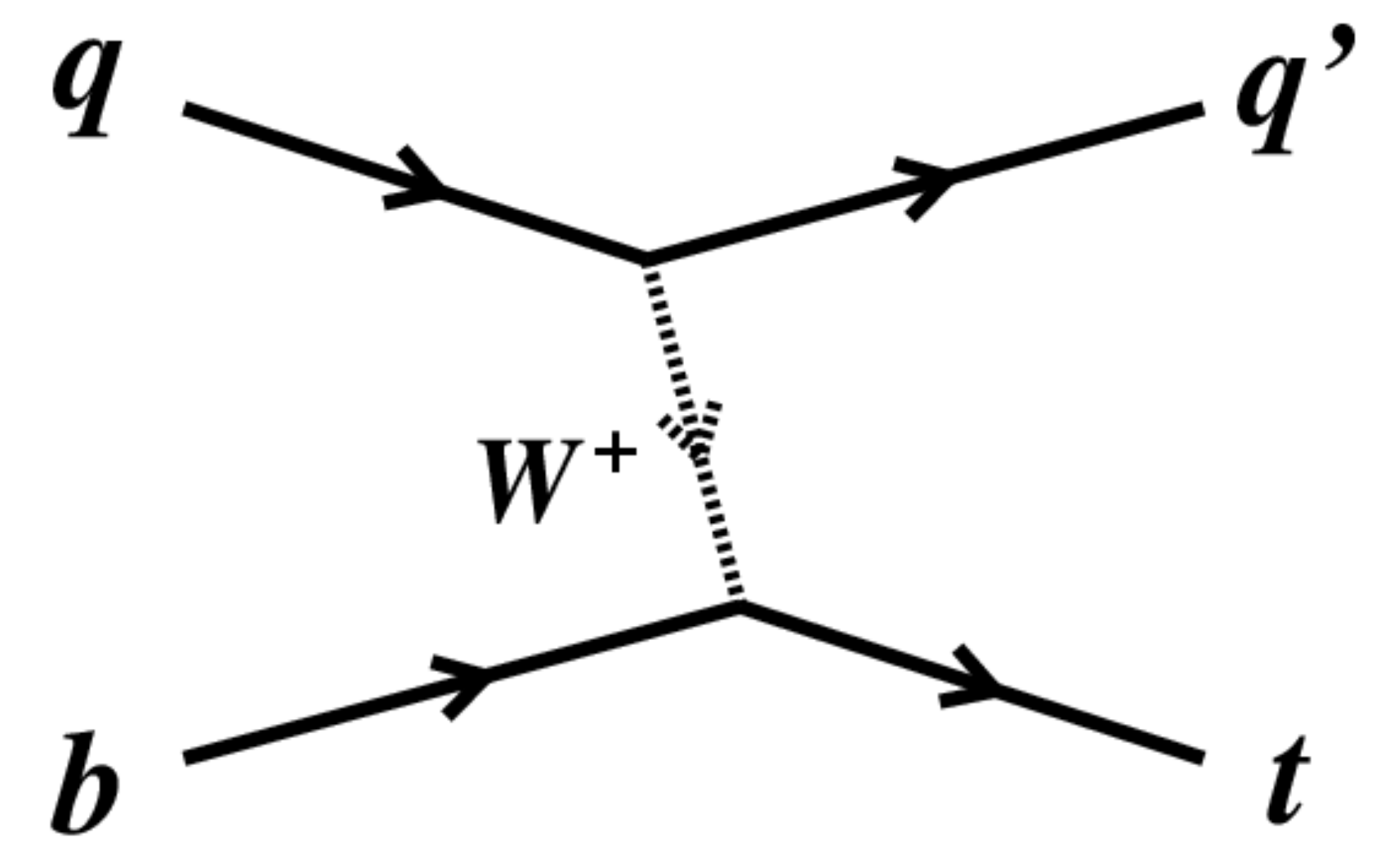}
}
\subfigure[]{
\includegraphics[width=0.23\textwidth]{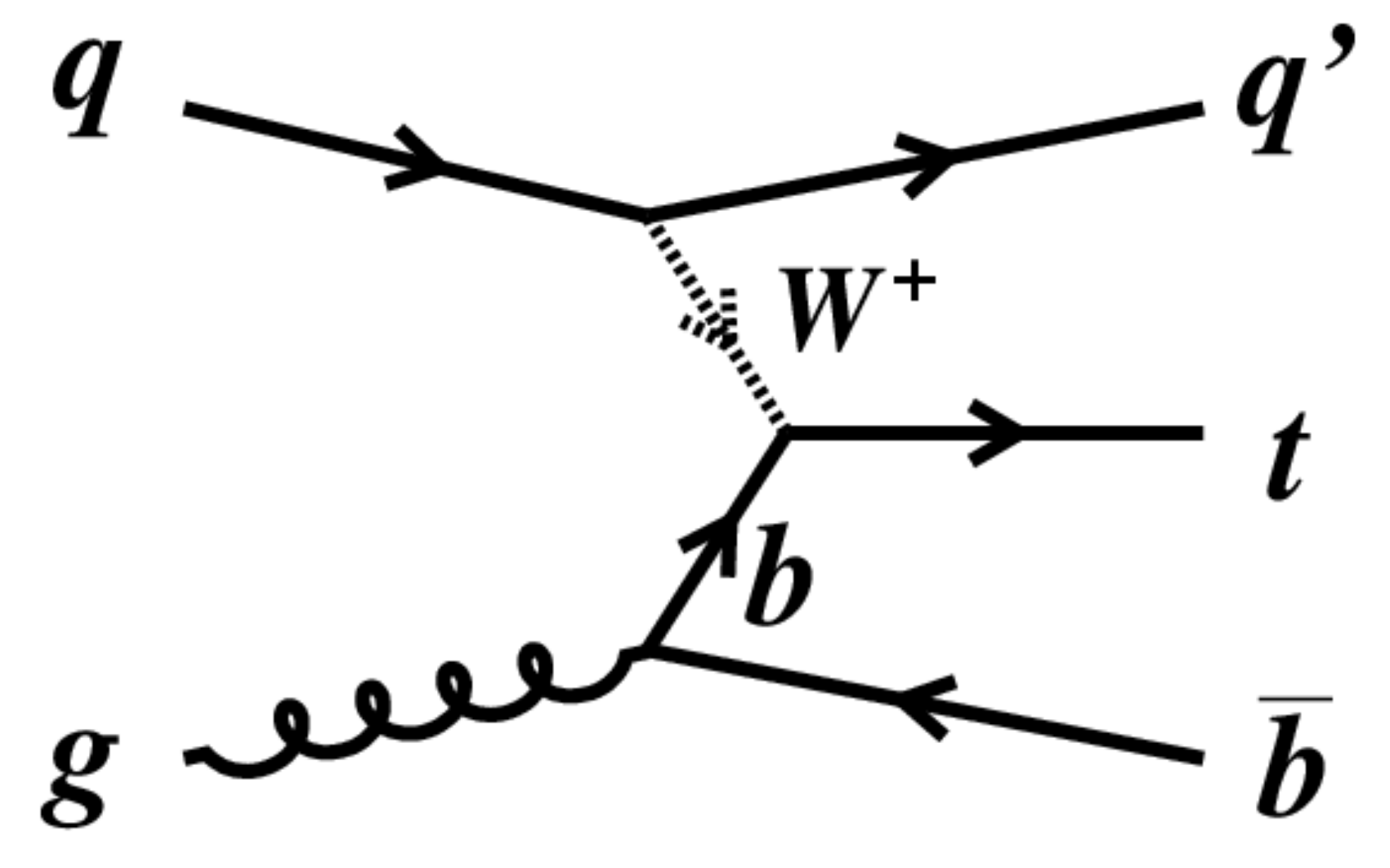}
}
\subfigure[]{
\includegraphics[width=0.23\textwidth]{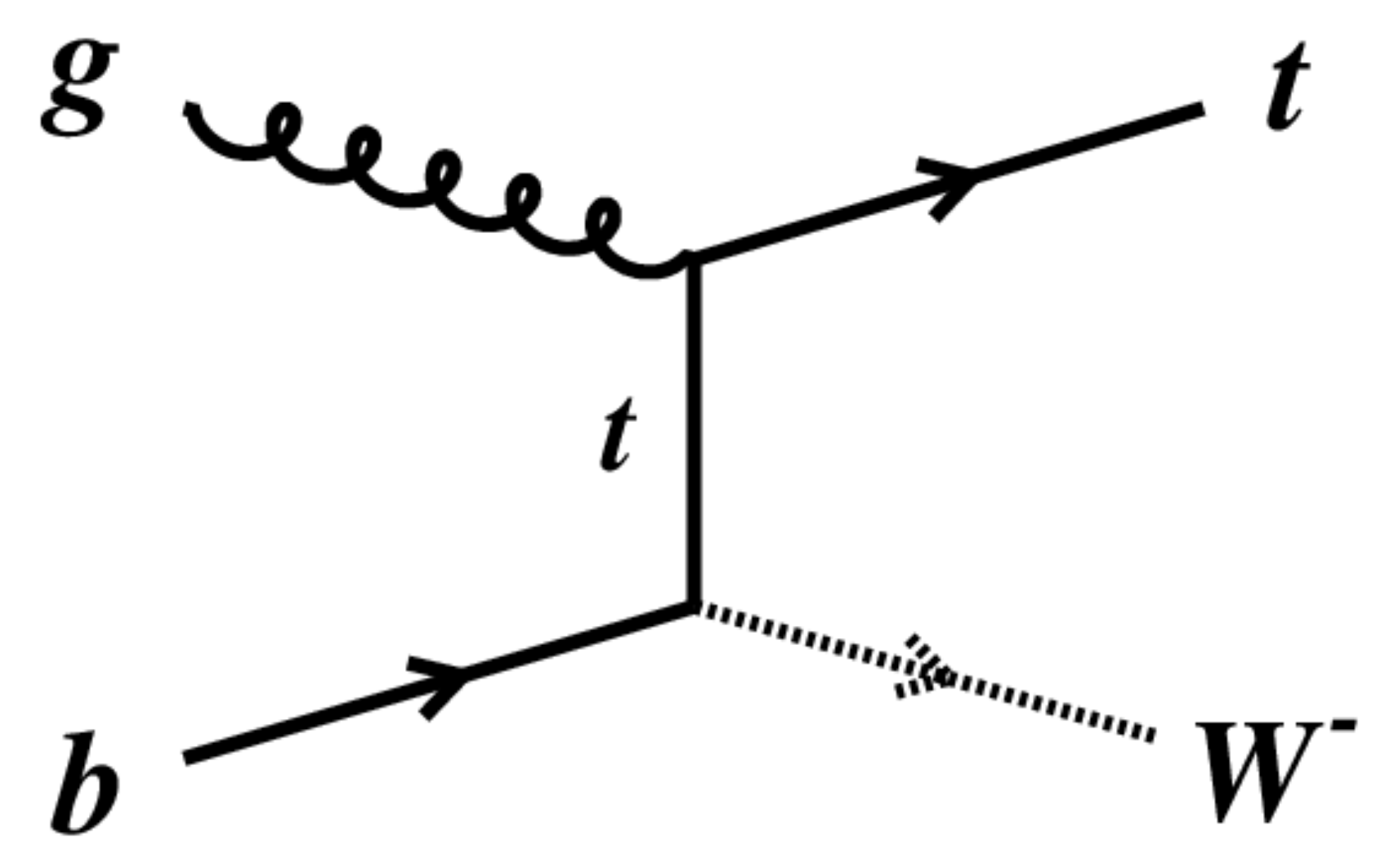}
}
\subfigure[]{
\includegraphics[width=0.23\textwidth]{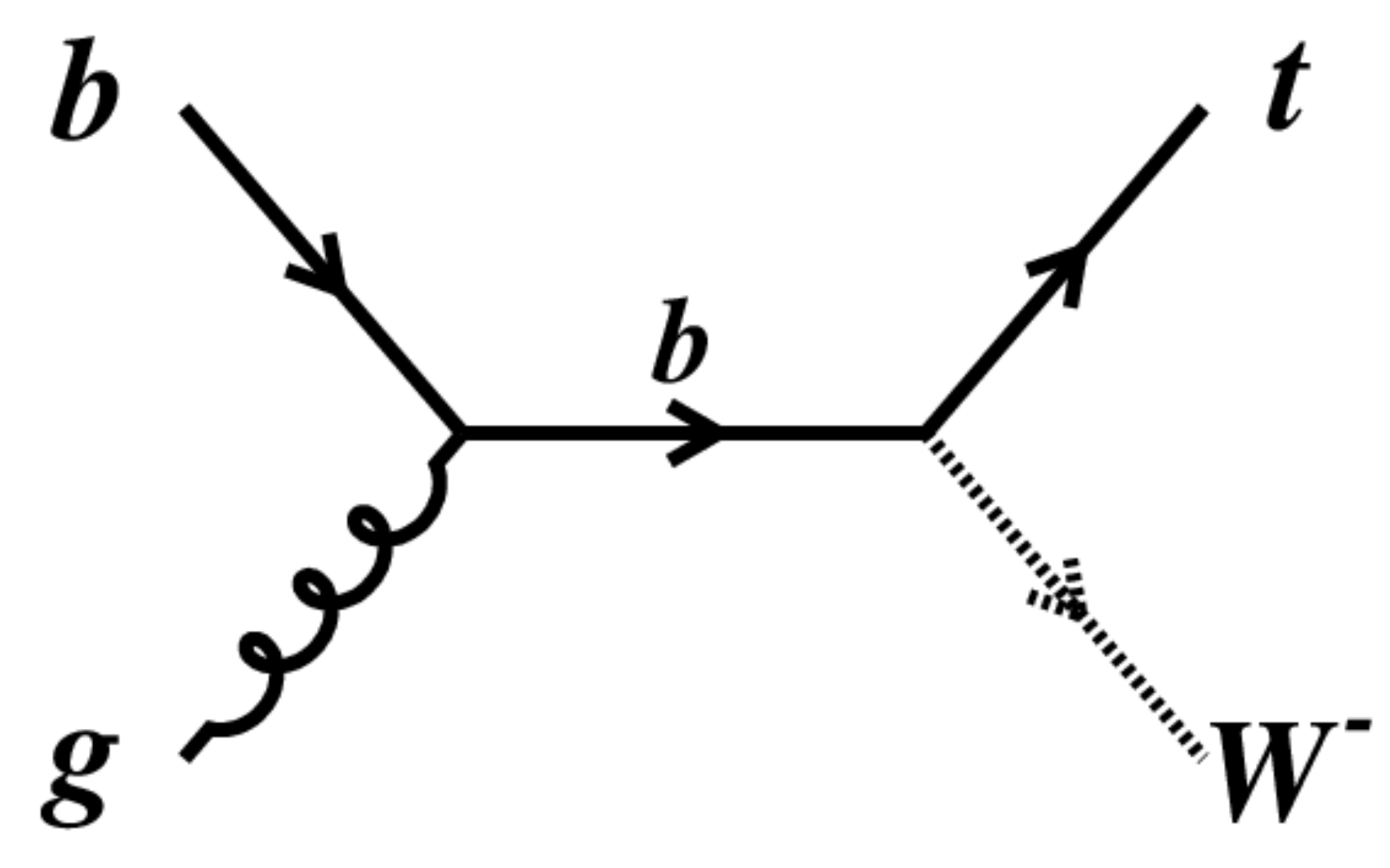}
}
\caption{Example Feynman diagrams of the production of single top quarks via the electroweak production at leading-order, 
with $q=u,c$ and $q'=d,s$. (a) s-channel, (b) $t$-channel as flavour excitation, (c) $t$-channel as $W$-gluon fusion, (d-e) 
$W$-associated single top quark production ($Wt$). In case of $t$-channel, processes induced by a light down-type anti-quark 
and a $b$-quark ($\bar{q}' b \rightarrow \bar{q} t$) or gluon ($\bar{q}' g \rightarrow \bar{b} \bar{q} t$) are not shown as 
these processes are only subdominant.}
\label{fig:singletopLO}
\end{figure}

In the SM three types of electroweak single top quark production modes exist that can be 
distinguished according to the virtuality ($Q^2=-q^2$) of the involved $W$-boson: $t$-channel 
(exchange of a space-like $W$-boson, $q^2<0$) and $s$-channel (exchange of a time-like $W$-boson, $q^2=\hat{s}$) 
processes, and $W$-associated single top quark production (real or close to real $W$-boson, $q^2=M_W^2$). 
Here, $q$ is the four-momentum of the $W$-boson, $\hat{s}$ the centre-of-mass energy of the partonic system and 
$M_W$ the mass of the $W$-boson. For all three production modes example Feynman diagrams at LO are presented in 
Figure~\ref{fig:singletopLO}.\\

The dominant contribution (see Table~\ref{tab:singletopAproxNNLO}) to the cross section is predicted to be from the $t$-channel 
process. While at the Tevatron $W$-associated single top quark production ($Wt$) is negligible, it gives the second largest 
contribution to the cross section at the LHC. The $s$-channel production gives the second largest contribution at the Tevatron, 
but at the LHC its relative contribution is very small.
Due to the valence content ($uud$) of the colliding protons the rate of $t$- and $s$-channel top quark production is at the 
$pp$ collider LHC roughly twice the rate of anti-top quark production. At the $p\bar{p}$ collider no difference in the inclusive 
rates of top quarks and anti-top quarks is observed.\\

Calculations containing full next-to-leading order corrections are available for all three single top quark production modes.\\ 

In the mid nineties first NLO calculations for the inclusive $t$-channel cross section~\cite{Bordes:1994ki,Stelzer:1997ns} appeared 
and in 2002 fully differential NLO results became available~\cite{Harris:2002md}. All these NLO calculations are based on the 
$2\rightarrow 2$ flavour excitation processes $qb\rightarrow q't$~\footnote{This notation includes not only $qb \rightarrow q't$ 
but also $\bar{q'}b \rightarrow \bar{q}t$ for top quark production and also the corresponding $2\rightarrow 2$ processes for 
anti-top quark production: $q'\bar{b} \rightarrow q\bar{t}$ and $\bar{q}\bar{b}\rightarrow \bar{q}'\bar{t}$, where $q=u,c$ and $q'=d,s$.} 
(Fig.~\ref{fig:singletopLO} (b)), where a $b$-quark ($\bar{b}$-quark in case of anti-top production) appears in the initial state. 
In this approach, referred to as five-flavor (5F) scheme, the $b$-quark is treated massless ($m_{b} = 0\;\mbox{GeV/c}^2$). 
Here, possibly large logarithms ($1/ln(\mu^2/m^2_b)$ where $\mu$ is the energy scale) caused by $g\rightarrow b\bar{b}$ splitting 
cases where the $\bar{b}$ ($b$ in case of anti-top production) in the final state is collinear to the initial gluon are consistently 
resummed into the $b$-quark parton distribution functions ($b$-PDF). This leads to an improved stability of the perturbative expansion. 
The $b$-PDF is usually obtained from the QCD evolution of the light quark and gluon PDFs, whereof the gluon PDF is most crucial. 
The $2 \rightarrow 3$ $W g$-fusion process $q g \rightarrow q't\bar{b}$~\footnote{This notation includes not only the process 
$qg \rightarrow q't\bar{b}$ but also $\bar{q'} g \rightarrow \bar{q} t \bar{b}$ for top quark production and also the corresponding 
$2\rightarrow 3$ processes for anti-top quark production: $q'g \rightarrow q\bar{t} b$ and $\bar{q} g\rightarrow \bar{q}'\bar{t} b$, 
where $q=u,c$ and $q'=d,s$.} (see Fig.~\ref{fig:singletopLO} (c) for an example), enters only at NLO meaning that quantities related 
to the ``spectator $b$'' ($\bar{b}$ in final state of  Fig.~\ref{fig:singletopLO} (c)) are computed effectively at LO in the 5F scheme approach. In case of $t$-channel single top quark production, NLO QCD corrections are small, of the order of a few percent~\cite{Harris:2002md} and the smallness of these corrections is due to strong cancellations between different sources of such corrections, e.g. partonic channels~\cite{Brucherseifer:2014ama}.\\
In 2009 NLO calculations based on the $2\rightarrow 3$ $W g$-fusion process $qg \rightarrow q't \bar{b}$ (Fig.~\ref{fig:singletopLO} (c))
and keeping a finite $b$-quark mass became available~\cite{Campbell:2009ss,Campbell:2009gj}. In this approach, referred to as four-flavor 
(4F) scheme, the $b$-quarks do not enter in the QCD evolution of the PDFs and of the strong coupling. 
Due to the inclusion of an additional parton in the final state and the presence of a further mass scale, the 4F scheme calculation 
is much more involved than the 5F scheme calculation. An advantage of the 4F scheme is, that quantities related to the ``spectator $b$'' 
are computed genuinely at NLO accuracy. The authors of Ref.~\cite{Campbell:2009gj} point out, that the calculations for the inclusive 
$t$-channel single top quark production cross sections in the two different schemes differ by 5\% or less at both the Tevatron and 
the LHC. They find, that the inclusive cross sections in both schemes are in substantial agreement if not only the scale uncertainties 
(uncertainty due to higher orders) and PDF uncertainties but also uncertainties on the top quark mass and in particular on the $b$-quark 
mass are considered. As demonstrated in~\cite{Campbell:2009gj} the uncertainty coming from higher orders (scale) is estimated to be much 
larger (about a factor 2 at the LHC) for the 4F scheme than for the 5F scheme. According to Ref.~\cite{Campbell:2009gj} this behaviour is somewhat expected, since the perturbative series for this process begins in the 4F scheme at $O(\alpha_s)$ rather than the purely electroweak leading order of the 5F scheme calculation.\\
Since 2004 calculations performed in the 5F scheme considering full NLO corrections in production and decay
became available.
First, calculations within the NWA~\cite{Campbell:2004ch,Cao:2004ky,Cao:2005pq} appeared, then those calculations have been extended to resonant top quark production, where the NWA was dropped and leading non-factorisable corrections have been included using an effective theory (ET) inspired method~\cite{Falgari:2010sf,Falgari:2011qa}. Finally the full off-shell (also non-resonant contributions) and interference effects at NLO~\cite{Papanastasiou:2013dta} have been computed in 2013. In general it was found, that finite top quark width effects are small for cross sections, but that they can be sizable large for specific differential distributions like the invariant mass of the $Wb$-system~\cite{Falgari:2011qa,Papanastasiou:2013dta}. Furthermore, it has been shown in Ref.~\cite{Papanastasiou:2013dta} that the ET approach of Ref.~\cite{Falgari:2011qa} is also for these variables very close to the full NLO result.\\

In 1996 NLO QCD ($O(\alpha_s)$) corrections and Yukawa ($O(\alpha_W m_t^2/M_W^2)$) corrections, 
arising from loops of Higgs bosons and the scalar components of virtual vector bosons, to the 
inclusive $s$-channel single top quark production cross section became available~\cite{Smith:1996ij}. 
While the Yukawa corrections were found to be never more than 1\% of the LO cross section, the NLO QCD 
corrections were found to be large, of the order of 50\% at both, Tevatron and LHC. Fully differential 
NLO results for $s$-channel single top quark production have been published in 2002~\cite{Harris:2002md}.\\ 
Calculations considering full NLO corrections in production and decay  considering on-shell top quarks (NWA) 
became available in 2004 and 2005~\cite{Campbell:2004ch,Cao:2004ky,Cao:2004ap} and have been extended to taking 
into account finite top quark width effects (resonant top quarks) and hence considering non-factorisable 
corrections using an ET approach~\cite{Falgari:2011qa}.\\

Full NLO corrections for the $Wt$ production appeared in 2002~\cite{Zhu:2002uj}. All calculations for $Wt$ 
production are performed in the 5F scheme and are hence based on the $2\rightarrow 2$ scattering process 
$g b \rightarrow Wt$~\footnote{This notation includes not only $g b \rightarrow W^{-}t$  but also the corresponding 
process for anti-top quark production $g \bar{b} \rightarrow W^{+} \bar{t}$.}, where a $b$-quark ($\bar{b}$-quark 
in case of $W\bar{t}$ production) appears in the initial state. Similar to the 5F NLO calculation of $t$-channel 
single top quark production, possible large logarithms are absorbed in the $b$-PDF.\\ 
At NLO, $Wt$ production interferes with LO top quark pair production and the subsequent decay $\bar{t}\rightarrow W^-\bar{b}$. 
The NLO correction to $Wt$, including $g g \rightarrow t\bar{t}^{*}\rightarrow t\bar{b}W^{-}$ and 
$q\bar{q}\rightarrow t\bar{t}^{*}\rightarrow t\bar{b}W^{-}$, represents a huge correction if the invariant mass of the final 
state $Wb$ system is close to the top mass. Effectively, this undermines the perturbative description of the $Wt$ mode. 
In order to keep the $Wt$ production a well-defined process at NLO a prescription how to deal with this interference is necessary. 
In Ref.~\cite{Zhu:2002uj} the method introduced in~\cite{Tait:1999cf} is followed where the total cross section is modified 
by subtracting a term which effectively removes the $t\bar{t}$-like contribution. In~\cite{Tait:1999cf} it was shown that 
this ansatz yields equal results compared to placing a cut on the invariant mass $M_{\bar{b}W}$ of the $\bar{b}W$ pair, 
$|M_{\bar{b}W}-m_t|>\kappa\Gamma_t$ (as proposed in Ref.~\cite{Belyaev:1998dn}), with $\Gamma_t$ being the width of the 
top quark, if the choice $\kappa\sim 15$ was made.\\ 
In 2005 a fully differential NLO calculation of $Wt$ production and decay in the context of the narrow width approximation 
appeared~\cite{Campbell:2005bb}. In this calculation, the suggestion was made to apply a maximum cut on the transverse 
momentum $p_{\mathrm{T}}$ of the ``spectator $b$'' (not coming from the top quark decay) of $50\;\mbox{GeV/c}$. 
The reasoning for this is, that the probability for two hard $b$-quarks is larger for diagrams with two top quarks 
being on-shell or close to on-shell than for cases where one of the top quarks is highly virtual. Hence, this cut reduces 
the interference of $Wt$ and $t\bar{t}$ production. Using a renormalisation and factorisation scale equal to the maximum 
$p_{\mathrm{T}}$ cut of the additional $b$-quark the NLO corrections are at the LHC of the order of 10\% compared to the 
LO result~\cite{Campbell:2005bb}.\\ 
In 2008 the first calculation of $Wt$ production at NLO QCD accuracy and interfaced with parton showers according to the 
MC@NLO~\cite{Frixione:2003ei,Frixione:2002ik} formalism appeared~\cite{Frixione:2008yi}. Here, two new prescriptions how 
to define $Wt$ production, which are well suited for event generators, are introduced: diagram removal (DR) and diagram 
subtraction (DS). In the DR approach all diagrams in the NLO $Wt$ amplitudes are removed that are doubly resonant 
(in the sense that the intermediate top quark can be on-shell), while in the DS approach the NLO $Wt$ cross section is 
modified by implementing a gauge invariant subtraction term that locally cancels the $t\bar{t}$ contribution. It has been 
demonstrated in~\cite{Frixione:2008yi} that although the DR approach violates gauge invariance this is not a problem in 
practice. The difference in the result is then mainly a measure of the interference between $t\bar{t}$ and $Wt$ production. 
In Ref.~\cite{White:2009yt} it has been shown that in a realistic analysis scenario to  measure $Wt$, kinematic distributions 
obtained in both approaches are in very good agreement demonstrating that interference in the selected phase space is small 
for such analyses.\\

The NLO calculations with stable top quarks and performed in the 5F approach in case of $t$-channel and $Wt$ production have been improved for all three single top quark production modes by considering threshold resummation.\\ 
First next-to-leading logarithmic (NLL) soft gluon corrections have been resummed to all orders of perturbation theory for $s$-, $t$-channel and $Wt$ production at the Tevatron~\cite{Kidonakis:2006bu} and the LHC~\cite{Kidonakis:2007ej}.\\ 
In 2010 and 2011 these calculations have been extended by two independent groups to threshold resummations at NNLL accuracy~\cite{Kidonakis:2010tc,Kidonakis:2010ux,Kidonakis:2011wy,Zhu:2010mr,Wang:2010ue}. The computations of the two groups differ by different resummation formalisms and by different choices for the threshold kinematics variables. By re-expanding the NNLL resummation N. Kidonakis obtains approximate fixed order NNLO cross sections for $s$-~\cite{Kidonakis:2010tc}, $t$-channel~\cite{Kidonakis:2011wy} and $Wt$ production~\cite{Kidonakis:2010ux}. In case of $t$-channel single top quark production not only soft gluon corrections but also collinear corrections were considered in Ref.~\cite{Kidonakis:2011wy} as it was found in Ref.~\cite{Kidonakis:2007ej} that in case of $t$-channel production at the LHC soft gluon resummation alone is not a good approximation. In Ref.~\cite{Kidonakis:2012rm} it is pointed out that the agreement between exact NLO and approximate NLO cross sections (based on LO cross section and NNLL resummation) for all three single-top channels is very good, which serves as the motivation to derive approximate NNLO results. In Ref.~\cite{Zhu:2010mr} the approximate NNLO cross section is given for the $s$-channel as by-product, while otherwise resummed cross sections are stated for $s$-channel~\cite{Zhu:2010mr} and $t$-channel~\cite{Wang:2010ue} production.\\ 
The inclusive cross sections computed by the two groups differ by about 10\% at the Tevatron and at the (1-3)\% level at the LHC.
Both groups observe a decrease of the scale dependence of the single top quark cross sections when resummation effects (NNLL accuracy) are included. Cross sections at LHC energies of 8 TeV have been reported in Ref.~\cite{Kidonakis:2012rm} for all three single top quark production modes.\\
\begin{table}[t]
\centering
\begin{tabular}{l|ccc}
\toprule
         & \multicolumn{3}{c}{cross section [pb]}\\
Collider & $t$-channel  & $Wt$ & $s$-channel\\ \midrule
Tevatron, $\sqrt{s}=1.96\,\mbox{TeV}$ & $2.08^{+0.00}_{-0.04}\pm{+0.12}$  & na & $1.046^{+0.002}_{-0.010}\,^{+ 0.060}_{-0.056}$ \\
\midrule
LHC, $\sqrt{s}=7\,\mbox{TeV}$ & $65.9^{+2.1}_{-0.7}\,^{+1.5}_{-1.7}$ & $15.6\pm 0.4 \pm 1.1$ & $4.56\pm 0.07^{+ 0.18}_{-0.17}$\\
LHC, $\sqrt{s}=8\,\mbox{TeV}$ & $87.2^{+2.8}_{-1.0}\,^{+2.0}_{-2.2}$ & $22.2\pm 0.6\pm 1.4$ & $5.55\pm 0.08\pm 0.21$ \\
LHC, $\sqrt{s}=14\,\mbox{TeV}$ & $243^{+6}_{-2}\,^{+5}_{-6}$ & $83.6\pm 2.0^{+ 3.0}_{-2.8}$ & $11.92\pm 0.19^{+ 0.45}_{-0.49}$ \\
\bottomrule
\end{tabular}
\caption{Approximate NNLO single top quark cross section predictions (based on threshold resummation at NNLL accuracy) for the 
$t$-channel, $s$-channel and $Wt$ production at the LHC and the Tevatron~\cite{Kidonakis:2010tc,Kidonakis:2011wy,Kidonakis:2012rm}. 
The cross sections given are the sum of $t$ and $\bar{t}$ production.
All numbers are computed using a top quark mass of $m_t=173\,\mbox{GeV}$ and the MSTW2008nnlo90cl pdf set~\cite{Martin:2009iq}. 
The first uncertainty indicates the uncertainty due to a variation of the scales between $m_t/2<\mu<2m_t$ and the second 
uncertainty is due to parton distribution functions (PDF) and is assessed using the MSTW2008 NNLO PDF sets at 90\% C.L.}
\label{tab:singletopAproxNNLO}
\end{table}
In Table~\ref{tab:singletopAproxNNLO} a summary of the approximate NNLO single top quark cross section predictions (based on threshold resummation at NNLL accuracy) for the $t$-channel, $s$-channel and $Wt$ production at the LHC and the Tevatron~\cite{Kidonakis:2010tc,Kidonakis:2011wy,Kidonakis:2012rm} is presented.\\

In 2014 parts of the full NNLO calculations for $t$-channel single top quark production (stable top quarks, 5F scheme) became available~\cite{Brucherseifer:2014ama,Assadsolimani:2014oga}. In Ref.~\cite{Brucherseifer:2014ama} the NNLO calculation is restricted to the vertex corrections and the related real corrections, meaning that corrections to the light- and heavy-quark lines are computed separately and that dynamical cross-talk between the two is neglected. The authors argue, that the neglected contribution, which appears at NNLO for the first time, is colour-suppressed and is expected to be subdominant. Within this approximation it is found in Ref.~\cite{Brucherseifer:2014ama} that the NNLO QCD corrections increase the NLO result by a few percent. Even if a cut on the transverse momentum of the top quark is applied the NNLO corrections stay small, in contrast to the case of NLO corrections.\\ 
As a further ingredient towards the full NNLO calculation the complete reduction of the two-loop amplitudes to a small set of master integrals has been presented in Ref.~\cite{Assadsolimani:2014oga}  for $t$-channel single top quark production.

\subsection{Event Signature of Top Quark Events}
\label{sec:topsignature}

The lifetime of the top quark is with $\tau_t \approx 0.5\cdot 10^{-24}\,\mbox{s}$~\cite{Agashe:2014kda} by far too short 
to produce a direct signal in the detector. Only the signals of the decay products can be observed in the detector. 
In the SM the top quark decays with almost 100\% via $t\rightarrow bW^{+}$ ($\bar{t}\rightarrow \bar{b}W^{-}$). 
The $b$-quark evolves into a jet of hadrons containing at least one $b$-hadron. $b$-hadrons have a mean lifetime with a 
$c\tau$ value of about $0.5\;\mbox{mm}$~\cite{Agashe:2014kda} and produce a secondary vertex that can be well reconstructed 
with silicon vertex detectors. The $W$-boson is also unstable ($\tau_W\approx 3.1\cdot 10^{-25}\mbox{s}$~\cite{Agashe:2014kda}~\footnote{The width $\Gamma_{W}$ of the $W$-boson is taken from~\cite{Agashe:2014kda} and converted into the the $W$-boson lifetime via $\tau = \hbar / \Gamma$.}) and decays either leptonically ($W^{\pm}\rightarrow \ell^{\pm} \nu$ with $\ell = e,\mu,\tau$) or hadronically ($W^+\rightarrow q\bar{q}'$ and $W^-\rightarrow \bar{q}q'$, $q=u,c$, $q'=s,d$). The quarks in the final state evolve into jets of hadrons.\\ 

In case of $t\bar{t}$ production the following three decay modes are distinguished~\cite{Agashe:2014kda}:
\begin{enumerate}
\item{\textit{All-hadronic channel (all-jets):}\\[0.1cm] 
\begin{tabular}{p{10cm}l}
$t\bar{t} \rightarrow W^{+}b\, W^{-}\bar{b} \rightarrow q\bar{q}' b \, q''\bar{q}'''\bar{b}$ & $BR=45.7\,\%$
\end{tabular}
}
\item{\textit{Lepton+jets channel ($\ell$+jets):}\\[0.1cm]  
\begin{tabular}{p{10cm}l}
$t\bar{t} \rightarrow W^{+}b\, W^{-}\bar{b} \rightarrow \ell^{+}\nu_{\ell} b \, q\bar{q}'\bar{b} \; + \;  q''\bar{q}''' b\, \ell^{-} \bar{\nu}_{\ell} \bar{b}$& $BR=43.8\,\%$
\end{tabular}
}
\item{\textit{Dilepton channel ($\ell\ell$):}\\[0.1cm]  
\begin{tabular}{p{10cm}l}
$t\bar{t} \rightarrow W^{+}b\, W^{-}\bar{b} \rightarrow \ell^{+}\nu_{\ell} b \, \ell^{-} \bar{\nu}_{\ell} \bar{b}$ & $BR=10.5\, \%$
\end{tabular}
}
\end{enumerate}
Their relative contributions~\cite{Agashe:2014kda} (branching ratios, indicated as $BR$) including hadronic corrections and assuming lepton universality are given. While in the above processes $\ell$ refers to $e,\mu,\tau$, most of the results to date rely on the $e$ and $\mu$ channels. 
The dilepton channel has small background (mainly Drell-Yan production in association with jets, $Z/ \gamma^{*}$+jets) but 
a small $BR$. The all-hadronic channel has a large $BR$ but suffers of a huge multi-jet background. As in the $e/\mu$+jets 
channel the background (mainly $W$-boson production in association with jets, $W$+jets, and multi-jet production) is moderate
and as the $BR$ is relative high most top quark property measurements to date employ this channel.\\

In case of single top quark $t$- and $s$-channel production only the lepton+jets channel 
($t\rightarrow W^+ b \rightarrow \ell^{+}\nu_{\ell} b$ and $\bar{t}\rightarrow W^- \bar{b} \rightarrow \ell^{-}\bar{\nu}_{\ell} \bar{b}$, 
$BR=32.7\%$~\cite{Agashe:2014kda}) has been considered and again almost all analyses employ only the $e$ and $\mu$ channels. 
The dominant backgrounds in the $e/\mu$+jets channel are $W$+jets production and top quark pair production 
(in particular at the LHC). In case of the single top quark $Wt$ production most analyses employ the dilepton channel 
($tW^{-}\rightarrow W^{+}b\, W^{-}\rightarrow \ell^{+}\nu_{\ell} b \, \ell^{-} \bar{\nu}_{\ell}$ and 
$\bar{t}W^{+}\rightarrow W^{-}\bar{b}\, W^{+}\rightarrow \ell^{-} \bar{\nu}_{\ell} \bar{b} \, \ell^{+}\nu_{\ell} $, 
$BR=10.5\%$~\cite{Agashe:2014kda}). So far, only measurements employing $e$ and $\mu$ are performed and the largest background 
in the channels is due to top quark pair production followed by $Z / \gamma^{*}$+jets production.


\section{Charge Asymmetry in Top Quark Pair Production}
\label{sec:asymmetry}

In this section a short introduction to the effect of the charge asymmetry in top quark pair production 
is given and the differences at the $p\bar{p}$ collider Tevatron and the $pp$ collider LHC are pointed out. 
Then, the measurements at the Tevatron experiment are discussed, where the first CDF measurement~\cite{Aaltonen:2008hc} 
is used
to explain the general ana\-ly\-sis strategy at the $p\bar{p}$ collider Tevatron. After that, the measurements at the
LHC are described using the CMS measurements~\cite{Chatrchyan:2011hk,Chatrchyan:2012cxa} for explaining the 
general analysis strategy at the $pp$ collider LHC.

\subsection[The Effect of the Charge Asymmetry in $t\bar{t}$ Production in the SM]{The Effect of the Charge Asymmetry in \boldmath{$t\bar{t}$} Production in the SM}
\label{sec:asy_theory}
In the SM, at leading order the relevant processes for top quark pair production, $q\bar{q}\rightarrow t\bar{t}$ and $gg\rightarrow t\bar{t}$, 
do not discriminate between the final top quark and anti-top quark, thus predicting identical differential distributions for both quarks. 
However, radiative corrections involving either virtual or real gluon emission lead to a difference between the kinematics of top quarks
and anti-top quarks. In the literature this effect, which occurs first at $O(\alpha_s^3)$, is known as ``charge asymmetry'' and has
been first predicted in 1987 for $t\bar{t}g$ production~\cite{Halzen:1987xd} and in 1998 also for the full inclusive $t\bar{t}+X$ 
production by J.~H.~K\"uhn and G. Rodrigo~\cite{Kuhn:1998jr,Kuhn:1998kw}. Please note, that although differential distributions of 
top quark and anti-top quark are different, the entire process of top quark pair production which is induced by the strong interaction 
is invariant under charge conjugation.\\ 

\begin{figure}[htb]
\subfigure[]{
\includegraphics[width=0.23\textwidth]{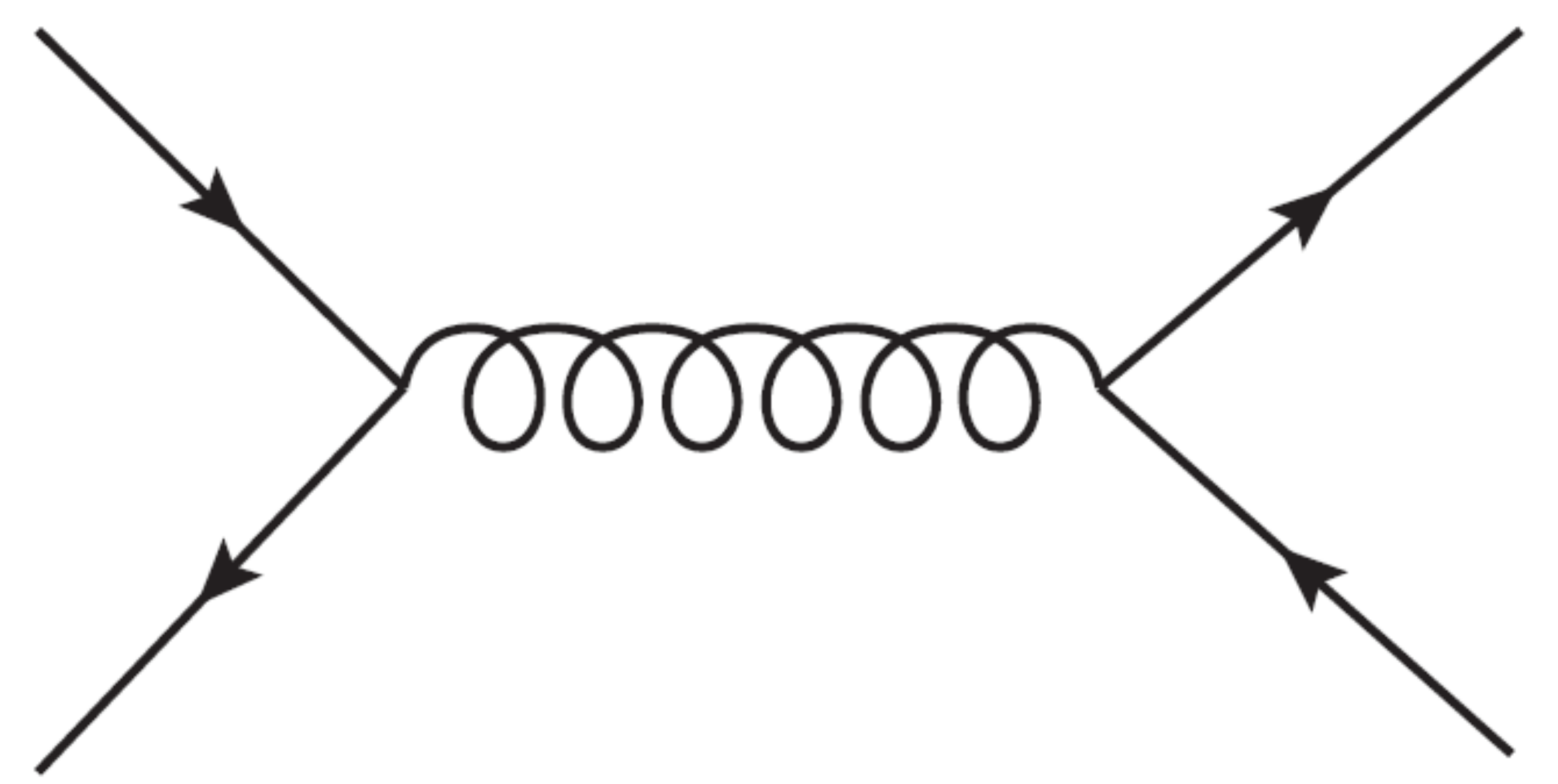}
}
\subfigure[]{
\includegraphics[width=0.23\textwidth]{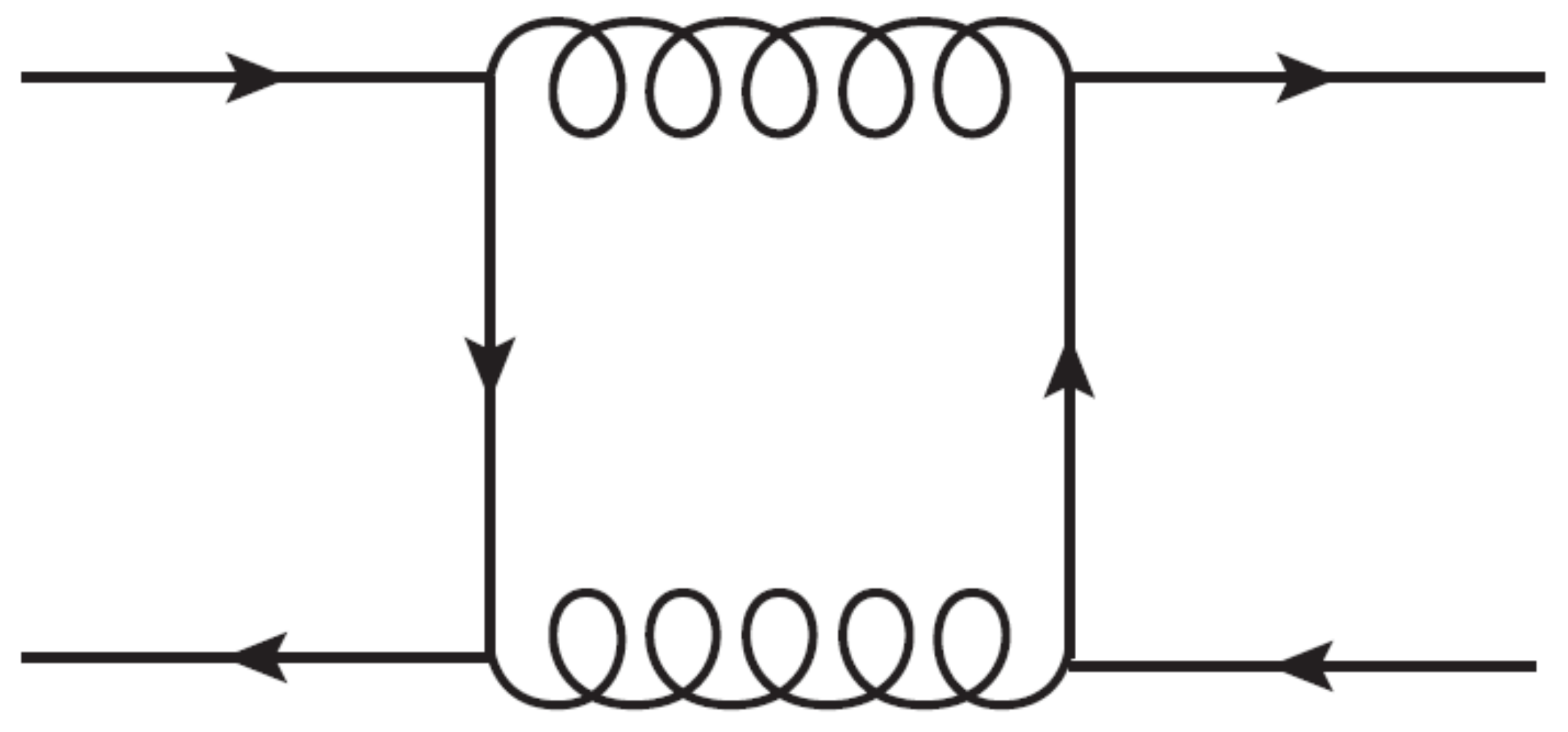}
}
\subfigure[]{
\includegraphics[width=0.23\textwidth]{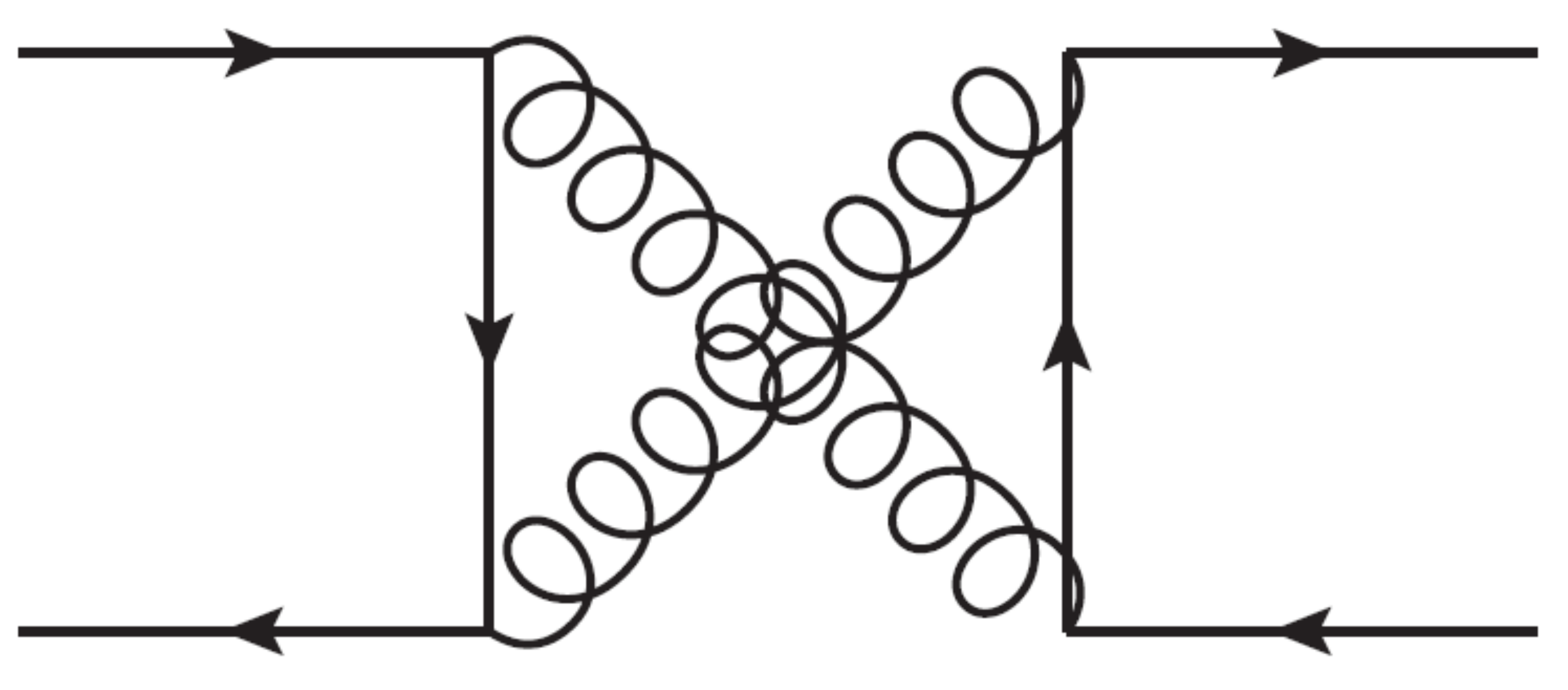}
}
\hspace{4cm}
\subfigure[]{
\includegraphics[width=0.23\textwidth]{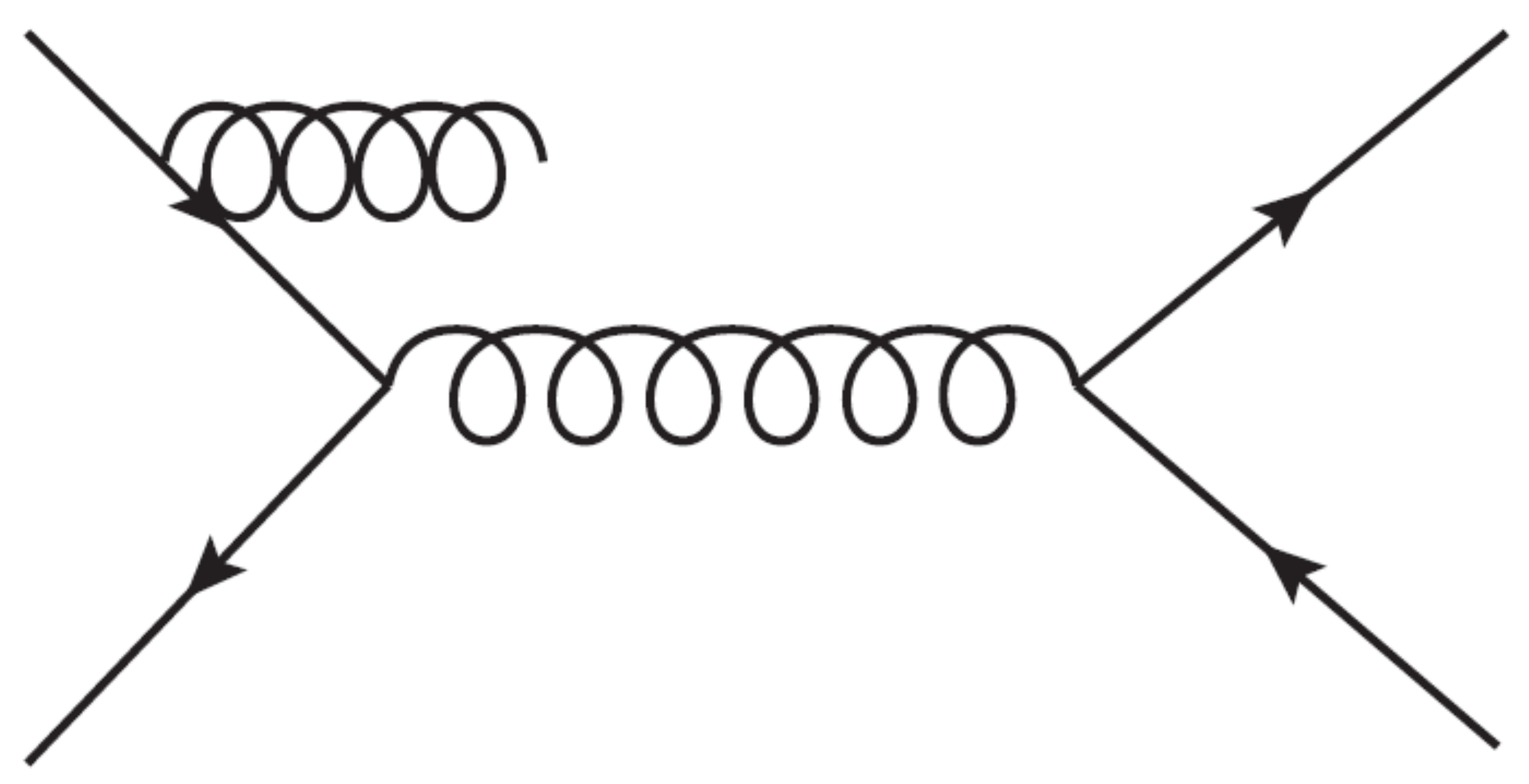}
}
\subfigure[]{
\includegraphics[width=0.23\textwidth]{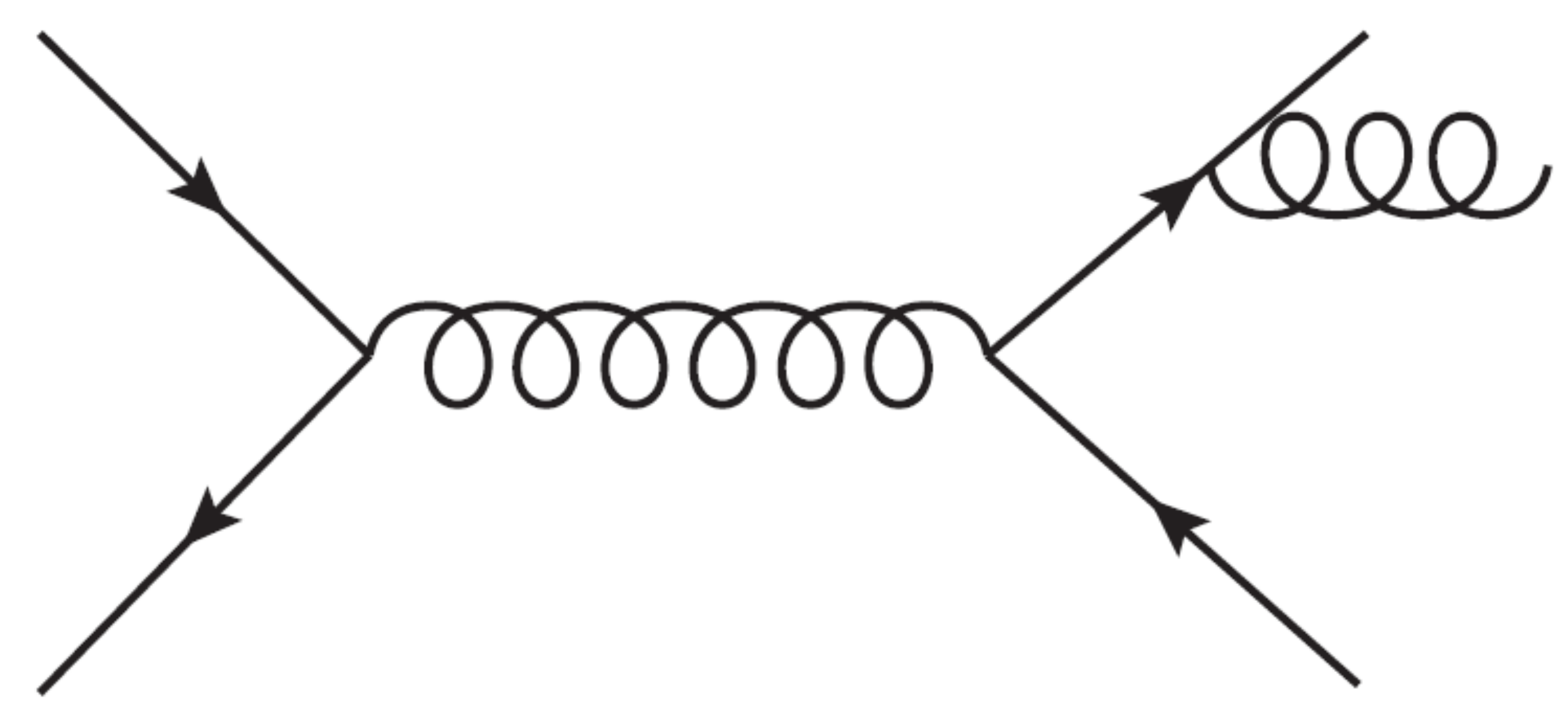}
}
\subfigure[]{
\includegraphics[width=0.23\textwidth]{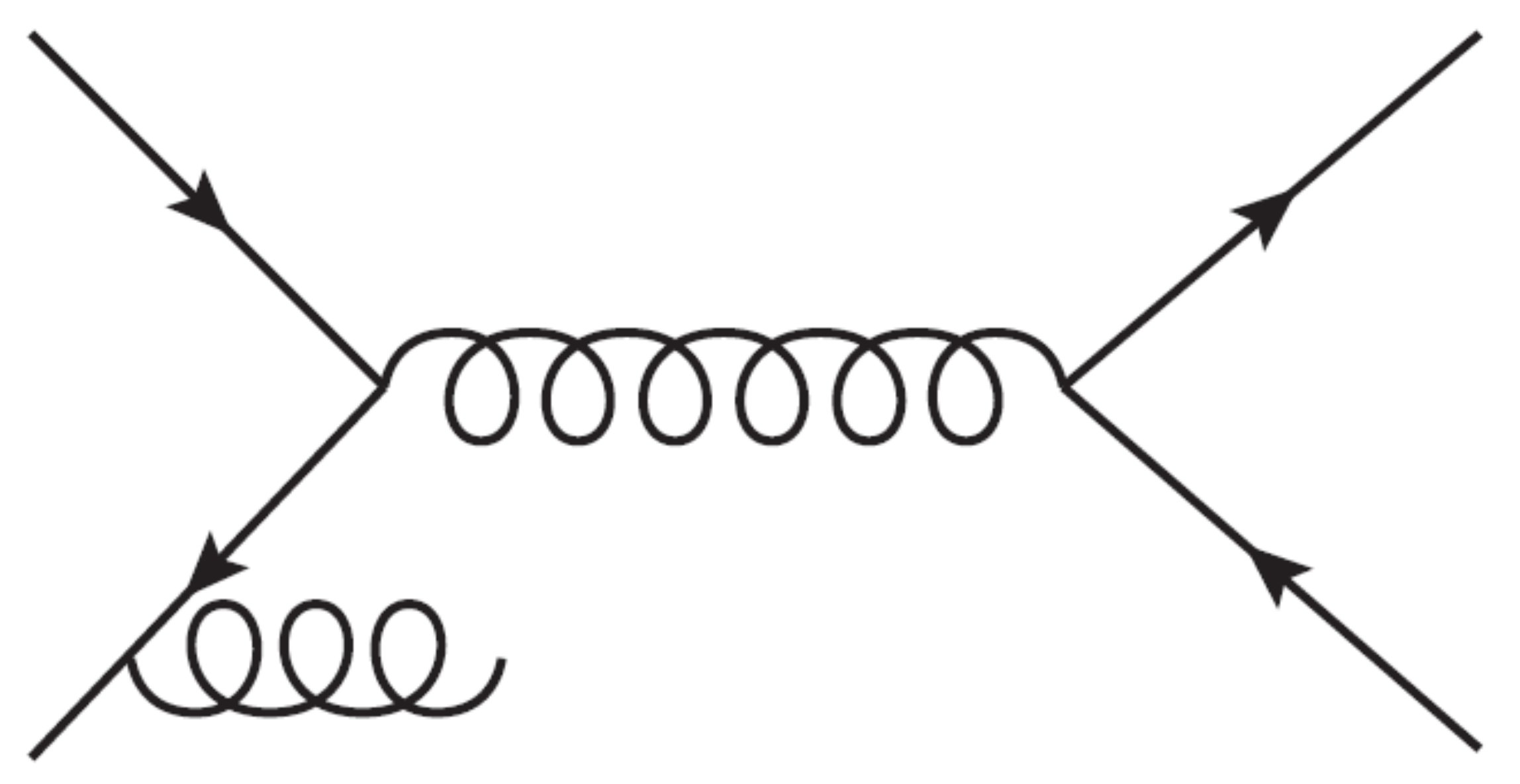}
}
\subfigure[]{
\includegraphics[width=0.23\textwidth]{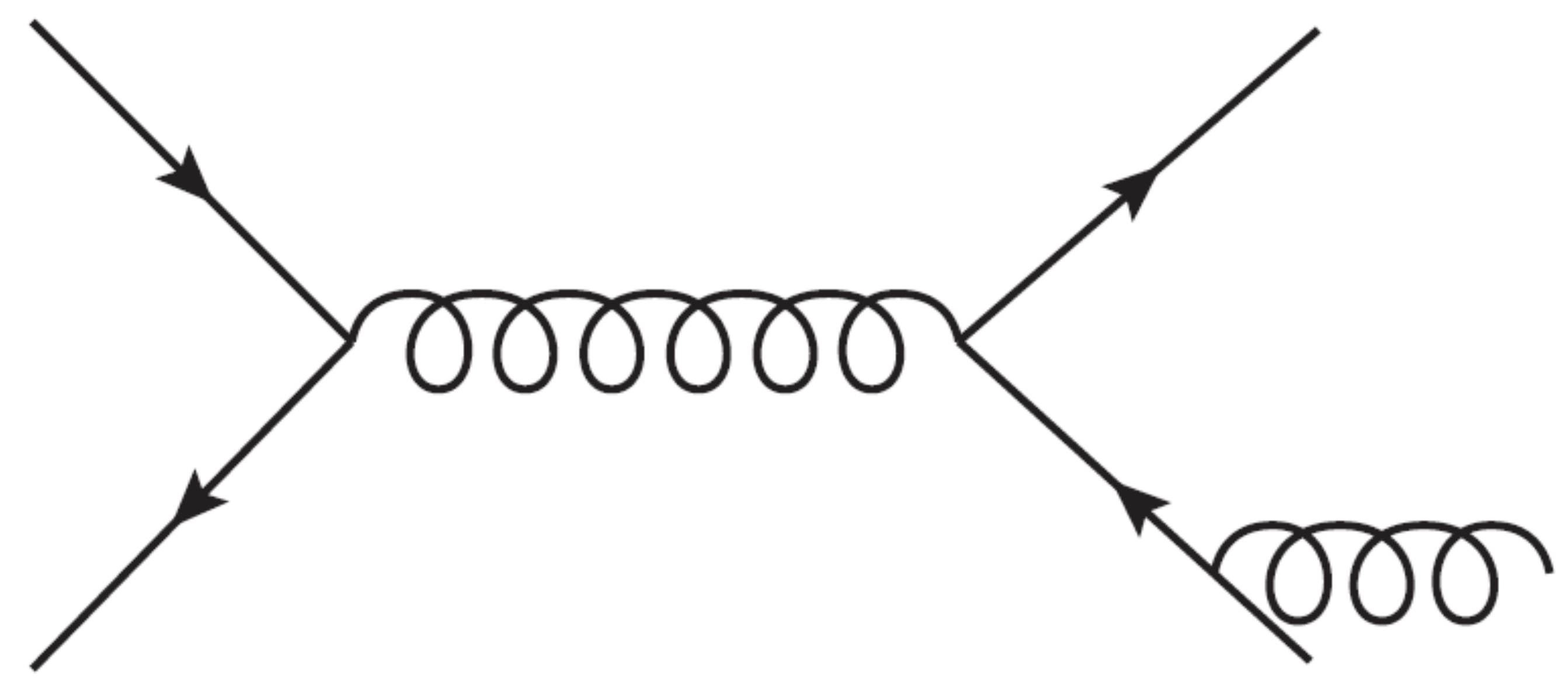}
}
\caption{Feynman diagrams originated by quark anti-quark annihilation that contribute to the QCD charge asymmetry in top quark pair 
production. The sketches are taken from~\cite{Ferrario2010}.}
\label{fig:Ac_qq_diagrams}
\end{figure}

As pointed out in~\cite{Kuhn:1998kw}, the charge asymmetry in QCD arises in analogy to the one in QED and can be traced to the interference
between amplitudes which are relatively odd under the exchange of $t$ and $\bar{t}$. The dominant contribution to the charge asymmetry in 
top quark pair production is due to radiative corrections to the quark anti-quark annihilation ($q\bar{q}$) process given by the interference 
of final-state (FS) with initial-state (IS) real gluon emission (Fig.~\ref{fig:Ac_qq_diagrams} (d+f) $\otimes$ (e+g)) and the interference of 
the double virtual gluon exchange with the Born amplitude (Fig.~\ref{fig:Ac_qq_diagrams} (a) $\otimes$ (b+c)). The charge asymmetry from the 
two different contributions has opposite sign and the latter one is positive and always larger than the former one.\\ 
In~\cite{Kuhn:1998kw} it has been shown that the interference between different graphs of $qg$ processes generates a contribution to the 
asymmetry which is much smaller than the one induced by $q\bar{q}$ processes. At the Tevatron the asymmetry from the $qg$ process is negligible, 
and at the LHC, it enhances the asymmetry in suitably chosen kinematic regions~\cite{Kuhn:1998kw}.\\
Gluon fusion processes do not induce a charge asymmetry as well as $q\bar{q}$ or $qg$ induced non-Abelian contributions~\cite{Kuhn:1998kw}, 
e.g. processes involving the triple gluon coupling.\\ 

Electroweak (EW) and mixed QCD-QED/EW contributions to the charge asymmetry have been carefully studied in Ref.~\cite{Hollik:2011ps}: 
At $O(\alpha^2)$ the interference of $q\bar{q}\rightarrow \gamma\rightarrow t\bar{t}$ and $q\bar{q}\rightarrow Z \rightarrow t\bar{t}$ 
yields a small contribution to the charge asymmetry, while its contribution to the total cross section is negligible. Mixed QCD-QED 
corrections involving the interference of $q\bar{q}\rightarrow g\rightarrow t\bar{t}$ and the Box diagram (+ crossed graph) containing 
one photon and one gluon, the interference of $q\bar{q}\rightarrow \gamma\rightarrow t\bar{t}$ and the Box diagram containing two gluons 
and the interference of $q\bar{q}\rightarrow g\rightarrow t\bar{t}g$ and $q\bar{q}\rightarrow \gamma \rightarrow t\bar{t}g$, yield a 
sizable contribution to the charge asymmetry. In contrast, mixed QCD-EW corrections, where the photon is replaced by the $Z$-boson, 
give only a small contribution to the asymmetry.\\ 

At the time my work on the measurement of the charge asymmetry started, the mixed QCD-QED/EW contribution to the charge asymmetry 
was estimated to increase the QCD induced charge asymmetry by a factor 1.09~\cite{Kuhn:1998kw}. In 2011 this contribution has been 
carefully investigated by W. Hollik and D. Pagani by evaluating all partonic channels that produce an asymmetry both at tree level 
and at NLO. This resulted in an enhancement factor of the QCD induced charge asymmetry of 1.2~\cite{Hollik:2011ps} and has been 
confirmed shortly afterwards in Ref.~\cite{Kuhn:2011ri}.\\ 

At partonic level, the differential charge asymmetry $A^i(\cos \hat{\theta})$ induced by 
$i=q\bar{q}, qg$ processes is defined as~\cite{Kuhn:1998kw}:
\begin{equation}
A^i(\cos \hat{\theta})=\frac{N_{t}(\cos \hat{\theta})-N_{\bar{t}}(\cos \hat{\theta})}{N_{t}(\cos \hat{\theta})+N_{\bar{t}}(\cos \hat{\theta})}
\end{equation}
with $\hat{\theta}$ being the angle between the produced top or anti-top quark and 
the incoming quark (not anti-quark) in the partonic rest-frame, and $N_{t(\bar{t})}$ is the number of 
top (anti-top) quarks produced with a certain angle $\hat{\theta}$.\\
 
As the strong interaction is invariant under charge conjugation, implying 
$N_{t}(\cos\hat{\theta})=N_{\bar{t}}(\cos$ $(\pi-\hat{\theta}))=N_{\bar{t}}(-\cos\hat{\theta})$ for 
the QCD induced $q\bar{q}$ process, the charge asymmetry induced by the $q\bar{q}$ process 
$A^{q\bar{q}}(\cos\hat{\theta})$ is equivalent to a forward-backward asymmetry $A_{\mathrm{FB}}$ of top quarks. 
So, top quarks are preferentially produced in the direction of the initial quark, 
while anti-top quarks are preferentially produced in the direction of the initial anti-quark. 
This behaviour is reflected in a prominent forward-backward asymmetry of 
$A^{q\bar{q}}(\cos\hat{\theta})$, depending linearly on the production angle 
$\cos\hat{\theta}$ for $\sqrt{\hat{s}}=400\,\mbox{GeV}$ with mild modifications 
towards larger $\sqrt{\hat{s}}$. In contrast, the differential partonic 
asymmetry $A^{qg}(\cos\hat{\theta})$ induced by $qg$ processes does not exhibit 
a marked forward-backward asymmetry~\cite{Kuhn:1998kw}.\\

The integrated charge asymmetry, also referred to as inclusive charge asymmetry, is given by~\cite{Kuhn:1998kw}:
\begin{equation}
A=\frac{N_{t}(\cos\hat{\theta}\ge 0)-N_{\bar{t}}(\cos\hat{\theta}\ge 0)}{N_{t}(\cos\hat{\theta}\ge 0)+N_{\bar{t}}(\cos\hat{\theta}\ge 0)}
\label{eq:AC}
\end{equation}
The $q\bar{q}$ induced integrated partonic asymmetry (normalised to the Born cross section) has its maximum at a partonic 
centre-of-mass energy of around $\sqrt{\hat{s}}\approx 500\,\mbox{GeV}$ and ranges between (6-8.5)\% for $\sqrt{\hat{s}}$ 
between 350 GeV and 2 TeV~\cite{Kuhn:1998kw}. The asymmetric cross section of the $qg$ process $\sigma_{\mathrm{A}}^{qg}$ vanishes 
for $\sqrt{\hat{s}}$ below 500 GeV and at about 600 GeV the asymmetric cross section of the $q\bar{q}$ process $\sigma_{\mathrm{A}}^{q\bar{q}}$ 
is about 300 times larger than $\sigma_{\mathrm{A}}^{qg}$. With larger $\sqrt{\hat{s}}$ the dominance of $\sigma_{\mathrm{A}}^{q\bar{q}}$ is reduced, 
but $\sigma_{\mathrm{A}}^{q\bar{q}}$ is still more than a factor of 5 larger than $\sigma_{\mathrm{A}}^{qg}$ at partonic centre-of-mass energies as 
large as 2 TeV~\cite{Kuhn:1998kw}. Even at the LHC with $\sqrt{s}=7\,\mbox{TeV}$ or 8 TeV  the largest bulk of top quark pair 
events are produced well below $\sqrt{\hat{s}}=2\,\mbox{TeV}$.\\ 

Due to the asymmetric initial state of proton anti-proton collisions this partonic charge asymmetry results at the Tevatron 
in an observable forward-backward asymmetry. Protons (anti-protons) constitute of valence quarks, $uud$ ($\bar{u}\bar{u}\bar{d}$), 
and sea quarks. As mostly, only valence quarks have a large enough proton momentum fraction to produce top quark pairs at the Tevatron, 
the proton (anti-proton) direction is mostly identical to the quark (anti-quark) direction. Therefore, the fact that in the by far 
dominant $q\bar{q}$ process top (anti-top) quarks are preferentially produced in the direction of quarks (anti-quarks) yields at 
the $p\bar{p}$ collider Tevatron top (anti-top) quarks preferentially produced in the direction of the proton (anti-proton).\\ 
For the Tevatron with a centre-of-mass energy of $1.96\;\mbox{TeV}$ an inclusive charge asymmetry in the laboratory frame ($p\bar{p}$-frame) 
of $A_{\mathrm{FB}}^{p\bar{p}}=(4.6-5.5)\%$ has been predicted in 1998~\cite{Kuhn:1998kw}, which is reduced by about 30\% compared to the 
partonic asymmetry as events where in the laboratory frame both $t$ and $\bar{t}$ quarks are produced with positive or negative 
rapidities do not contribute to the asymmetry~\cite{Antunano:2007da}. To estimate the uncertainty on the charge asymmetry different 
choices of the structure function and different choices of the factorisation and renormalisation scale ranging between $\mu=m_t/2$ and 
$\mu_t=2m_t$, have been considered. Furthermore, an enhancement factor of 1.09 for mixed QCD-QED/EW contributions has been included 
in the calculation. It is emphasised in~\cite{Kuhn:1998kw} that the numerator and denominator in the charge asymmetry formula Eq.(~\ref{eq:AC}), 
are evaluated in leading order (LO). As the asymmetry occurs first at NLO $t\bar{t}X$ production and thus in LO, there exist no
NLO corrections to the numerator (which would require NNLO $t\bar{t}X$).
The authors argue that it gives a more reliable result, when in the denominator also the LO prediction for the 
total $t\bar{t}$ cross section is used although it is known that NLO corrections to the total $t\bar{t}$ cross section are of the order 
of 30\%. The authors also state that from a more conservative point of view an uncertainty of around 30\% has to be assigned to the 
prediction of the asymmetry.\\

\begin{figure}[tb]
\subfigure[]{
\includegraphics[width=0.23\textwidth]{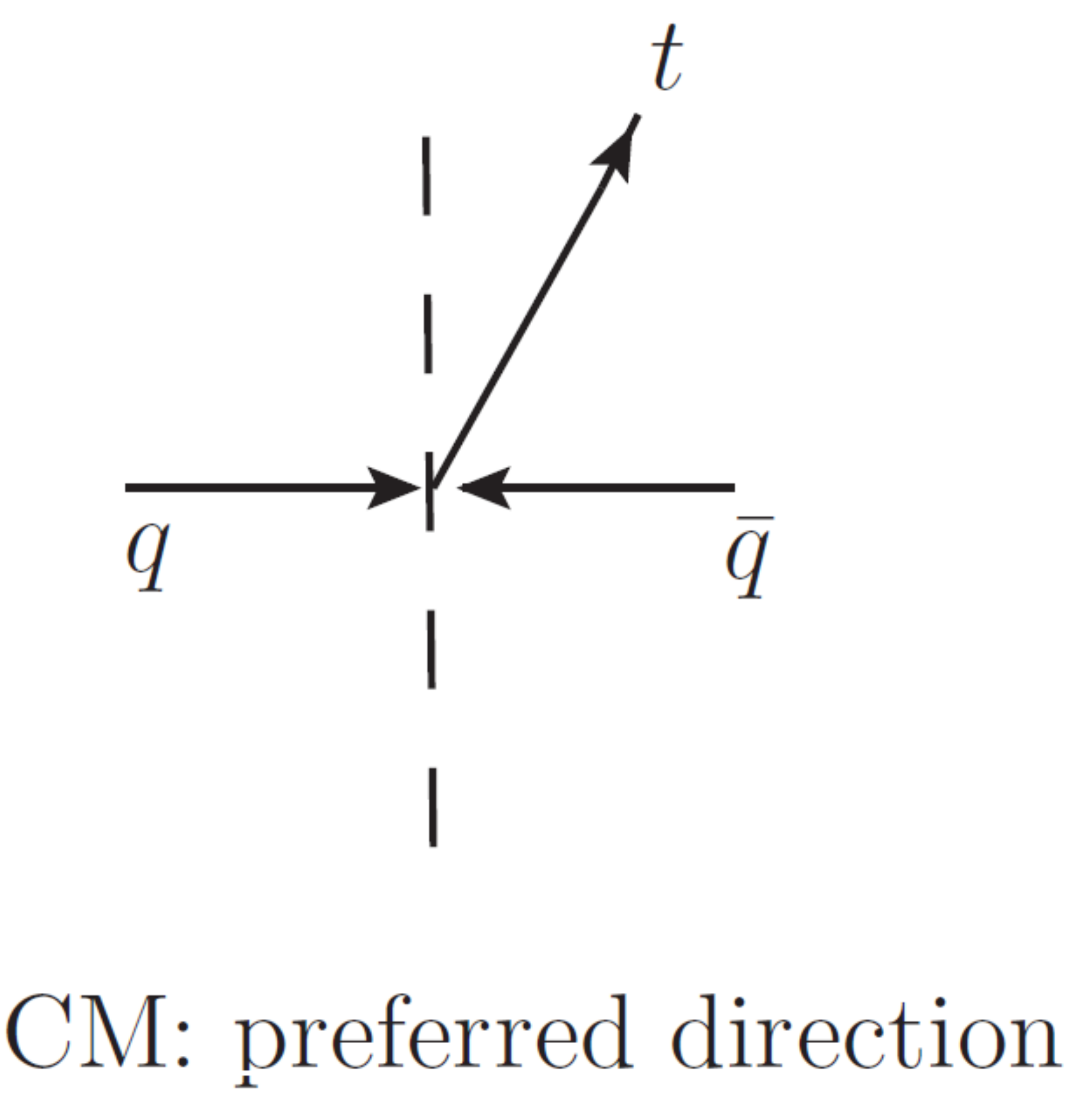}
}
\subfigure[]{
\includegraphics[width=0.23\textwidth]{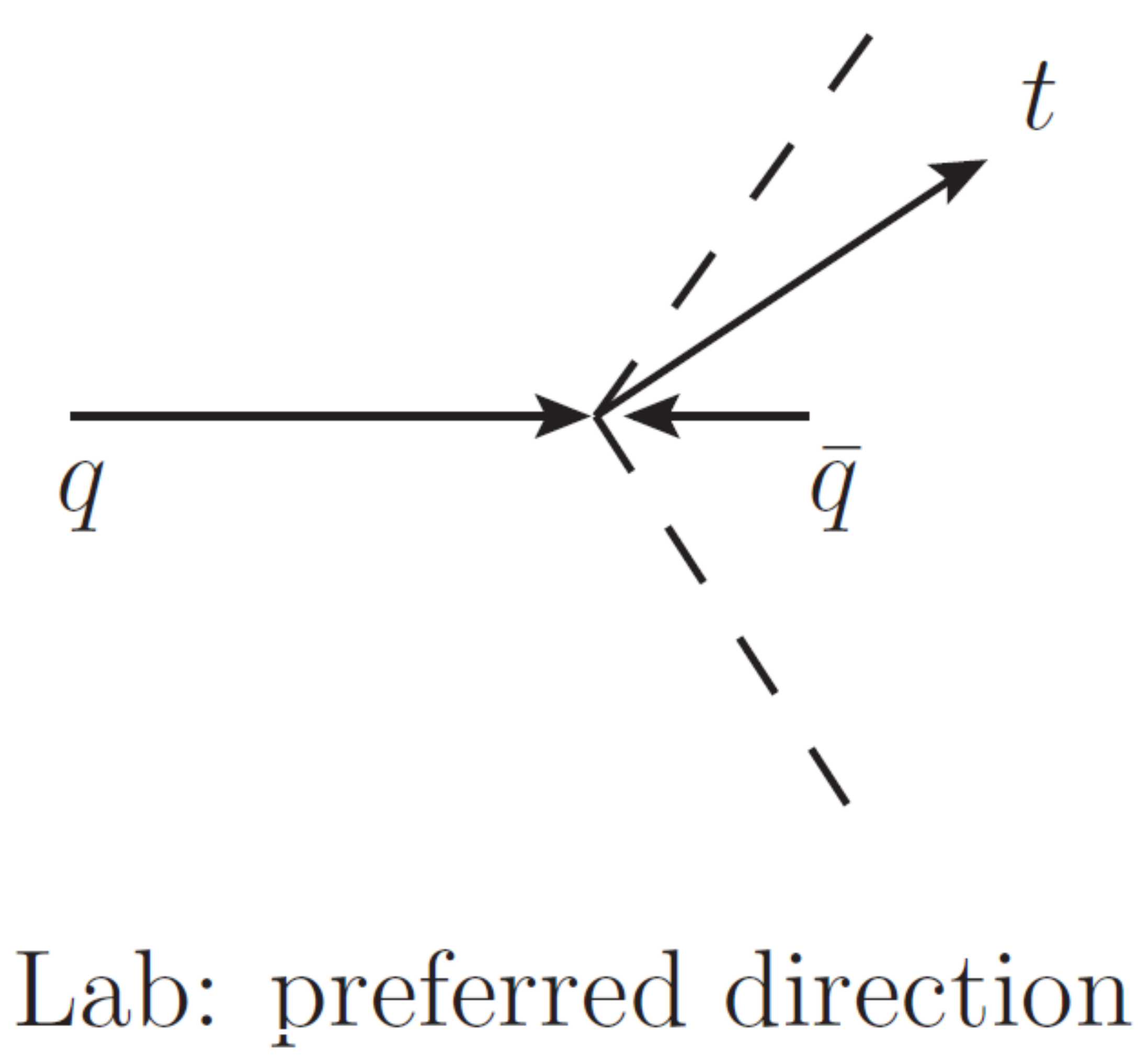}
}
\subfigure[]{
\includegraphics[width=0.35\textwidth]{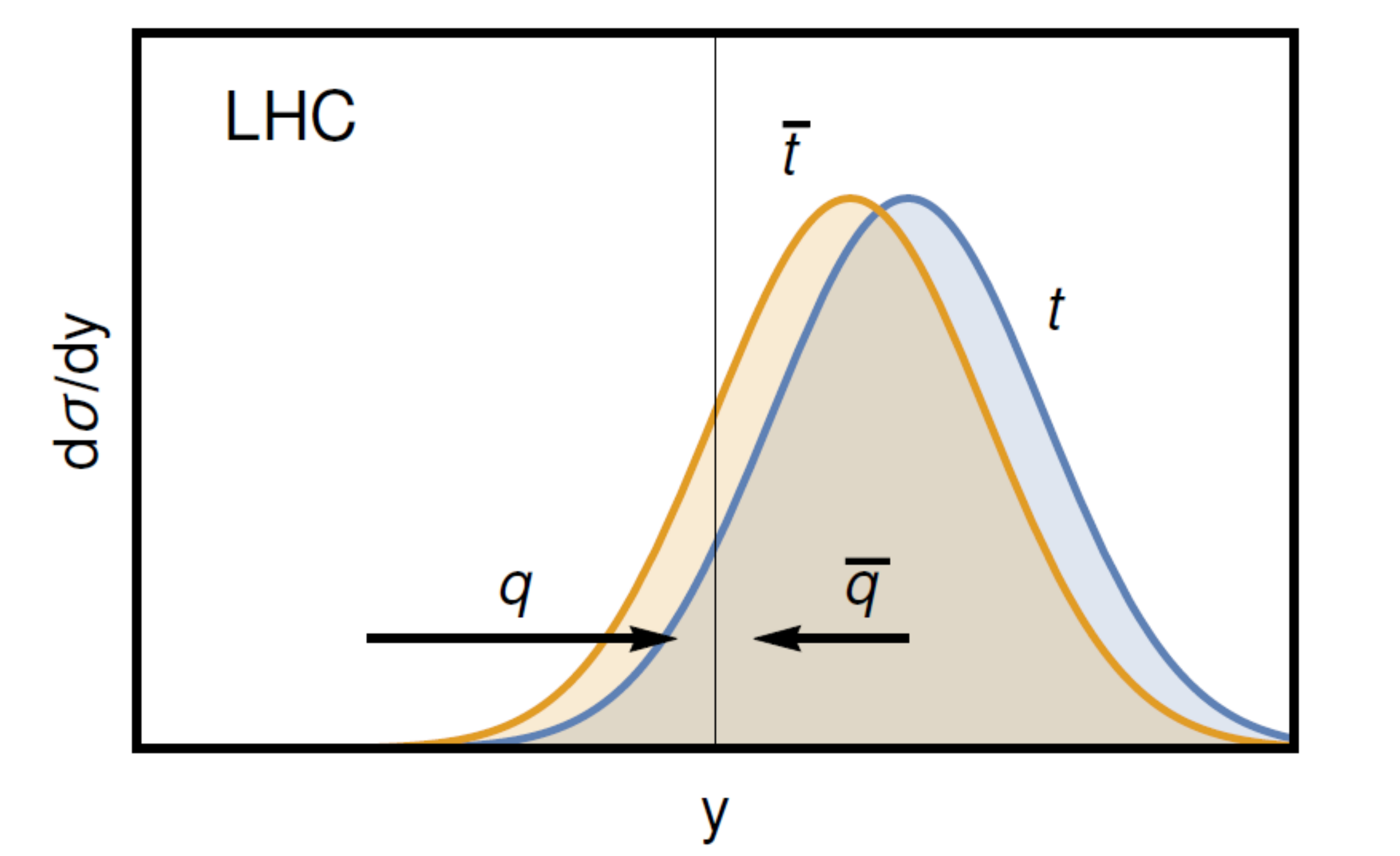}
}
\hspace{3cm}
\subfigure[]{
\includegraphics[width=0.23\textwidth]{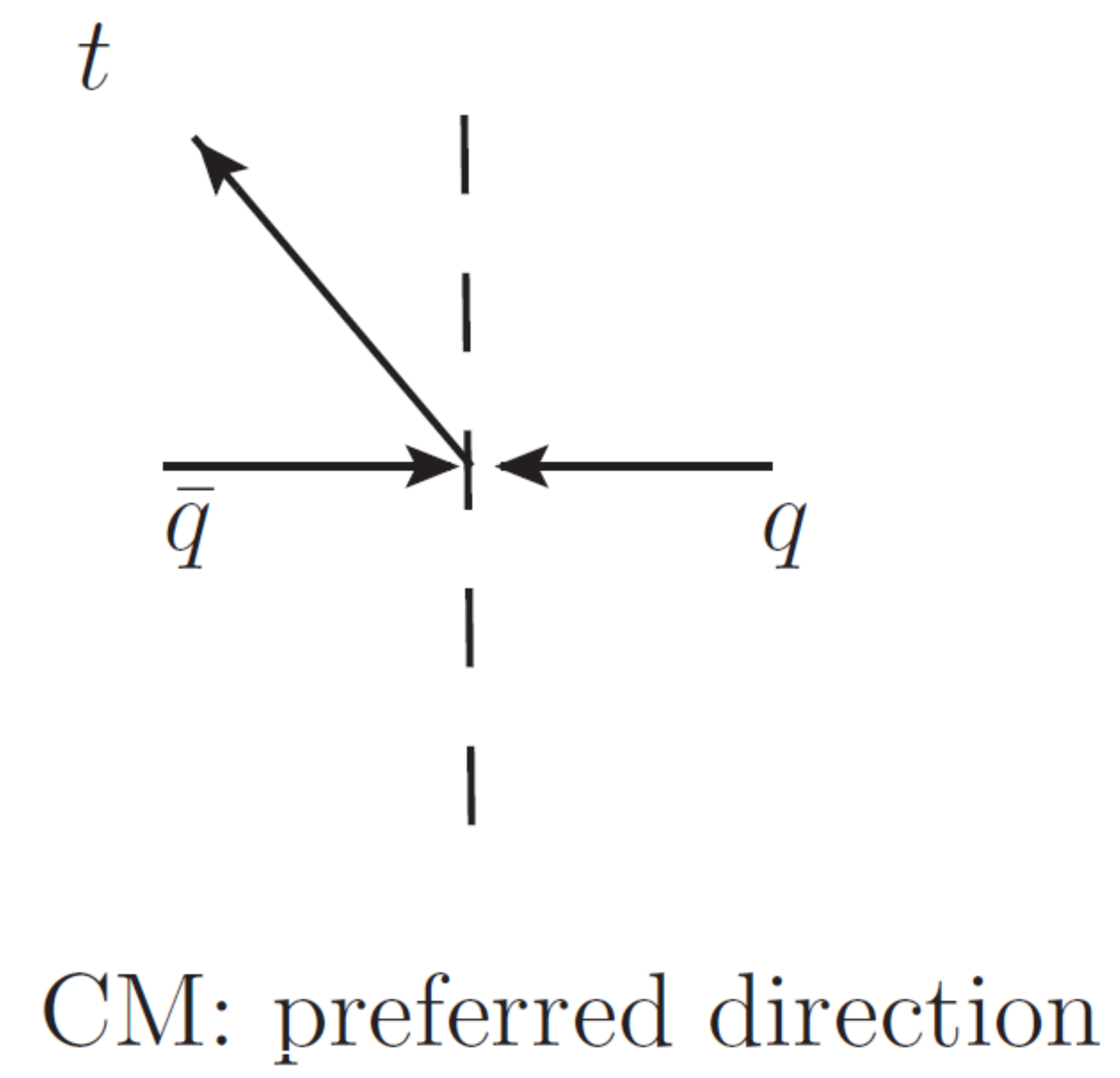}
}
\subfigure[]{
\includegraphics[width=0.23\textwidth]{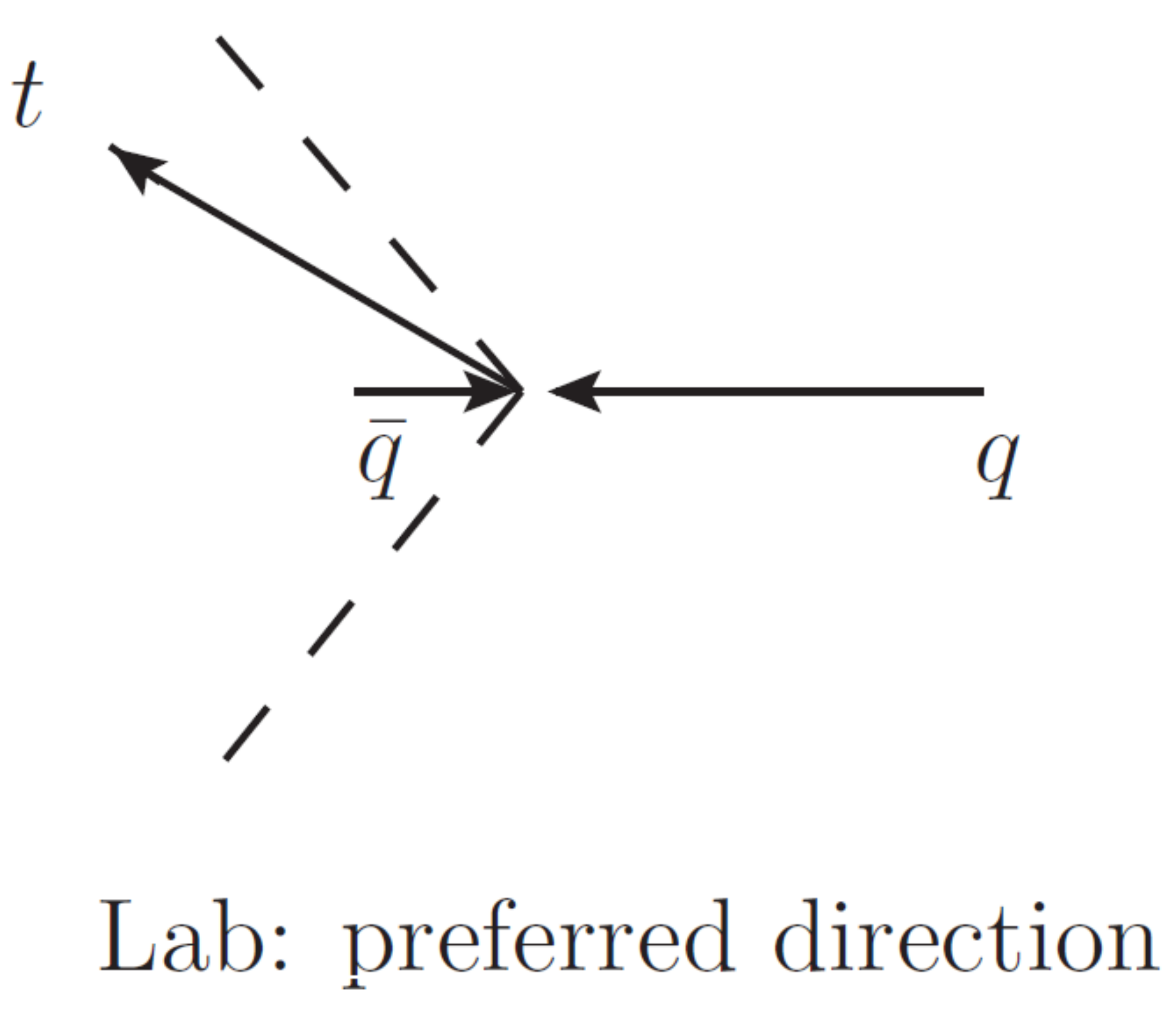}
}
\subfigure[]{
\includegraphics[width=0.35\textwidth]{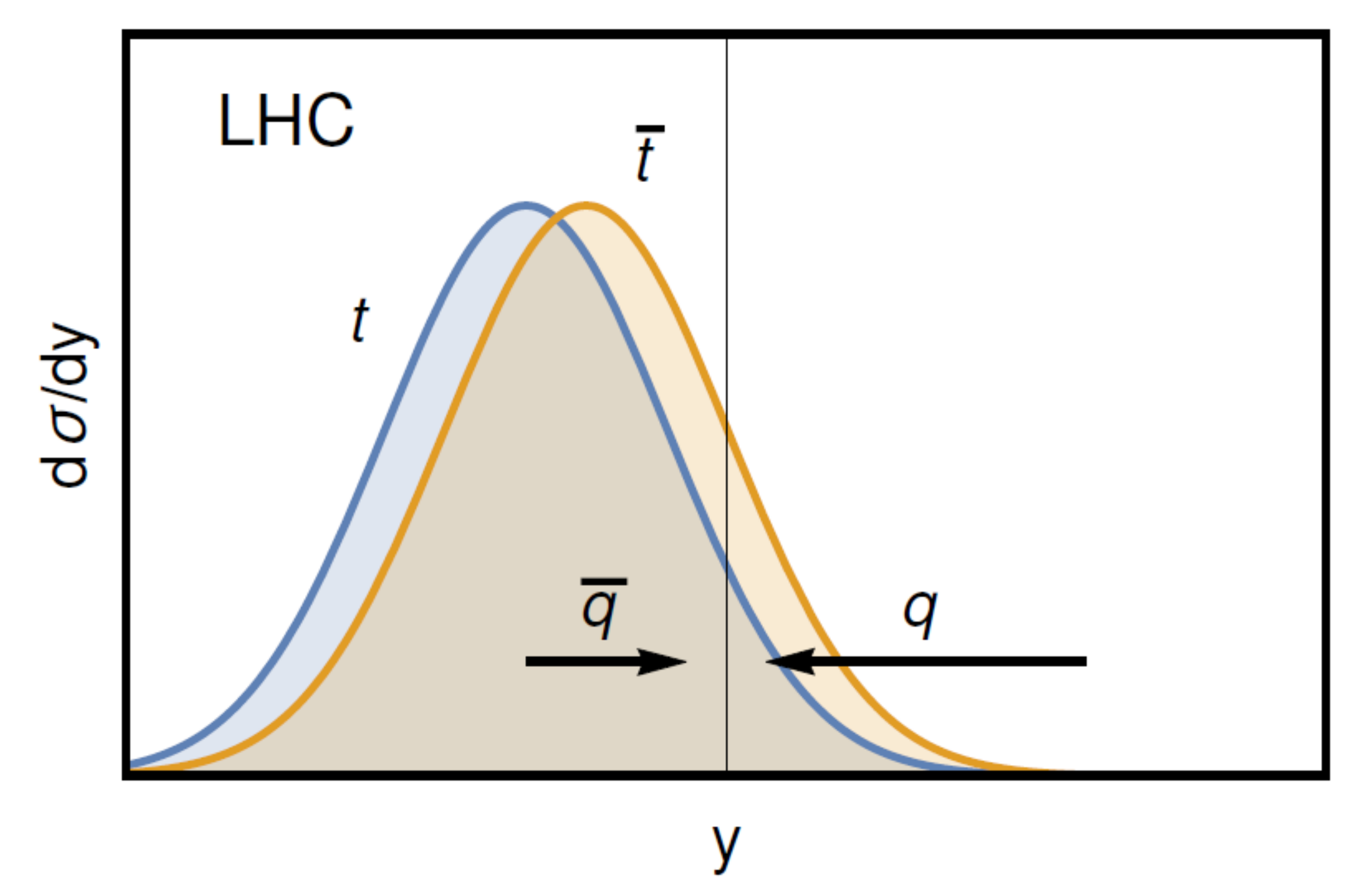}
}
\hspace{3cm}
\caption{Illustration of the origin of the QCD induced charge asymmetry at the $pp$ collider LHC. (a)+(d): Charge asymmetry in the 
centre-of-mass $q\bar{q}$ rest frame (CM). Top quarks are preferentially produced in the direction of the incoming quark (a,d), 
anti-top quarks are preferentially produced in the direction of the incoming anti-quark (not shown). (b)+(c) and (e)+(f): After a 
boost into the laboratory frame, the top quarks are more likely produced in forward (b,c) or backward (e,f) direction. In (c,f) y 
is the rapidity of the top quark in the laboratory frame. The sketches are taken from~\cite{Ferrario2010} and ~\cite{Kuehn:2014rla}.}
\label{fig:AcLHC}
\end{figure}
\begin{figure}[tb]
\begin{center}
\subfigure[]{
\includegraphics[width=0.38\textwidth]{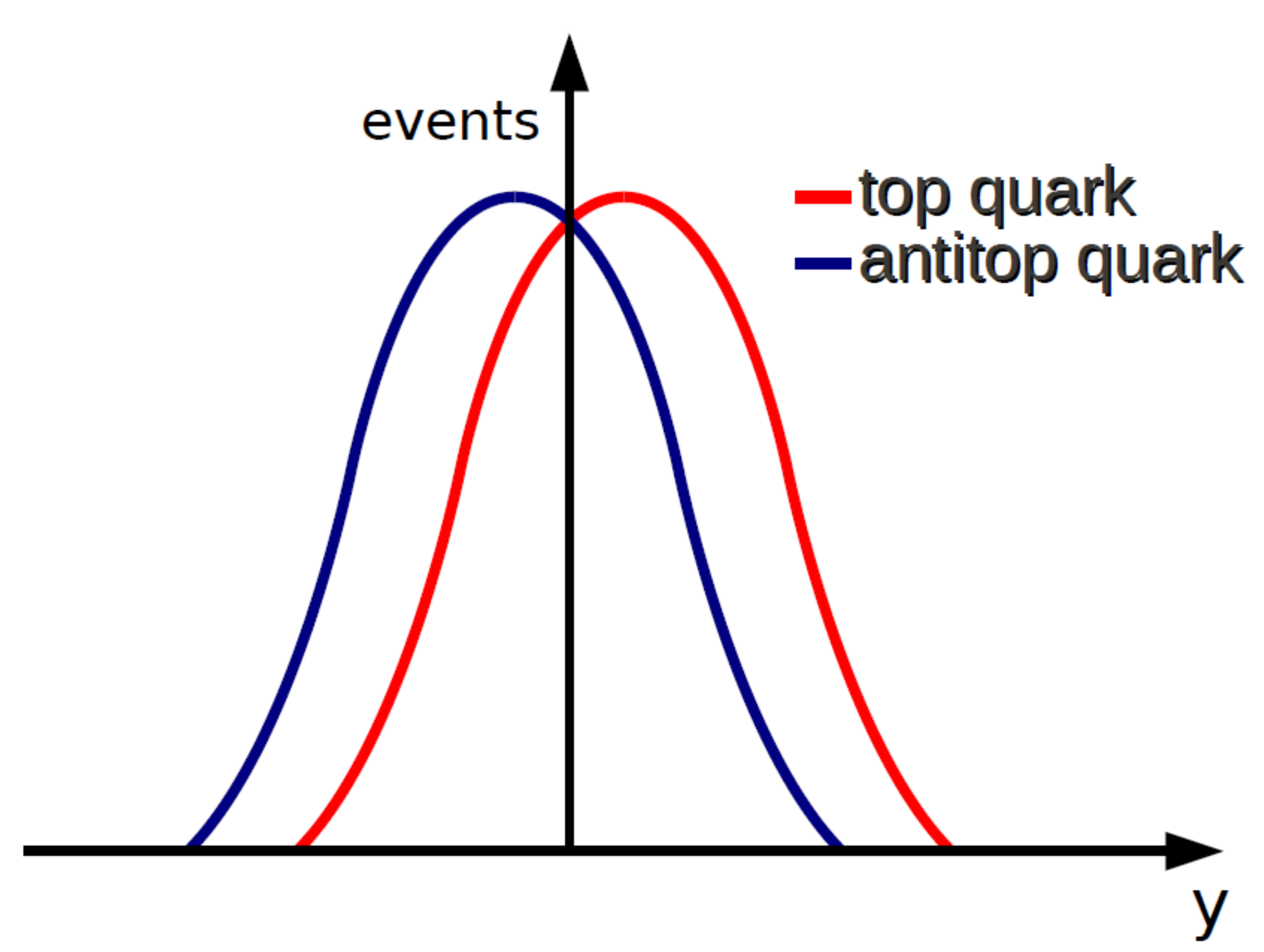}
}
\hspace{1.5cm}
\subfigure[]{
\includegraphics[width=0.38\textwidth]{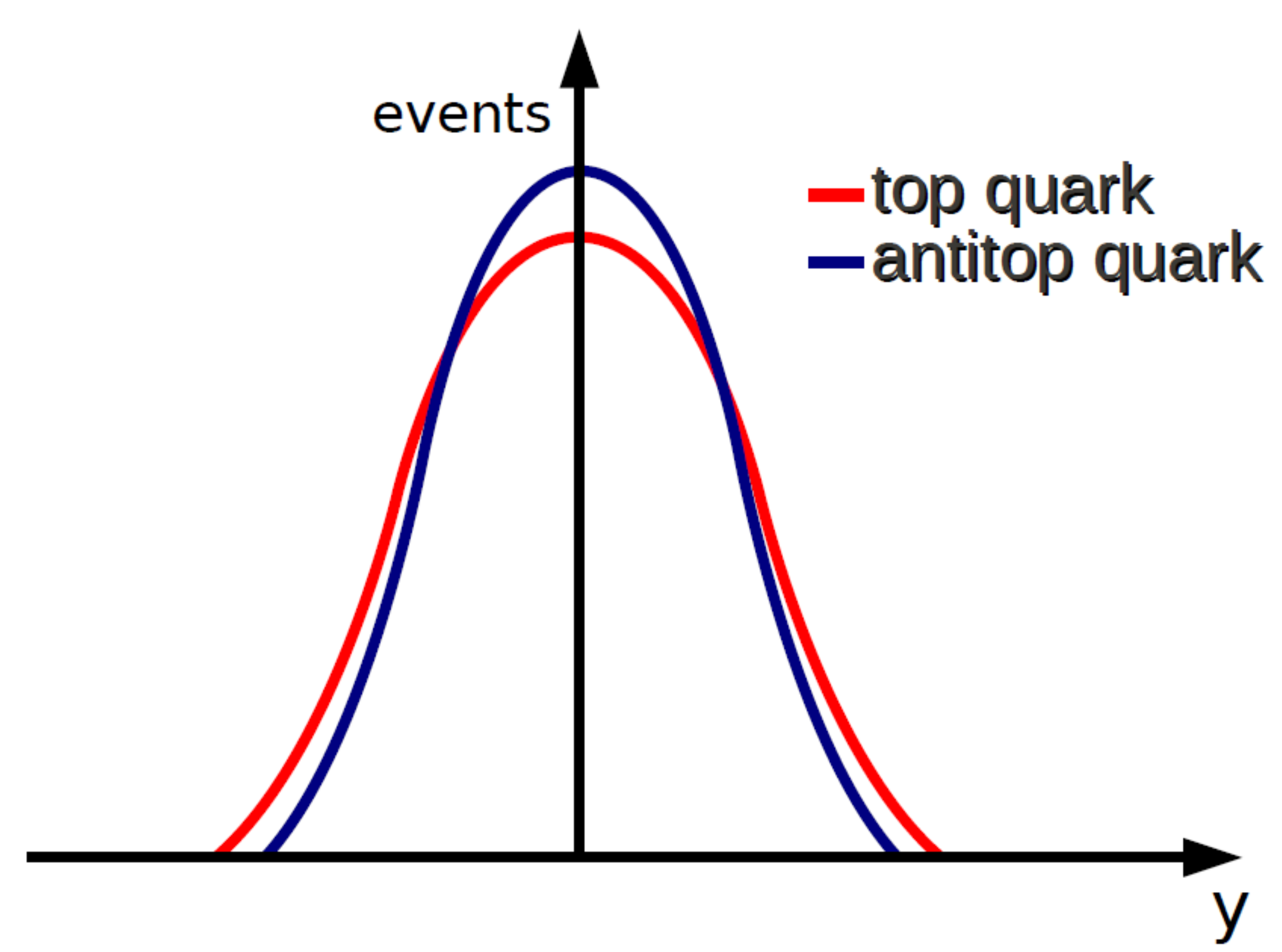}
}
\end{center}
\caption{Illustration of the effect of the charge asymmetry in the SM at the $p\bar{p}$ collider Tevatron (a) and at the $pp$ collider LHC (b). 
Shown are the rapidity distributions for top and anti-top quarks. Sketches are taken from~\cite{Boser:2011rca}.}
\label{fig:AC_Tev_vs_LHC}
\end{figure}

Compared to the Tevatron, the fraction of $q\bar{q}$ and $gg$ processes is roughly reversed at the LHC (see section~\ref{sec:ttbar}) 
resulting in a substantially smaller effect of the charge asymmetry at the LHC than at the Tevatron. Furthermore, at the LHC the partonic 
charge asymmetry does not result in an observable forward-backward asymmetry due to the symmetric colliding $pp$ initial state, meaning 
that quarks and anti-quarks, respectively, don't have a preferred direction.\\ 
However, as the proton consists of valence quarks ($uud$) with relative large proton momentum fraction $x$ and of sea-quarks and gluons 
with substantially smaller $x$, $q\bar{q} \rightarrow t\bar{t}(g)$ production is dominated by initial quarks with large $x$ and anti-quarks 
with small $x$. Because of the partonic charge asymmetry, top (anti-top) quarks are preferentially produced in the direction of the incoming 
quark (anti-quark) in the partonic rest frame (see Fig.~\ref{fig:AcLHC} (a) and (d)). The boost into the laboratory frame 
($pp$ frame) ``squeezes'' the top quark mainly in the forward and backward directions, while anti-top quarks are left more 
abundant in the central region (see Figure~\ref{fig:AcLHC} (b,c) and (e,f)). Hence, at the LHC the partonic charge asymmetry 
turns into a broader rapidity $y$ distribution for top quarks than for anti-top quarks, with $y=\frac{1}{2}\frac{E-p_z}{E+p_z}$ 
and $E$ and $p_z$ being the energy and $p_z$ the momentum of the top (anti-top) quark relative to the beam axis.
Note that the rapidity $y$ is invariant under Lorentz transformations longitudinal to the beam axis except for a constant. 
For massless particles the pseudorapidity $\eta=-\ln(\tan \theta/2)$, where $\theta$ is the polar angle relative to the beam 
axis, is identical to the rapidity.\\ 
Since, $y$ preserves the sign of the production angle, an asymmetry in $y$ is identical to the asymmetry in the top quark 
production angle. Figure~\ref{fig:AC_Tev_vs_LHC} shows qualitatively the effect of the charge asymmetry visible in the 
rapidity distributions for top and anti-top quark production at the Tevatron and LHC, respectively.\\

In order to quantify the difference in the rapidity distribution at the LHC globally a definition of a central 
charge asymmetry $A_{\mathrm{C}}$ where only the central region (rapidities below the cut value $y_{\mathrm{C}}$) is taken into account 
has been proposed in 2008 in Ref.~\cite{Antunano:2007da}:
\begin{equation}
\label{eq:centralAsy}
A_{\mathrm{C}}(y_{\mathrm{C}})=\frac{N_t(|y|<y_{\mathrm{C}})-N_{\bar{t}}(|y|<y_{\mathrm{C}})}{N_t(|y|<y_{\mathrm{C}})+N_{\bar{t}}(|y|<y_{\mathrm{C}})}
\end{equation}
$A_{\mathrm{C}}$ vanishes if the whole rapidity spectrum is integrated over and it was argued that the maximum of 
$A_{\mathrm{C}}$ is reached for $y_{\mathrm{C}}=1$. At the LHC with $\sqrt{s}=14\,\mbox{TeV}$, the QCD induced asymmetry is 
for $\sqrt{\hat{s}}=1\,\mbox{TeV}$ $A_{\mathrm{C}}(y_{\mathrm{C}}=1)=-0.86(4)\%$~\cite{Antunano:2007da}, where $\sqrt{\hat{s}}$ 
could be interpreted as invariant mass of the $t\bar{t}g$ system. With larger $\sqrt{\hat{s}}$ the central 
QCD charge asymmetry increases first up to -2\% due to the suppression of $gg$ processes, and is then 
reduced for $\sqrt{\hat{s}}=2\,\mbox{TeV}$ due to a larger contribution of $qg$ processes that 
partly compensate the asymmetry generated by $q\bar{q}$ events.


\subsection{Measurement of the Forward-Backward Asymmetry at the Tevatron}


As mentioned in section~\ref{sec:asy_theory} the charge asymmetry at the $p\bar{p}$ collider Tevatron results in a forward-backward asymmetry in the laboratory frame, which is reduced compared to the asymmetry in the experimentally inaccessible partonic frame. In order to reconstruct the charge asymmetry in the partonic rest frame as closely as possible other sensitive variables than the top quark production angle in the laboratory frame have been studied.\\ 

For $t\bar{t}$ events with $e/\mu$+jets signature (see chapter~\ref{sec:topsignature}) 
the rapidity difference $\Delta y$ of the top and anti-top quark measured in the laboratory frame, 
$\Delta y = y_t-y_{\bar{t}}= Q_\ell\cdot (y_{t_\ell}-y_{t_h})$, has been introduced in the PhD thesis 
of Ref.~\cite{Hirschbuehl:2005bj} in 2005. Here, $y_{t_\ell}$ and $y_{t_h}$ indicate the rapidities 
of the top (or anti-top) quark, that decays leptonically ($t_\ell \rightarrow b\ell \nu$) and 
hadronically ($t_h\rightarrow b q\bar{q}'$), respectively, and $Q_\ell$ is the charge of the lepton 
($e$ or $\mu$) originating from either the top or anti-top quark and is equal to the sign of the 
charge of the top quark it comes from. After multiplication of $(y_{t_\ell}-y_{t_h})$ by the lepton 
charge $Q_\ell$ the difference between the top and anti-top quark rapidities $\Delta y$ is obtained. 
The rapidity difference $\Delta y$ is related to the rapidity of the top quark in the $t\bar{t}$ rest frame, $y_t^{t\bar{t}}$, 
through $\Delta y= y_t -y_{\bar{t}}=y_t^{t\bar{t}}-y_{\bar{t}}^{t\bar{t}}= y_t^{t\bar{t}}+y_{t}^{t\bar{t}}=2\cdot y_t^{t\bar{t}}$ 
exploiting the invariance of rapidity differences under Lorentz transformations longitudinal to 
the beam axis and assuming that boosts transverse to the beam axis are negligible. 
Hence, the asymmetry $A_{\Delta y}$ defined as:
\begin{equation}
\label{eq:Adeltay}
A_{\Delta y} = A_{\mathrm{FB}}^{t\bar{t}} = \frac{N(\Delta y > 0) - N(\Delta y <0)}{N(\Delta y > 0) - N(\Delta y <0)}
\end{equation}
measures the charge asymmetry in the approximate (LO) $t\bar{t}$ rest frame and it is referred to as asymmetry in the $t\bar{t}$ rest frame $A_{\mathrm{FB}}^{t\bar{t}}$ in most of the literature.\\

A few years later, theorists have introduced in 2007 the differential pair asymmetry~\cite{Antunano:2007da} $A(Y)$:
\begin{equation}
A(Y)=\frac{N(y_t>y_{\bar{t}}) - N(y_t<y_{\bar{t}}) }{N(y_t>y_{\bar{t}}) + N(y_t<y_{\bar{t}})}
\end{equation}
with a fixed average rapidity $Y=1/2\cdot (y_t+y_{\bar{t}})$. It has been pointed out by the authors that 
after integrating over $Y$ the integrated pair asymmetry is equivalent to the definition of the asymmetry 
$A_{\Delta y} = A_{\mathrm{FB}}^{t\bar{t}}$ given in equation Eq. (\ref{eq:Adeltay}). The prediction for the QCD induced 
asymmetry $A_{\Delta y} = A_{\mathrm{FB}}^{t\bar{t}}$ including an estimation of the mixed QCD-QED/EW corrections 
from Ref.~\cite{Kuhn:1998kw} and evaluating numerator and denominator at LO is 
$A^{\Delta y}=A_{\mathrm{FB}}^{t\bar{t}}=(7.8 \pm 0.9)\%$~\cite{Antunano:2007da}, which is indeed substantially 
larger than the asymmetry in the laboratory frame of about 5\%~\cite{Kuhn:1998kw}.\\

In a subsequent experimental MC based study performed in the diploma thesis of Ref.~\cite{Weinelt:2006mh} it 
has been shown that the reconstructed asymmetry after applying an event selection to enrich $t\bar{t}$ events 
with $e/\mu$+jets signature is strongly reduced because the asymmetry decreases linearly with the number of 
reconstructed jets. This is understandable as the asymmetry induced from the interference of $q\bar{q}$ induced 
$t\bar{t}g$ diagrams gives a negative contribution to the charge asymmetry, while the interference of the 
$q\bar{q}$ induced born and box diagrams yields a positive contribution. Requiring more reconstructed jets 
implies that the contribution from $t\bar{t}g$ events giving a negative contribution to the inclusive 
asymmetry increases.\\ 
At that time, it was surprising that this qualitative behaviour was observed not only for NLO Monte-Carlo generators 
like MC@NLO~\cite{Frixione:2002ik,Frixione:2003ei}, but also for LO+parton shower (PS) Monte-Carlo generators 
like PYTHIA~\cite{Sjostrand:2000wi} or HERWIG~\cite{Corcella:2000bw,Corcella:2002jc}. 
The only difference found, was an offset coming from the missing part of the interference of the born and box diagrams in 
LO+PS MC generators, while the interference of the $t\bar{t}g$ diagrams is partially modelled in LO+PS MC generators by 
the angular ordering of gluon radiation implemented in the parton showering.\\ 

In 2012, hence about six years later, theorists carefully studied the asymmetry predictions of LO+PS MC 
generators ~\cite{Skands:2012mm}. So, the effect described above arises from valid physics built into event 
generators, namely coherent parton showers which are imple\-men\-ted through di\-pole showering and angular 
ordering. In the hard process $q\bar{q}\rightarrow t\bar{t}$, the colour flows from the incoming quark to 
the top quark and from the incoming anti-quark to the anti-top quark. This leads to a more violent acceleration 
of colour, and consequently more QCD radiation, when the top quark is produced backwards in the $q\bar{q}$ rest 
frame than when it goes forwards. Hence, in $t\bar{t}g$ events the top quark is preferentially produced in 
backward direction. Furthermore, the authors demonstrate that even LO+PS generators can also generate a net 
inclusive asymmetry, if the shower kinematics allow for migration between positive and negative $\Delta y$ regions. 
In particular, they find, that the asymmetries in HERWIG++~\cite{Bahr:2008pv} and SHERPA~\cite{Gleisberg:2008ta} 
are comparable to the LO perturbative results obtained from NLO $t\bar{t}$ calculations.\\

Based on these two preparatory studies the charge asymmetry in top quark pair production has been measured 
at CDF in the $t\bar{t}$ rest-frame~\cite{Aaltonen:2008hc} $A_{\mathrm{FB}}^{t\bar{t}}$ using $\Delta y$ as sensitive 
variable and using data corresponding to an integrated luminosity of $\mathcal{L}=1.9\;\mbox{fb}^{-1}$. 
Besides the asymmetry in the $t\bar{t}$ rest frame a complementary measurement of the asymmetry 
$A^{p\bar{p}}_{\mathrm{FB}}$in the laboratory frame has been performed~\cite{Schwarz:2006ud} and is 
documented in the same journal paper~\cite{Aaltonen:2008hc}.\\ 

Both analyses of Ref.~\cite{Aaltonen:2008hc} select $t\bar{t}$ events with $e/\mu$+jets signature and 
make use of the ``standard'' CDF event selection. Here, events are selected with an isolated electron 
or muon with $|\eta|\le 1.0$ and $E_{\mathrm{T}}\ge 20\,\mbox{GeV}$ or $p_{\mathrm{T}}\ge 20\,\mbox{GeV/c}$, respectively, 
missing transverse energy $E_{\mathrm{T}}^{mis}\ge 20\,\mbox{GeV}$ (the imbalance in transverse momentum in an event) 
consistent with a neutrino from a $W$ boson decay, and at least four hadronic jets with $|\eta|\le 2.0$ and 
$E_{\mathrm{T}}\ge 20\,\mbox{GeV}$, whereof at least one jet needs to be identified as $b$-jet containing a 
reconstructed secondary vertex consistent with the decay of a bottom hadron in the jet~\cite{Acosta:2004hw}. 
Jets are reconstructed using a cone algorithm by summing the transverse energy $E_{\mathrm{T}}$ within a cone of 
radius $\Delta R = 0.4$ and jet energies are corrected to parton-level values~\cite{Bhatti:2005ai}. 
In total 484 event candidates are selected with a background contamination of about 18\%.\\ 

The $t\bar{t}$ signal events are modelled using PYTHIA, HERWIG and MC@NLO event generators and 
the top quark mass is set equal to $m_t=175\;\mbox{GeV/c}^2$.\\
The largest background is due to $W$-boson events in association with jets. Other contributing 
backgrounds are multi-jet events and electroweak processes like diboson events ($WW$, $WZ$, $ZZ$) 
and singly produced top quark events. Backgrounds are estimated based on the method in~\cite{Abulencia:2006in} 
using partially data.\\ 
The normalisation and shape of multi-jet events is exclusively modelled using data. Events with five jets 
where one jet models a misreconstructed electron are used as multi-jet sample. This sample is also used for 
muon events after correcting for the different lepton acceptance. The normalisation of the multi-jet events is 
then determined from a fit to the missing transverse energy distribution.\\ 
$W$+jets events with tagged heavy flavor or mistagged light partons are modelled using the MC generator 
ALPGEN~\cite{Mangano:2002ea} interfaced to PYTHIA parton showering. The $b$-tagging efficiencies in MC 
are corrected with scale factors to account for small deviations between MC and data seen in jet events. 
The mistagging rates are fully determined from jet data, parametrised in several quantities and applied 
to the used MCs. Furthermore, K-factors for the fractions of $W$+$c$, $W$+$cc$ and $W$+$b\bar{b}$ events 
in the MC have been determined from multi-jet data.
The small electroweak backgrounds, $WW$, $WZ$, $ZZ$ and single-top are modelled with PYTHIA and with 
MADEVENT~\cite{Alwall:2007st}.\\ 

\begin{figure}[t]
\begin{center}
\subfigure[]{
\includegraphics[width=0.38\textwidth]{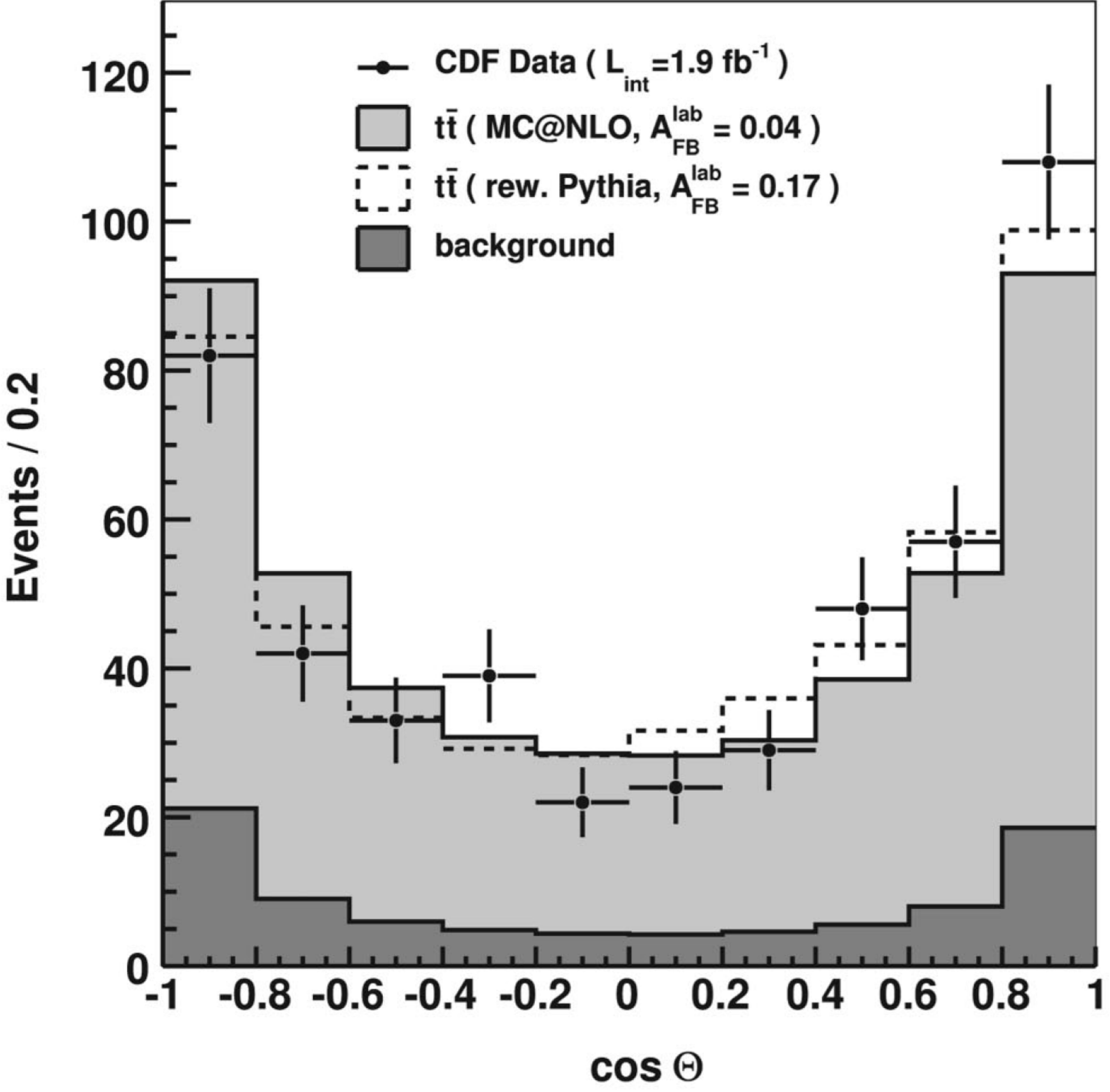}
}
\hspace*{1cm}
\subfigure[]{
\includegraphics[width=0.38\textwidth]{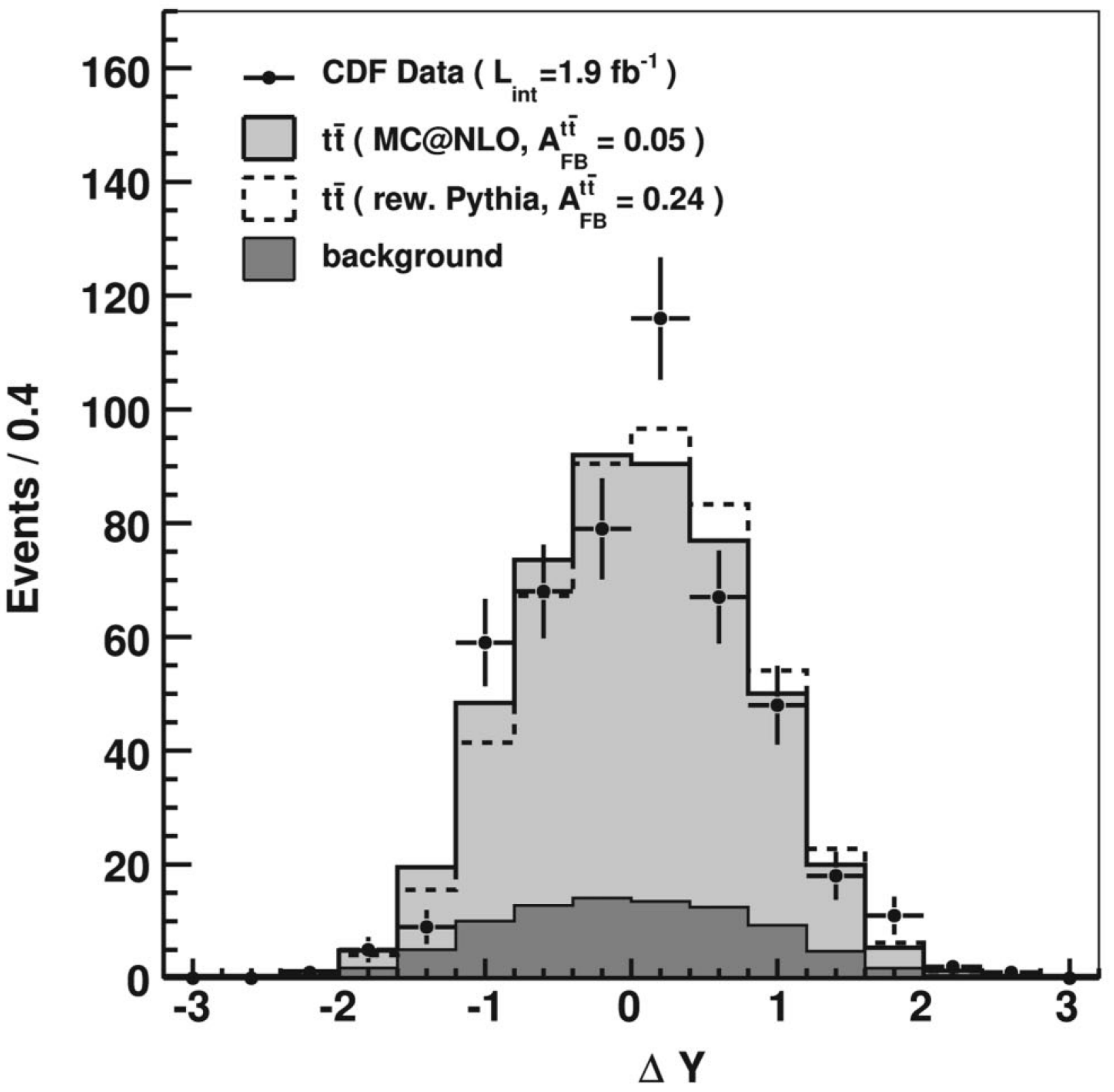}
}
\end{center}
\vspace*{-0.5cm}
\caption{$A_{\mathrm{FB}}$ in $t\bar{t}$ production as measured by the CDF 
collaboration in the year 2008 using data at $\sqrt{s}=1.96\;\mbox{TeV}$ corresponding 
to an integrated luminosity of $1.9\;\mbox{fb}^{-1}$~\cite{Aaltonen:2008hc}. Top quark pair
events with $e/\mu$+jets signature are employed.
(a) $A_{\mathrm{FB}}$ measured in the laboratory frame using as sensitive variable the cosine of the 
polar angle $\cos\Theta$ between the top quark and the proton beam. 
(b) $A_{\mathrm{FB}}$ measured in the $t\bar{t}$ frame using as sensitive variable 
$\Delta Y=\Delta y = y_t - y_{\bar{t}}$ (see text).
The solid line is the prediction for $t\bar{t}$ with MC@NLO
model of the QCD induced charge asymmetry and $\sigma_{t\bar{t}}=8.2\,\mbox{pb}$, 
plus the expected non-$t\bar{t}$ backgrounds. The dashed curve shows the
prediction when $t\bar{t}$ is reweighted using the measured $A_{\mathrm{FB}}$ value
and assuming a linear dependence of the asymmetry on the top quark production angle.}
\label{fig:AFB_CDF_first}
\end{figure}

In order to measure the asymmetry in the laboratory frame and the $t\bar{t}$ rest frame the 
kinematics of the hadronically decaying top quark $t_h$ and of both top quarks, respectively, 
has to be reconstructed.
For the asymmetry measurement in the laboratory frame the algorithm employed in the top quark mass 
measurement of Ref.~\cite{Abulencia:2005ak} is used, while for the asymmetry in the $t\bar{t}$ rest 
frame another technique has been developed in~\cite{Chwalek:2006bh}, which has been employed to measure 
the $W$-boson helicity fractions in top quark pair events~\cite{Abulencia:2006ei}. 
In this technique constraints on the $W$ boson masses, the difference of the reconstructed top and 
anti-top quark mass (but not $m_t$), the total transverse energy and the probability of jets 
to be a $b$-jet~\cite{Abulencia:2006kv} are used. In contrast to the first method, all jets 
in the event are considered and not only the four leading jets.\\
Figure~\ref{fig:AFB_CDF_first} compares the data distribution of the reconstructed sensitive 
variables to the SM prediction obtained by the MC@NLO generator, showing that the data 
feature a larger asymmetry than predicted by MC@NLO.\\

Some of the backgrounds are asymmetric weak processes, whereof the $W$-boson production in association with jets
originating from heavy quarks is the most asymmetric one. The modelling of the backgrounds, in particular in the sensitive variables, 
has been checked carefully in control samples enriched with different background processes. 
No deviations between data and MC have been observed.\\ 
The background-corrected distributions of the sensitive variables are distorted from their 
parton-level shapes by an acceptance bias (as observed in~\cite{Weinelt:2006mh}) and the 
imperfect reconstruction of the $t\bar{t}$ kinematics. To account for this a matrix inversion 
technique ($4x4$ matrix) is used to derive the parton-level distributions in the $t\bar{t}$ rest frame and the laboratory frame.\\ 
Using pseudo experiments, the extraction of the corrected asymmetry has been carefully checked 
and systematic uncertainties have been estimated. For the asymmetry in the laboratory frame 
the background normalisation is the largest uncertainty, while for the asymmetry in the 
$t\bar{t}$ frame the uncertainty on a possible bias of the corrected result due to the shape 
uncertainty of the parton-level distribution of the sensitive variable is largest.\\ 

The corrected asymmetries measured by CDF are:
\begin{eqnarray*}
A_{\mathrm{FB}}^{t\bar{t}} & = &  0.24 \pm 0.13\; \mbox{(stat.)} \pm 0.04\; \mbox{(syst.)}  = 0.24 \pm 0.14\\
A^{p\bar{p}}_{\mathrm{FB}} & = & 0.17 \pm 0.07\; \mbox{(stat.)} \pm 0.04\; \mbox{(syst.)} = 0.17 \pm 0.08
\end{eqnarray*}
The results show the expected frame dependence. Compared to the QCD+EW NLO predictions at that time of 
$A^{t\bar{t}}_{\mathrm{FB}}=A_{\Delta y}= 0.078 \pm 0.009$ and $A^{p\bar{p}}_{\mathrm{FB}}= 0.051\pm 0.006$~\cite{Antunano:2007da} 
the measured asymmetries are larger but consistent with these SM predictions within two Gaussian standard deviations 
($2\sigma$) and disfavour exotic sources of top quark production with significant negative $A_{\mathrm{FB}}$~\cite{Antunano:2007da}.\\ 
Besides the inclusive asymmetry, also the background corrected asymmetry for events with no additional hard jet 
($N_{jets}=4$, 85\% $t\bar{t}$) and events with at least one additional hard jet ($N_{jets}\ge 5$, 56\% $t\bar{t}g$) 
have been measured. For $A_{\mathrm{FB}}^{p\bar{p}}$ no trend is seen, but for $A_{\mathrm{FB}}^{t\bar{t}}$ the expected decrease of 
the asymmetry with the number of jets is observed.\\ 

A few months earlier the $D0$ collaboration published a charge asymmetry me\-asure\-ment using data corresponding 
to an integrated luminosity of $0.9\;\mbox{fb}^{-1}$~\cite{Abazov:2007ab}. In this analysis the background corrected 
asymmetry (no correction for acceptance and reconstruction effects applied) $A_{\mathrm{FB}}^{t\bar{t}}$ has been measured in $t\bar{t}$ 
events with $e/\mu$+jets signature using $\Delta y$. 
The background corrected inclusive asymmetries of $0.12\pm 0.08$~\cite{Abazov:2007ab} and $0.12\pm 0.06$~\cite{Aaltonen:2008hc} 
obtained by D0 and CDF after applying an event selection (similar but not identical) are well consistent.\\ 
To handle the shift in the measured background corrected asymmetry due to acceptance effects and reconstruction effects the 
D0 collaboration followed a different ansatz. Instead of correcting the data a $\Delta y$ dependent correction or dilution 
function valid within a visible phase space has been given with the idea that theoretical predictions for $A_{\mathrm{FB}}^{t\bar{t}}$, 
calculated differentially in $\Delta y$ and performed in this visible phase space, could be compared to the data by applying 
this dilution function to the prediction. Unfortunately, no theoretical group has chosen to go this way.\\ 

As a deviation from the SM prediction has been observed in the first asymmetry measurements at CDF and D0, the community was 
highly interested in new results from the Tevatron based on a substantial larger data set.\\ 

In June 2011 the CDF collaboration published a new measurement of the asymmetry in $t\bar{t}$ events using data corresponding 
to an integrated luminosity of $5.3\;\mbox{fb}^{-1}$\cite{Aaltonen:2011kc}. Again, $t\bar{t}$ events with $e/\mu$+jets 
signature have been considered to determine corrected values of $A_{\mathrm{FB}}^{p\bar{p}}$ and $A_{\mathrm{FB}}^{t\bar{t}}$ inclusively 
as well as of $A_{\mathrm{FB}}^{t\bar{t}}$ depending (two bins) on $\Delta y$ and the invariant mass of the top quark pair $m_{t\bar{t}}$.\\ 
The measured inclusive asymmetries of $A_{\mathrm{FB}}^{p\bar{p}}=0.150\pm 0.055\;\mbox{(stat.+syst.)}$ and $A_{\mathrm{FB}}^{t\bar{t}}=0.158\pm 0.075\;\mbox{(stat.+syst.)}$ 
are compared to the SM predictions computed with MCFM\cite{Campbell:1999ah} version 5.7: $0.038\pm 0.06$ and $0.058\pm 0.009$, respectively. 
Please note, that MCFM predicts smaller asymmetries than those predicted in Ref.~\cite{Antunano:2007da}. There are two reasons for this. 
First, MCFM predicts only pure QCD induced asymmetries, EW effects are completely neglected. Secondly, as denominator the NLO $t\bar{t}$ 
cross section is used instead of the LO cross section. The latter is considered to be theoretically more reasonable as it gives the asymmetry 
at a fixed order. The authors state that the asymmetry measured in the laboratory frame is two standard deviations above the value predicted 
by MCFM. Comparing it to the QCD+EW NLO prediction in Ref.~\cite{Antunano:2007da} the deviation is below $2\sigma$.\\ 
Furthermore, the CDF collaboration found that the measured $A_{\mathrm{FB}}^{t\bar{t}}$ rises with $m_{t\bar{t}}$ and that the asymmetry measured 
in events with $m_{t\bar{t}}\ge 450\;\mbox{GeV/c}^2$ is $3.4\sigma$ above the SM prediction by MCFM and at he $3\sigma$ boundary if it is 
compared to the QCD+EW NLO prediction in Ref.~\cite{Hollik:2011ps,Kuhn:2011ri}, which became available shortly after the CDF publication appeared 
and includes the full revised EW corrections (see discussion above).\\ 

Only a few months later the D0 collaboration published a new measurement~\cite{Abazov:2011rq} of the asymmetry in $t\bar{t}$ events using 
data corresponding to an integrated luminosity of $5.4\;\mbox{fb}^{-1}$. As in the previous measurements by CDF and D0 $t\bar{t}$ 
events with $e/\mu$+jets signature have been used. This time, the D0 collaboration corrected the asymmetry on reconstruction 
level (background corrected) for detector and acceptance effects using a fine binned unfolding with regularisation~\cite{Bohm2010,TUnfold}. 
Compared to the $4x4$ matrix inversion used by CDF the usage of a more sophisticated regularised unfolding results in a reduced 
uncertainty (sum of stat. and syst.) of about 20\%.\\ 
The corrected asymmetry of $A_{\mathrm{FB}}^{t\bar{t}}= 0.196\pm 0.065\; \mbox{(stat.+syst.)}$ measured by the D0 collaboration is 
well consistent with the previous corrected measurements from CDF~\cite{Aaltonen:2008hc,Aaltonen:2011kc}. Comparing the D0 
measurement to MC@NLO, which has the same weakening as MCFM compared to the QCD+EW NLO calculation from~\cite{Hollik:2011ps,Kuhn:2011ri}, 
the statistical significance of the upward deviation is $2.4\sigma$. In contrast to the CDF publication~\cite{Aaltonen:2011kc} no 
statistically significant enhancement of the asymmetry for high $m_{t\bar{t}}$ values has been found on reconstruction level.\\ 
Furthermore, the selected $t\bar{t}$ events with $e/\mu$+jets signature have been used to measure for the first time the leptonic 
asymmetry defined as
\begin{equation}
\label{eq:Afbl}
A_{\mathrm{FB}}^{\ell}=\frac{N(Q_{\ell}y_{\ell}>0)-N(Q_{\ell}y_{\ell}<0)}{N(Q_{\ell}y_{\ell}>0)+N(Q_{\ell}y_{\ell}<0)}
\end{equation}
where $y_{\ell}$ and $Q_{\ell}$ are the rapidity measured in the laboratory frame and the charge of the lepton (electron or muon), 
respectively. The leptonic asymmetry represents an alternative approach that does not depend on the full reconstruction of the $t\bar{t}$ 
system. The leptonic asymmetry has been measured at the reconstruction level and after unfolding. In this case unfolding corrects only 
minimally for the migration in $Q_{\ell}y_{\ell}$ but mainly for acceptance effects. To avoid large acceptance corrections only the 
region $|y_{\ell}|<1.5$ has been used. The measured leptonic asymmetries are compared with the asymmetries computed using MC@NLO. 
Again, the measured asymmetries are significantly larger than the MC@NLO prediction. So, the deviation of the measured corrected 
leptonic asymmetry $A_{\mathrm{FB}}^{\ell}=0.152\pm 0.04\;\mbox{(stat.+syst.)}$ from the MC@NLO prediction is $3.3\sigma$. The QCD+EW NLO 
prediction including the top quark decay in the NWA approximation and $t\bar{t}$ spin correlation of 
$A_{\mathrm{FB}}^{\ell}=(3.4^{+0.5}_{-0.2})\%$~\cite{Bernreuther:2010ny} is unfortunately not directly comparable with the measurement of 
the D0 collaboration as D0 has chosen different fiducial cuts.\\

In summary, in 2011 the charge asymmetry measurements from the CDF and D0 collaborations show discrepancies with the SM 
prediction~\cite{Kuhn:1998jr,Kuhn:1998kw,Antunano:2007da,Kuhn:2011ri,Hollik:2011ps,Bernreuther:2010ny} of the order of two 
standard deviations and even more in certain phase space regions. This has generated a large number of theoretical explanations 
that attribute them to contributions from physics beyond the SM. An overview of the variety of theoretical explanations from 
the year 2011 can be found, e.g. in Ref.~\cite{AguilarSaavedra:2011hz} and references therein.


\subsection{Measurement of the Charge Asymmetry at the LHC}

In the PhD thesis of Ref.~\cite{Peiffer2011} and motivated by the large charge asymmetry observed at the Tevatron a first 
preliminary measurement of the charge asymmetry has been performed at the $pp$ collider LHC using CMS data at 
$\sqrt{s}=7\;\mbox{TeV}$ corresponding to an integrated luminosity of $36\mbox{pb}^{-1}$. In this thesis, $t\bar{t}$ events 
with $e/\mu$+jets signature are selected.\\

At the beginning of the PhD thesis of Ref.~\cite{Peiffer2011} only the central asymmetry~\cite{Antunano:2007da} 
(see Eq. (\ref{eq:centralAsy})) has been proposed by theorists as measurable quantity in 2007, which has been further 
developed in 2011 to the edge asymmetry~\cite{Xiao:2011kp}, where only events with an absolute rapidity value above a certain
cut value are considered. As the central asymmetry proposal suffers of some experimental caveats a different approach has been
invented in this thesis. First, the proposed asymmetry (the same is also true for the edge asymmetry) uses only a subset of 
$t\bar{t}$ events, resulting in a limited statistical sensitivity, in particular for small data sets. Secondly, the central 
and edge asymmetry definitions proposed don't reveal the asymmetry in one single sensitive variable that can be calculated 
event-wise.\\
As alternative approach the variable $|\eta_t|-|\eta_{\bar{t}}|$, with $\eta_t$ ($\eta_{\bar{t}}$) being the pseudorapidity 
measured in the laboratory frame of the top (anti-top) quark, has been invented in this study. This variable gives the event-wise 
information whether the anti-top quark is produced more centrally than the top quark or not. The cut-independent charge asymmetry 
$A_{\mathrm{C}}$ at the LHC is then computed from the number of $t\bar{t}$ events $N$ with positive and negative value of 
$\Delta |\eta|=|\eta_t|-|\eta_{\bar{t}}|=Q_{\ell}\cdot(|\eta_{t_l}|-|\eta_{t_h}|)$ and is defined as~\cite{Peiffer2011}:
\begin{equation}
A_{\mathrm{C}}^{\eta} = \frac{N(\Delta |\eta |>0)-N(\Delta |\eta |<0)}{N(\Delta |\eta |>0)+N(\Delta |\eta |<0)}\;\;\;.
\end{equation}
In 2009 exactly this asymmetry definition has been proposed in Ref.~\cite{Diener:2009ee} in the context of improving
the existing forward-backward asymmetry variable to distinguish possible $s$-channel resonances, decaying to fermions.\\

With this $\Delta|\eta|$ variable, the $A_{\mathrm{C}}^{\eta}$ measurement at the LHC has been conducted in a very similar way 
as the asymmetry measurements at the CDF experiment. As the event selection has been adopted from the $t\bar{t}$ 
cross section measurement~\cite{Chatrchyan:2011ew} the event yield of the backgrounds could be directly taken from 
the $t\bar{t}$ cross section measurement.\\

An important ingredient towards the first published charge asymmetry measurement at the LHC using CMS 
data~\cite{Chatrchyan:2011hk} was the implementation of an unfolding procedure using generalised 
matrix inversion with regularisation~\cite{Peiffer2011,Blobel:2002pu,Tikhonov:1963,Phillips:1962}. 
Extensive studies on this containing the comparison of different unfolding routines are presented 
in the PhD thesis of Ref.~\cite{Peiffer2011}. As the expected charge asymmetry at the LHC is 
substantially smaller than that at the Tevatron it was important to avoid substantial systematic 
uncertainties due to the use of a simple matrix inversion, in particular because the D0 collaboration 
has already demonstrated the advantages of regularised unfolding~\cite{Abazov:2011rq}.\\

In a subsequent diploma thesis~\cite{Boser:2011rca}, which is the basis for the published charge 
asymmetry measurement from the CMS collaboration using data at $\sqrt{s}=7\;\mbox{TeV}$ corresponding 
to an integrated luminosity of $1.09\;\mbox{fb}^{-1}$~\cite{Chatrchyan:2011hk}, $t\bar{t}$ events with 
$e/\mu$+jets signature are used, again.\\

The events are triggered via $e/\mu$+jets triggers, that require an electron or muon with 
$p_{\mathrm{T}}>25\,\mbox{GeV/c}$ or $p_{\mathrm{T}}>17\,\mbox{GeV/c}$ and at least three jets, each with $E_{\mathrm{T}}>30\,\mbox{GeV}$. 
As in the measurement the same jet cut is applied in the offline selection as in the online trigger requirement 
and as the jet definitions are different it can happen that the trigger accepts an event and the offline selection 
does not accept it or vice versa. In the theses of Ref.~\cite{Boser:2011rca, Roscher2012} a method to account 
for this effect has been established. First, so-called trigger turn-on curves as a function of $p_{\mathrm{T}}$ and in few 
bins of $\eta$ of the jet have been determined for a single offline reconstructed jet. Then, the probability 
that an offline selected event is accepted by the three jet trigger path is computed by combining the probabilities 
of the individual jets to be accepted.\\
In the offline selection the particle flow algorithm~\cite{CMS:2009nxa} is used to reconstruct electrons, muons, 
jets and and any imbalance in transverse momentum due to neutrinos. This algorithm aims to reconstruct the entire 
event by combining information from all sub-detectors, including tracks of charged particles in the tracker and 
muon system, and energy depositions in the electromagnetic and hadronic calorimeters.\\
Top quark pair candidate events are selected with an isolated electron (muon) with $|\eta|\le 2.5$ ($|\eta|\le 2.1$) 
and $p_{\mathrm{T}}\ge 30\,\mbox{GeV/c}$ ($p_{\mathrm{T}}\ge 20\,\mbox{GeV/c}$), and at least four hadronic jets with $|\eta|\le 2.4$ 
and $E_{\mathrm{T}}\ge 30\,\mbox{GeV}$, whereof at least one jet needs to be identified as $b$-jet by an algorithm that 
orders the tracks in impact parameter significance and discriminates using the track with the second highest 
significance~\cite{CMS:2010hua,CMS:2011cra}. Jets are reconstructed using the anti-$k_\mathrm{T}$ 
algorithm~\cite{Cacciari:2008gp,Cacciari:2005hq,Cacciari:2011ma} with the distance parameter $R=0.5$ and jet 
energies are corrected for additional contributions from multiple interactions, as well as for $\eta$ and 
$p_{\mathrm{T}}$-dependent detector response.\\ 

Top quark pair events are generated with the tree-level matrix-element generator MADGRAPH version 5~\cite{Alwall:2011uj}, 
interfaced to PHYTHIA version 6.4~\cite{Sjostrand:2006za} for the parton showering, and with a top quark mass of 
$m_t=172.5\;\mbox{GeV/c}^2$. The MLM algorithm~\cite{Mangano:2006rw} is used for the parton shower/matrix element matching. 
Higher order gluon or quark production is described by the matrix elements with up to three extra partons beyond the 
$t\bar{t}$ system.\\ 
The weak vector boson production in association with up to four jets and single top quark production 
(generated with $m_t=172.5\;\mbox{GeV/c}^2$) is simulated using the same combination of MADGRAPH and PYTHIA programs. 
Multi-jet events are entirely modelled from data using events which pass looser isolation requirements.\\ 

\begin{figure}[t]
\begin{center}
\subfigure[]{
\includegraphics[width=0.45\textwidth]{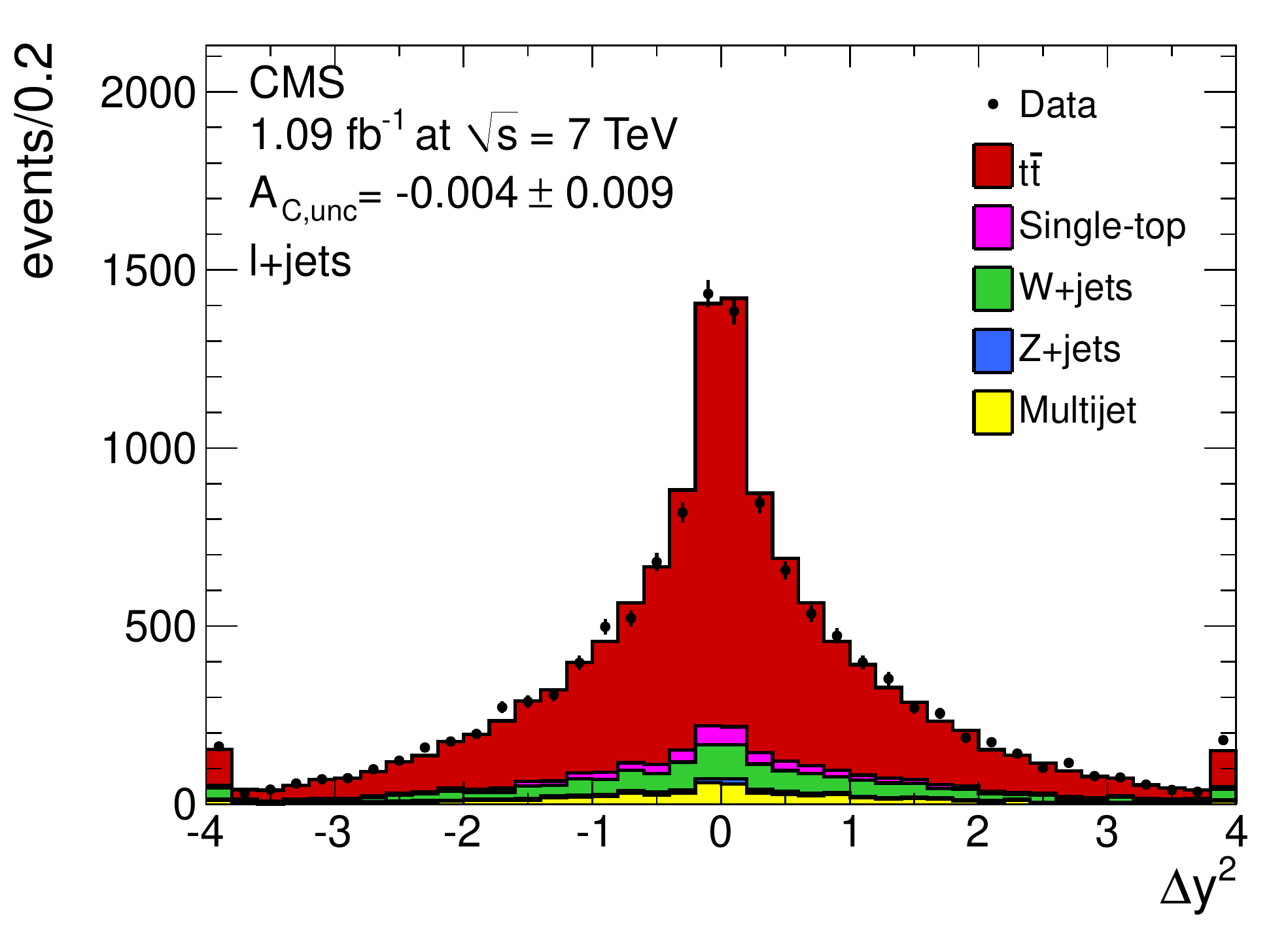}
}
\hspace*{0.5cm}
\subfigure[]{
\includegraphics[width=0.45\textwidth]{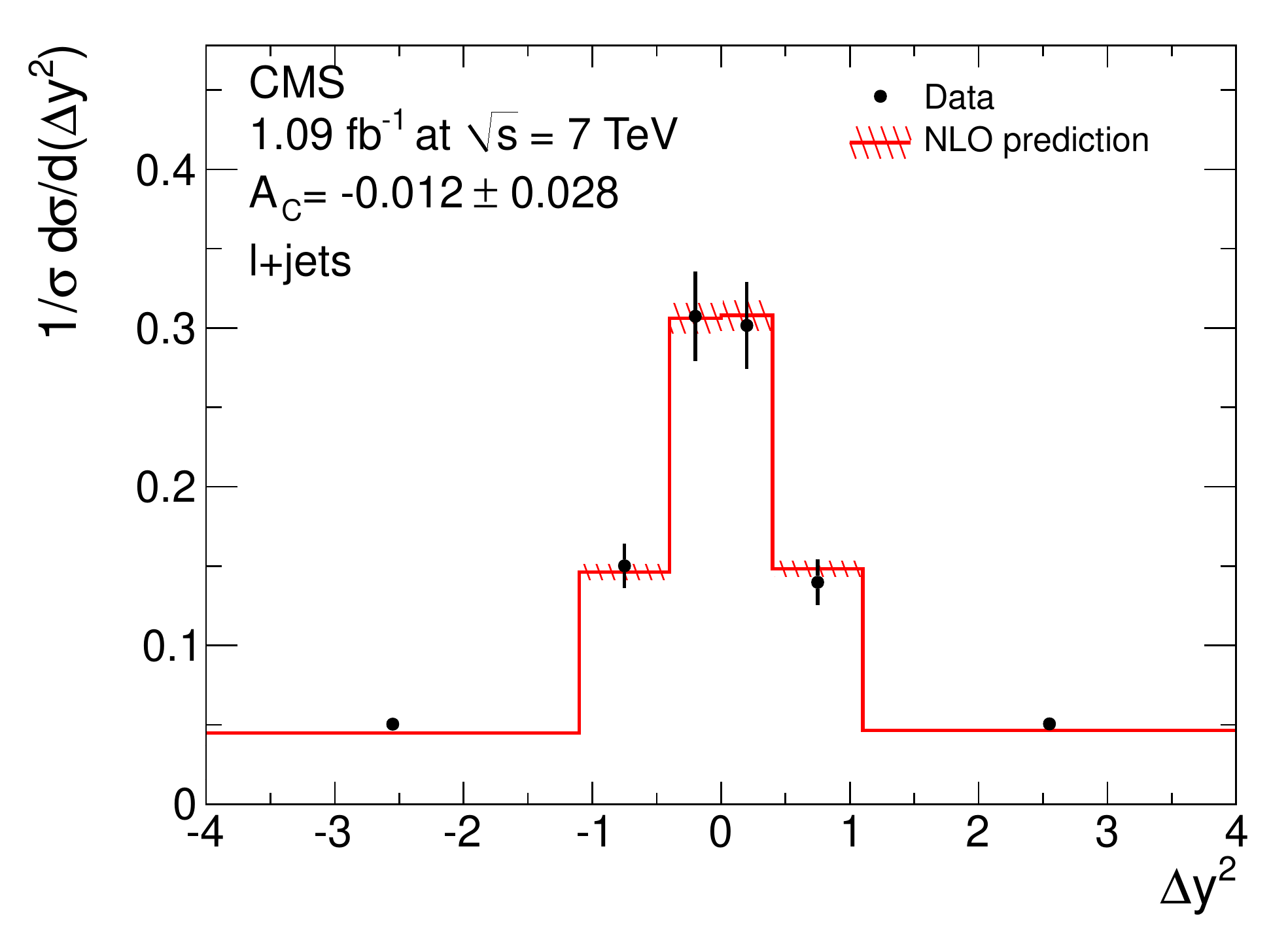}
}
\end{center}
\vspace*{-0.5cm}
\caption{Inclusive charge asymmetry $A_{\mathrm{C}}$ in $t\bar{t}$ production as measured by the CMS 
collaboration using data at $\sqrt{s}=7\;\mbox{TeV}$ corresponding 
to an integrated luminosity of $1.09\;\mbox{fb}^{-1}$~\cite{Chatrchyan:2011hk}. 
Top quark pair events with $e/\mu$+jets signature are employed.
(a) Reconstructed sensitive variable $\Delta y^2=(y_t-y_{\bar{t}})\cdot (y_t+y_{\bar{t}})$ 
(see text). The last bin include the sum of all contributions for $\Delta y^2$. 
The signal and background contributions are normalised to the results of the background estimation
described in the text.
(b) Unfolded $\Delta y^2$ normalised spectra. The SM QCD+EW NLO prediction is based on the calculations
of Ref.~\cite{Kuhn:2011ri}. The last bin include the sum of all contributions $|\Delta y^2|>4.0$ 
The uncertainties shown on the data are statistical, while the uncertainties on the prediction
account also for the dependence on the top-quark mass, parton distribution functions (PDF), 
and factorisation and renormalisation scales.
}
\label{fig:AC_CMS_first}
\end{figure}
In total 12757 event candidates are selected with a background contamination of about 20\%. The background estimation has 
been conducted with a similar method as used in~\cite{Chatrchyan:2011ew} but has been optimised and adjusted for the charge 
asymmetry measurement in the context of the diploma thesis of Ref.~\cite{Roscher2012}. A binned likelihood fit is conducted 
exploiting the discrimination power of the missing transverse energy $E_{\mathrm{T}}^{mis}$ and the $M3$ variable, where $M3$ is the 
invariant mass of the three jets in an event with the largest vectorially summed transverse momentum. As the $W$+jets 
background is asymmetric at the LHC and as more $W^+$ than $W^-$ bosons are produced the method has been adjusted. The 
data set have been split into events with positively and negatively charged leptons and in a simultaneous fit to all data 
subsets the background for $W^+$+jets and $W^-$+jets is determined besides the rate for the multi-jet background and 
other minor backgrounds.\\ 
The basic idea of the reconstruction of the $t\bar{t}$ kinematics as employed in the first $A_{\mathrm{FB}}^{t\bar{t}}$ measurement 
from the CDF collaboration~\cite{Aaltonen:2008hc} has been adopted in the first CMS stu\-dy~\cite{Peiffer2011} but 
adjustments to account for differences in the experiments have been made. In the diploma thesis~\cite{Roscher2012} 
this method has been improved by transforming the hitherto existing criterion into a likelihood criterion and by 
decorrelating variables.\\ 

In 2011 it was proposed to measure the charge asymmetry at the LHC using 
$\Delta y^2=(y_t-y_{\bar{t}})\cdot (y_t+y_{\bar{t}})$~\cite{Jung:2011zv} with 
$y_t$ ($y_{\bar{t}}$) being the rapidity of the top (anti-top) quark measured in the 
laboratory frame. The variable $\Delta y^2$ can be rewritten as 
$\Delta y^2=Q_\ell\cdot(y_{t_l}-y_{t_h})(y_{t_l}+y_{t_h})$. 
Hence, the charge asymmetry has been measured in both variables, 
$\Delta |\eta|$ and $\Delta y^2$.\\
Figure~\ref{fig:AC_CMS_first} (a) shows the reconstructed $\Delta y^2$ distribution, and in 
Figure~\ref{fig:AC_CMS_first} (b) the $\Delta y^2$ distribution after unfolding is presented, 
which is used to determine the corrected charge asymmetry $A_{\mathrm{C}}$.\\

The corrected asymmetry values measured by the CMS collaboration~\cite{Chatrchyan:2011hk} 
of $A_{\mathrm{C}}^\eta=-0.017$ $\pm 0.032\;\mbox{(stat.)}\;^{+0.025}_{-0.036}$ $\mbox{(syst.)}$ and 
$A_{\mathrm{C}}^{y}=-0.013\pm 0.028\;\mbox{(stat.)}\;^{+0.029}_{-0.031}\;\mbox{(syst.)}$ are consistent with 
the SM QCD+(full)EW NLO prediction of $A_{\mathrm{C}}^\eta(\mbox{theory})=0.0136\pm 0.0008$~\cite{Kuhn:2011ri} 
and $A_{\mathrm{C}}^y(\mbox{theory})=0.0115\pm 0.0006$~\cite{Kuhn:2011ri}.
Furthermore, background-subtracted asymmetries as a function of the reconstructed mass of the 
top quark pair $m_{t\bar{t}}$ have been measured and no statistically significant dependency 
on $m_{t\bar{t}}$ has been found.\\ 

In 2012 the ATLAS collaboration published a measurement of the charge asymmetry~\cite{ATLAS:2012an} 
using data at $\sqrt{s}=7\;
\mbox{TeV}$ corresponding to an integrated luminosity of $1.04\;\mbox{fb}^{-1}$. Here, the variable 
$\Delta |y|= |y_t|-|y_{\bar{t}}| = Q_{\ell}\cdot(|y_{t_l}|-|y_{t_h}|)$ is used instead of $\Delta y^2$. 
As the inclusive asymmetry in $\Delta |y|$ and $\Delta y^2$ is identical a direct comparison of both results
is feasible. The ATLAS collaboration measures a corrected inclusive asymmetry of 
$A_{\mathrm{C}}^y=-0.019 \pm 0.028 \mbox{(stat.)} \pm 0.024\mbox{(syst.)}$, which is well consistent with the CMS result~\cite{Chatrchyan:2011hk}.
Furthermore, the ATLAS collaboration measures the corrected asymmetry for two bins in $m_{t\bar{t}}$, 
where the same bin boarder of $450\;\mbox{GeV/c}^2$ as used at the Tevatron has been employed. No deviation
from the SM prediction computed with MC@NLO is observed.\\

Using CMS data at $\sqrt{s}=7\;\mbox{TeV}$ corresponding to an integrated luminosity of 
$5.0\;\mbox{fb}^{-1}$ the charge asymmetry measurement has been updated and further 
developed in the di\-plo\-ma thesis of Ref.~\cite{Roscher2012}. 
This thesis is the basis for the inclusive and differential measurements of the 
$t\bar{t}$ charge asymmetry by the CMS collaboration~\cite{Chatrchyan:2012cxa} published in 2012. 
In this updated measurement the event selection, background estimation, and reconstruction of the $t\bar{t}$ 
system have been adopted from the previous CMS measurement~\cite{Chatrchyan:2011hk}. To be consistent with the 
previous charge asymmetry measurement from the ATLAS collaboration~\cite{ATLAS:2012an}, the charge asymmetry 
is extracted using the variable $\Delta |y|$.\\

This time, not only an inclusive measurement is performed but also three differential measurements as a 
function of the rapidity $y_{t\bar{t}}$, transverse momentum $p_{\mathrm{T}}^{t\bar{t}}$, and invariant mass 
$m_{t\bar{t}}$ of the top quark pair system. Each of these three variables is sensitive to a certain 
aspect of the $t\bar{t}$ charge asymmetry. As the fraction of $q\bar{q}$ induced $t\bar{t}$ events 
rises with $|y_{t\bar{t}}|$, the charge asymmetry is enhanced with increasing $|y_{t\bar{t}}|$~\cite{Kuhn:2011ri}. 
As non-zero $p_{\mathrm{T}}^{t\bar{t}}$ is related to additional hard gluon radiation, the negative contribution of 
the charge asymmetry is enhanced due to the interference of FS with IS radiation~\cite{Kuhn:2011ri}. 
Furthermore, the charge asymmetry is expected to depend on $m_{t\bar{t}}$ because the fraction of $q\bar{q}$ induced 
$t\bar{t}$ events rises with $m_{t\bar{t}}$ and because potential new physics contributions would result via the 
interference with the SM process in an asymmetry increasing with $m_{t\bar{t}}$~\cite{Rodrigo:2010gm}.\\ 

To extract asymmetries as a function of a certain variable $V_i=y_{t\bar{t}},p_{\mathrm{T}}^{t\bar{t}},m_{t\bar{t}}$ the 
unfolding used in~\cite{Chatrchyan:2011hk} has been generalised to deal with two dimensional distributions in 
the diploma thesis of Ref.~\cite{Roscher2012}. For the differential measurement the efficiency depending on 
$\Delta |y|$ and $V_i$ has to be considered and in the migration matrices not only the migration between bins 
of the sensitive variable $\Delta |y|$ but also between bins of $V_i$ has to be taken into account. In order 
to stabilise the unfolding procedure and to avoid a loss of resolution huge effort has been put in the binning 
choice of the two-dimensional unfolding procedure resulting in a different $\Delta|y|$ binning for each bin of $V_i$. 
Due to this the vertical overlap between horizontally neighbouring bins is different and the regularisation 
procedure has been extended to cope with this.\\

In this updated analysis the $t\bar{t}$ signal is modelled with the NLO generator POWHEG~\cite{Alioli:2010xd} 
interfaced to PYTHIA version 6.4.24~\cite{Sjostrand:2006za} for the parton shower. Also the EW single top quark 
$t$-channel and $Wt$ production are simulated with the combination of POWHEG and PYTHIA. The top quark mass is 
set to $m_t=172.5\;\mbox{GeV/c}^2$ in all MC samples containing top quarks.
After applying the event selection, 57697 $t\bar{t}$ candidate events with $e/\mu$+jets signature are accepted 
with a background contamination of about 20\%. After the full reconstruction of the $t\bar{t}$ system, the 
one-dimensional $\Delta |y|$ distribution as well as two-dimensional distributions of $\Delta |y|$ and of one 
of the three variables $V_i$ are constructed on reconstruction level.\\
\begin{figure}[h]
\begin{center}
\subfigure[]{
\includegraphics[width=0.45\textwidth]{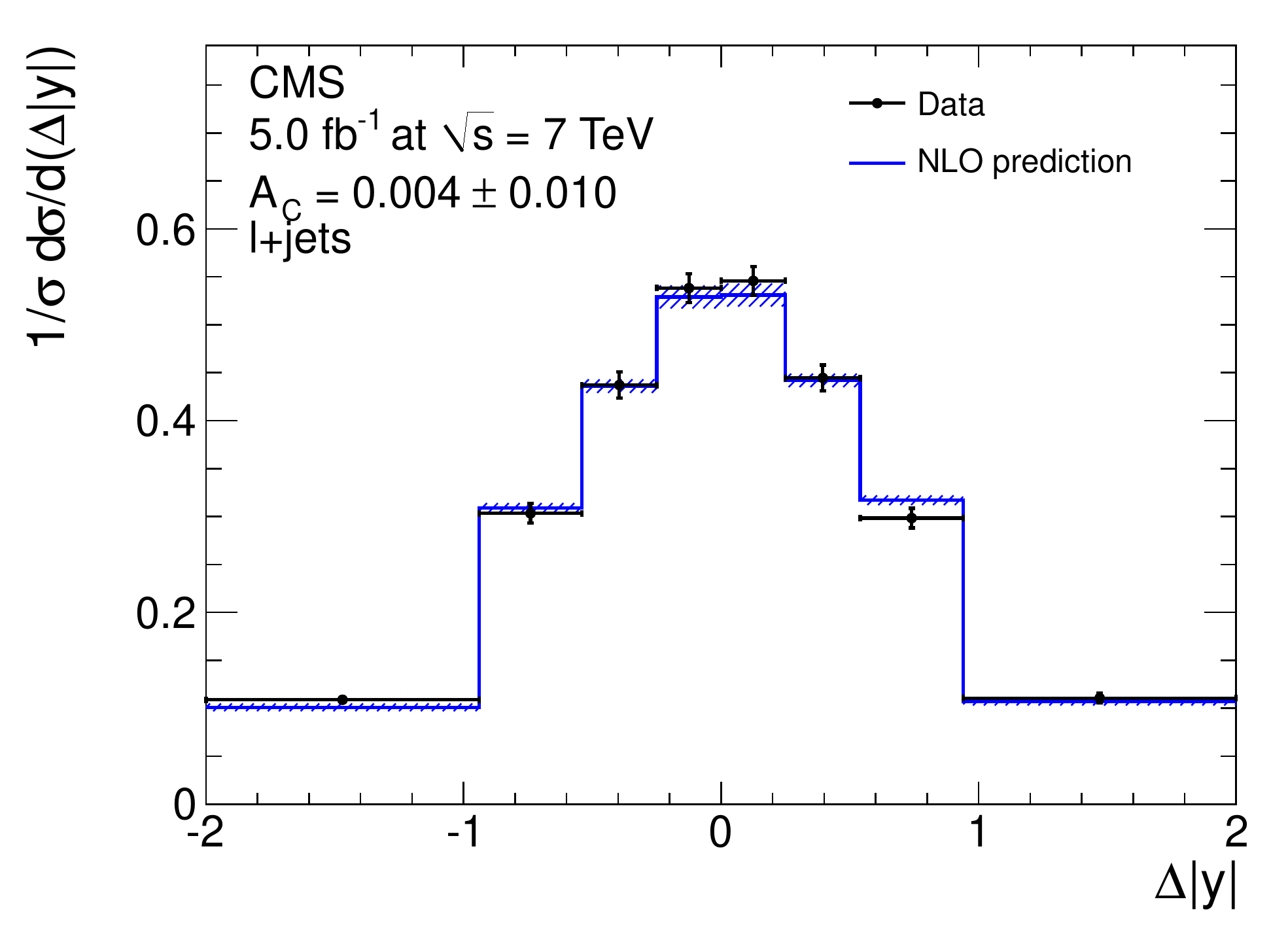}
}
\hspace*{0.5cm}
\subfigure[]{
\includegraphics[width=0.45\textwidth]{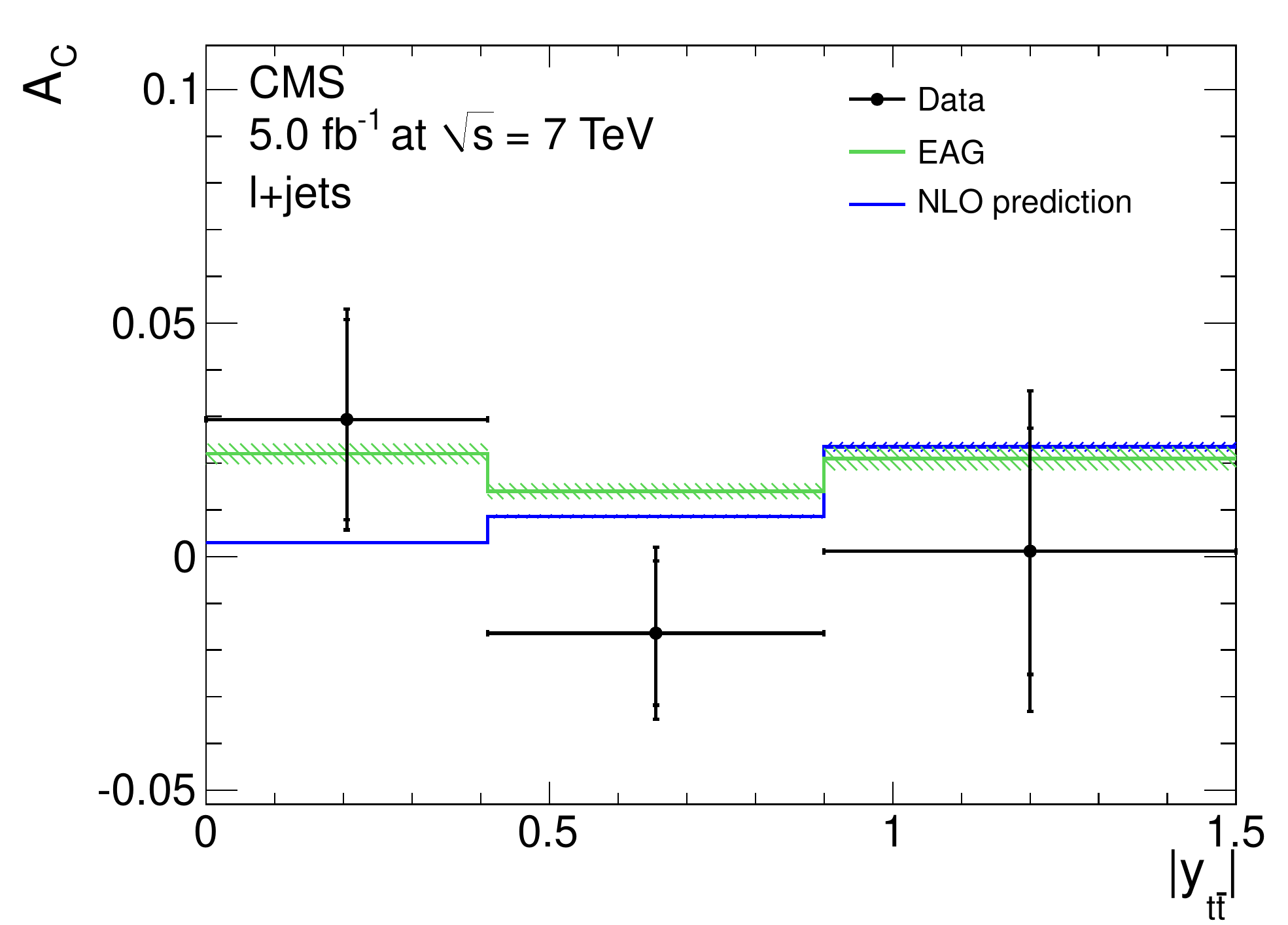}
}
\end{center}
\begin{center}
\subfigure[]{
\includegraphics[width=0.45\textwidth]{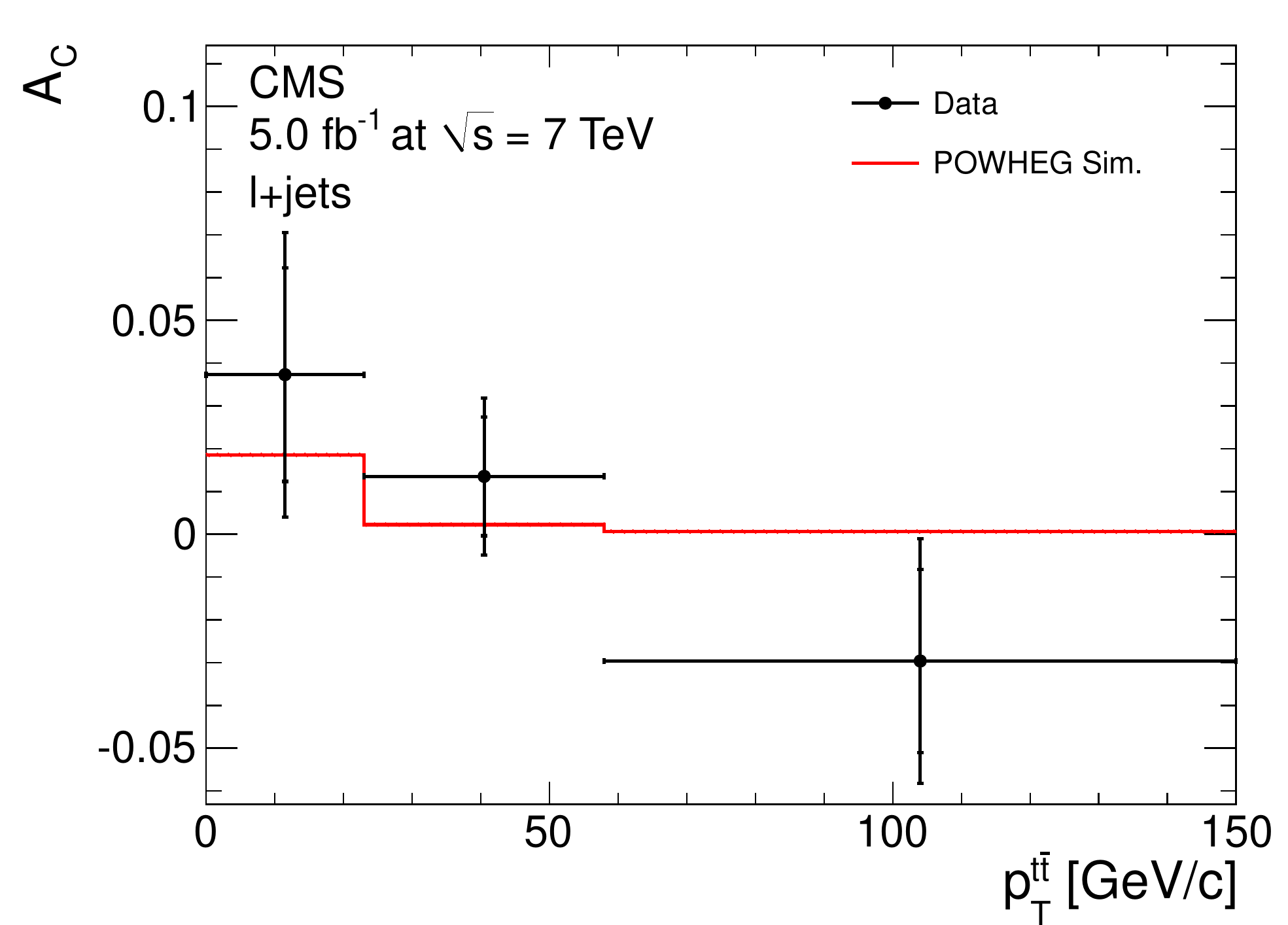}
}
\hspace*{0.5cm}
\subfigure[]{
\includegraphics[width=0.45\textwidth]{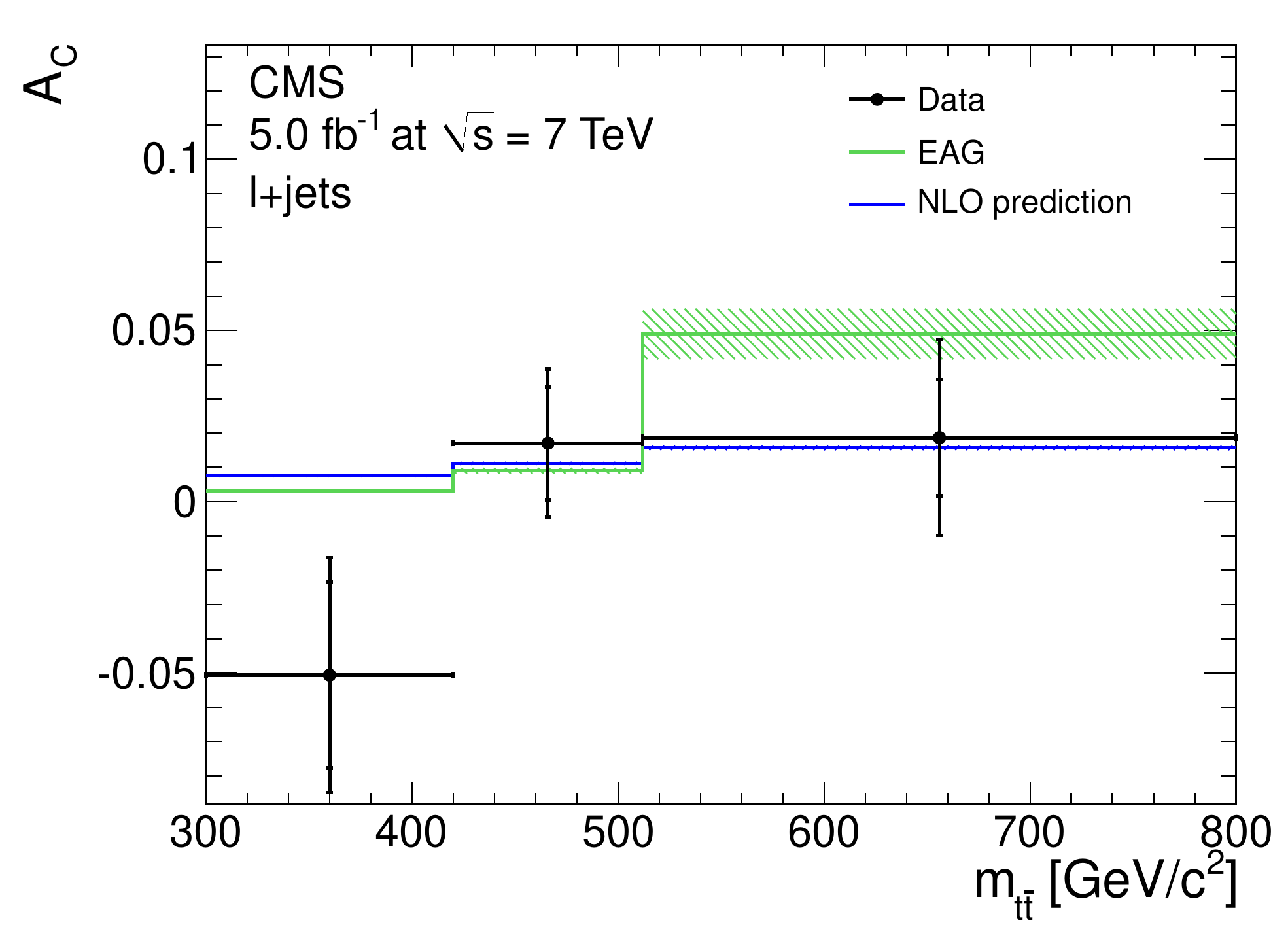}
}
\end{center}
\vspace*{-0.5cm}
\caption{Inclusive and differential charge asymmetry $A_{\mathrm{C}}$ in $t\bar{t}$ production as 
measured by the CMS collaboration using data at $\sqrt{s}=7\;\mbox{TeV}$ corresponding 
to an integrated luminosity of $5.0\;\mbox{fb}^{-1}$~\cite{Chatrchyan:2012cxa}. 
Top quark pair events with $e/\mu$+jets signature are employed.
(a) Unfolded inclusive $\Delta |y|$ distribution used to extract the inclusive asymmetry. 
(b-d) Corrected asymmetry as a function of $|y_{t\bar{t}}|$
(b), $p_{\mathrm{T}}^{t\bar{t}}$ (c), and $m_{t\bar{t}}$ (d). The measured values
are compared to the SM QCD+EW NLO predictions based on the calculations of Ref.~\cite{Kuhn:2011ri} 
and to the predictions of a model featuring an effective axial-vector coupling of
the gluon (EAG)~\cite{Brooijmans:2012yi}. The error bars on the differential asymmetry values indicate the 
statistical and total uncertainties, determined by adding statistical and systematic
uncertainties in quadrature. The shaded areas indicate the theoretical uncertainties on 
the NLO calculations.
}
\label{fig:AC_CMS_second}
\end{figure}

Figure~\ref{fig:AC_CMS_second} (a) shows the normalised $\Delta |y|$ spectrum after subtracting 
backgrounds and correcting for detector and efficiency effects using unfolding. Using this spectrum
to extract the corrected inclusive asymmetry a value of 
$A_{\mathrm{C}}^y=0.004\pm 0.010\;\mbox{(stat.)}\pm 0.011\;\mbox{(syst.)}$ is obtained by the CMS 
collaboration~\cite{Chatrchyan:2012cxa} which is consistent with the previous 
charge asymmetry measurements by the ATLAS~\cite{ATLAS:2012an} and CMS~\cite{Chatrchyan:2011hk} 
collaborations and which is well in agreement with the SM QCD+EW NLO prediction of Ref.~\cite{Kuhn:2011ri}. 
Furthermore, the corrected asymmetries as a function of $y_{t\bar{t}}$, $p_{\mathrm{T}}^{t\bar{t}}$, and 
$m_{t\bar{t}}$ (see Figure~\ref{fig:AC_CMS_second} (b)-(d)) are measured. The differential results 
are consistent with the SM QCD+EW NLO prediction ($y_{t\bar{t}}$ and $m_{t\bar{t}}$) and with 
the POWHEG computation ($p_{\mathrm{T}}^{t\bar{t}}$; no QCD+EW NLO prediction of Ref.~\cite{Kuhn:2011ri} exists as the 
LO $t\bar{t}$ cross section in the denominator is used which is zero for $p_{\mathrm{T}}^{t\bar{t}}>0$), respectively. 
Please note that the SM prediction from POWHEG has the same weakening as those from MCFM and MC@NLO 
compared to the QCD+EW NLO calculation from~\cite{Kuhn:2011ri}.\\
Although the data disfavour large deviations from the SM there is room for contributions from new 
physics beyond the SM (BSM) within the uncertainties. Further explorations with more statistics 
will be essential for a stronger conclusion.\\

In 2014, the ATLAS collaboration published an updated and extended charge asymmetry measurement using data 
at $\sqrt{s}=7\;\mbox{TeV}$ corresponding to an integrated luminosity of $4.7\;\mbox{fb}^{-1}$~\cite{Aad:2013cea}. 
The charge asymmetry is measured in $t\bar{t}$ events with $e/\mu$+jets signature and as sensitive 
variable to extract the asymmetry $\Delta |y|$ is used, again.\\ 
The corrected inclusive $t\bar{t}$ charge asymmetry is measured to be $A_{\mathrm{C}}^y=0.006\pm 0.010\,$(stat.\,+ syst.) 
and is consistent with the SM QCD+EW NLO predictions from Ref.~\cite{Bernreuther:2012sx}, which became 
available recently after the first publication on the charge asymmetry from CMS, of 
$A_{\mathrm{C}}^y(\mbox{theory})=0.0123\pm 0.005$ and from Ref.~\cite{Kuhn:2011ri} of $A_{\mathrm{C}}^y(\mbox{theory})=0.0115\pm 0.006$.
Furthermore, the ATLAS collaboration measured the corrected charge asymmetry for events with 
$m_{t\bar{t}}>600\;\mbox{GeV/c}^2$, as well as a function of $y_{t\bar{t}}$, $p_{\mathrm{T}}^{t\bar{t}}$, $m_{t\bar{t}}$. 
To enhance the sensitivity to BSM effects~\cite{AguilarSaavedra:2011cp} the inclusive $A_{\mathrm{C}}$ result and the 
differential result as a function of $m_{t\bar{t}}$ are also measured with the additional requirement of a 
minimum velocity $\beta_{z,t\bar{t}}$ of the $t\bar{t}$-system along the beam axis ($\beta_{z,t\bar{t}}>0.6$). 
Here, no unfolding of $\beta_{z,t\bar{t}}$ is performed as the authors found that resolution effects on the 
reconstructed $\beta_{z,t\bar{t}}$ variable do not introduce any bias in the measurement.\\ 
All these measurements are statistically limited and are found to be compatible with the SM prediction 
within the uncertainties.\\

\begin{table}[t]
\centering
\begin{tabular}{l|c}
\toprule
   & Asymmetry \\ \midrule
CMS, $\ell$+j, $L=1.09\;\mbox{fb}^{-1}$, \cite{Chatrchyan:2011hk} & $A_{\mathrm{C}}^{y}=-0.013^{+0.040}_{-0.042}$ \\[0.1cm]
ATLAS, $\ell$+j, $L=1.04\;\mbox{fb}^{-1}$, \cite{ATLAS:2012an} &  $A_{\mathrm{C}}^y=-0.019 \pm 0.037$\\[0.1cm]
CMS, $\ell$+j, $L=5.0\;\mbox{fb}^{-1}$, \cite{Chatrchyan:2012cxa} & $A_{\mathrm{C}}^y=0.004\pm 0.015$ \\[0.1cm]
ATLAS,  $\ell$+j,  $L=4.7\;\mbox{fb}^{-1}$, \cite{Aad:2013cea} & $A_{\mathrm{C}}^y=0.006\pm 0.010$ \\ \midrule
CMS, $\ell\ell$, $L=5.0\;\mbox{fb}^{-1}$, \cite{Chatrchyan:2014yta} & $A_{\mathrm{C}}^y=-0.010\pm 0.019$ \\[0.1cm]
ATLAS, $\ell\ell$, $L=4.6\;\mbox{fb}^{-1}$, \cite{Aad:2015jfa} & $A_{\mathrm{C}}^y=0.021\pm 0.030$ \\ \midrule
Theory (NLO, QCD+EW), \cite{Bernreuther:2012sx} & $A_{\mathrm{C}}^y= 0.0123\pm 0.005$\\[0.1cm]
Theory (NLO, QCD+EW), \cite{Kuhn:2011ri} & $A_{\mathrm{C}}^y=0.0115\pm 0.006$\\ \midrule\midrule
CMS, $\ell\ell$, $L=5.0\;\mbox{fb}^{-1}$, \cite{Chatrchyan:2014yta} & $A_{\mathrm{C}}^{\ell\ell}=0.009\pm 0.012$ \\[0.1cm]
ATLAS, $\ell\ell$, $L=5.0\;\mbox{fb}^{-1}$, \cite{Aad:2015jfa} & $A_{\mathrm{C}}^{\ell\ell}=0.024\pm 0.017$ \\  \midrule
Theory (NLO, QCD+EW), \cite{Bernreuther:2012sx} & $A_{\mathrm{C}}^{\ell\ell}=0.007\pm 0.003$\\ \bottomrule
\end{tabular}
\caption{Summary of the measured inclusive charge asymmetries in $t\bar{t}$ production at the $pp$ collider LHC 
with $\sqrt{s}=7\;\mbox{TeV}$ that have been published in a journal (status May 2015). $A_{\mathrm{C}}^y$ is the asymmetry 
in the variable $\Delta |y|=|y_t|-|y_{\bar{t}}|$, $A_{\mathrm{C}}^{\ell\ell}$ is the asymmetry in the variable 
$\Delta \eta^{\ell\ell}=|\eta_{\ell^+}|-|\eta_{\ell^-}|$. $\ell$+j and $\ell\ell$ indicate the $e/\mu$+jets and 
dilepton ($\ell = e,\mu$) $t\bar{t}$ decay channels. The SM calculations are performed by K\"uhn and Rodrigo~\cite{Kuhn:2011ri} 
and by Bernreuther and Si~\cite{Bernreuther:2012sx}.}
\label{tab:Ac_LHC}
\end{table}
At both LHC experiments the $t\bar{t}$ charge asymmetry has been recently measured also in the dilepton decay channel 
($\ell\ell$ with $\ell=e,\mu$). Beside the $t\bar{t}$ charge asymmetry extracted from the variable $\Delta |y|$ also 
the lepton-based charge asymmetry $A_{\mathrm{C}}^{\ell\ell}$ defined as:
\begin{equation}
A_{\mathrm{C}}^{\ell\ell} = \frac{N(\Delta |\eta^{\ell\ell}| >0) - N(\Delta |\eta^{\ell\ell}| <0)}{N(\Delta |\eta^{\ell\ell}| >0) + N(\Delta |\eta^{\ell\ell}| <0)}
\end{equation}
is measured, where $\Delta |\eta^{\ell\ell}|=|\eta_{\ell^+}|-|\eta_{\ell^-}|$ is used as sensitive variable. 
In this definition, $\eta^{\ell^+}$ ($\eta^{\ell^-}$) is the pseudorapidity of the positively (negatively) 
charged lepton and $N$ is the number of events with positive or negative $\Delta |\eta^{\ell\ell}|$.
Both asymmetries are corrected for detector, efficiency and acceptance effects. In addition to the inclusive 
asymmetries the CMS collaboration measures the corrected $A_{\mathrm{C}}^{\ell\ell}$ as a function of $y_{t\bar{t}}$, $p_{\mathrm{T}}^{t\bar{t}}$, 
and $m_{t\bar{t}}$. All LHC charge asymmetry measurements in the dilepton decay channel~\cite{Chatrchyan:2014yta,Aad:2015jfa} 
are consistent with the SM QCD+EW NLO prediction of Ref.~\cite{Bernreuther:2012sx}. Table~\ref{tab:Ac_LHC} summarises 
the status as of May 2015 of the measured inclusive corrected charge asymmetries in $t\bar{t}$ production at the LHC. 
All these measurements are consistent with the SM prediction.\\

\begin{table}[tb]
\centering
\begin{tabular}{l|c}
\toprule
   & Asymmetry \\ \midrule
CDF, $\ell$+j, $L=1.9\;\mbox{fb}^{-1}$, \cite{Aaltonen:2008hc} & $A_{\mathrm{FB}}^{t\bar{t}}=0.24\pm 0.14$ \\ 
CDF, $\ell$+j, $L=5.3\;\mbox{fb}^{-1}$, \cite{Aaltonen:2011kc} &  $A_{\mathrm{FB}}^{t\bar{t}}=0.158\pm 0.075$\\
D0, $\ell$+j, $L=5.4\;\mbox{fb}^{-1}$, \cite{Abazov:2011rq} & $A_{\mathrm{FB}}^{t\bar{t}}=0.196\pm 0.065$ \\
CDF, $\ell$+j, $L=9.4\;\mbox{fb}^{-1}$, \cite{Aaltonen:2012it} &  $A_{\mathrm{FB}}^{t\bar{t}}=0.164\pm 0.045$\\
D0, $\ell$+j, $L=9.7\;\mbox{fb}^{-1}$, \cite{Abazov:2014cca} & $A_{\mathrm{FB}}^{t\bar{t}}=0.106\pm 0.030$ \\ \midrule
Theory (NLO, QCD+EW), \cite{Hollik:2011ps} & $A_{\mathrm{FB}}^{t\bar{t}}= 0.089^{+0.008}_{-0.006}$\\
Theory (NLO, QCD+EW), \cite{Kuhn:2011ri} & $A_{\mathrm{FB}}^{t\bar{t}}= 0.087\pm 0.010$\\
Theory (NLO, QCD+EW), \cite{Bernreuther:2012sx} & $A_{\mathrm{FB}}^{t\bar{t}}=0.0875^{+0.0058}_{-0.0048}$\\
Theory (NNLO, QCD+EW), \cite{Czakon:2014xsa} & $A_{\mathrm{FB}}^{t\bar{t}}= 0.095\pm 0.007$\\ \midrule \midrule
CDF, $\ell\ell$ and $\ell$+j, up to $L=9.4\;\mbox{fb}^{-1}$, \cite{Aaltonen:2014eva} & $A_{\mathrm{FB}}^{\ell}= 0.090^{+0.028}_{-0.026}$ \\ 
D0, $\ell\ell$ and $\ell$+j, $L=9.7\;\mbox{fb}^{-1}$, \cite{Abazov:2014oea} & $A_{\mathrm{FB}}^{\ell}= 0.047\pm 0.027$ \\ \midrule
Theory (NLO, QCD+EW), \cite{Bernreuther:2012sx} & $A_{\mathrm{FB}}^{\ell}= 0.038\pm 0.003$\\ \midrule \midrule
CDF, $\ell\ell$, $L=9.1\;\mbox{fb}^{-1}$, \cite{Aaltonen:2014eva} & $A_{\mathrm{FB}}^{\ell\ell}= 0.076\pm 0.082$ \\ 
D0, $\ell\ell$, $L=9.7\;\mbox{fb}^{-1}$, \cite{Abazov:2013wxa} & $A_{\mathrm{FB}}^{\ell\ell}=0.123\pm 0.056$ \\  \midrule
Theory (NLO, QCD+EW), \cite{Bernreuther:2012sx} & $A_{\mathrm{FB}}^{\ell\ell}= 0.048\pm 0.004$\\ \bottomrule\bottomrule
\end{tabular}
\caption{Summary of the measured inclusive corrected charge asymmetries in $t\bar{t}$ production (extrapolated to full phase space) 
at the $p\bar{p}$ collider Tevatron with $\sqrt{s}=1.96\;\mbox{TeV}$ that have been published in a journal (status March 2015). 
$A_{\mathrm{FB}}^{t\bar{t}}$ is the asymmetry determined in the $t\bar{t}$ rest frame measured using the variable $\Delta y=y_t-y_{\bar{t}}$, $A_{\mathrm{FB}}^{\ell}$ 
is the asymmetry in the variable $Q_{\ell}\cdot\eta_{\ell}$, and $A_{\mathrm{FB}}^{\ell\ell}$ is the asymmetry in the variable 
$\Delta\eta^{ll}=\eta_{l^+}-\eta_{l^-}$. $\ell$+j and $\ell\ell$ indicate the $e/\mu$+jets and dilepton ($\ell=e,\mu$) $t\bar{t}$ 
decay channels, respectively. The SM calculations are performed by Hollik and Pagani~\cite{Hollik:2011ps}, by K\"uhn and 
Rodrigo~\cite{Kuhn:2011ri}, by Bernreuther and Si~\cite{Bernreuther:2012sx}, and by Czakon, Fiedler and Mitov~\cite{Czakon:2014xsa}.}
\label{tab:Ac_Tevatron}
\end{table}
Also the charge asymmetry measurements in top quark pair production at the CDF 
experiment~\cite{Aaltonen:2012it,CDF:2013gna,Aaltonen:2013vaf,Aaltonen:2014eva} and the D0 
experiment~\cite{Abazov:2012oxa,Abazov:2013wxa,Abazov:2014oea,Abazov:2014cca} have been updated and extended. 
In particular also the dilepton $t\bar{t}$ decay channel has been explored with emphasis on leptonic asymmetries. 
A summary of the corrected inclusive asymmetries extrapolated to the full phase space measured at the Tevatron is 
presented in Table~\ref{tab:Ac_Tevatron}. All measured asymmetries are above the predicted value, but in particular 
the leptonic asymmetries $A_{\mathrm{FB}}^{\ell}$ (see Eq. (\ref{eq:Afbl})) and $A_{\mathrm{FB}}^{\ell\ell}$ are consistent with the 
SM prediction. Here, $A_{\mathrm{FB}}^{\ell\ell}$ indicates the asymmetry in the variable $\Delta\eta^{\ell\ell}=\eta_{\ell^{+}}-\eta_{\ell^{-}}$ 
and is defined as:
\begin{equation}
A_{\mathrm{FB}}^{\ell\ell}=\frac{N(\Delta\eta^{\ell\ell}>0)-N(\Delta\eta^{\ell\ell}<0)}{N(\Delta\eta^{\ell\ell}>0)+N(\Delta\eta^{\ell\ell}<0)}
\end{equation}
with $\eta^{\ell^{+}}$ ($\eta^{\ell^{-}}$) being the pseudorapidity of the positively (negatively) charged lepton measured in the laboratory frame.\\

The most recent inclusive measurement of $A_{\mathrm{FB}}^{t\bar{t}}$ from the CDF collaboration~\cite{Aaltonen:2012it} is compatible within $1.5\sigma$ 
with the NNLO prediction~\cite{Czakon:2014xsa} while the most recent measurement of the D0 collaboration~\cite{Abazov:2014cca} is in perfect 
agreement with this new calculation but also in good agreement with older NLO (QCD+EW) calculations~\cite{Kuhn:2011ri,Bernreuther:2012sx}. 
In contrast to CDF, the D0 collaboration increased the sensitivity to the charge asymmetry in the most recent publication in the $e/\mu$+jets 
channel~\cite{Abazov:2014cca} by adding also events with three reconstructed jets. This change reduced also the acceptance corrections.\\
%
The corrected asymmetries in the $t\bar{t}$ rest frame, $A_{\mathrm{FB}}^{t\bar{t}}$, measured by both Tevatron experiments in the $e/\mu$+jets channel 
increases with $m_{t\bar{t}}$~\cite{Aaltonen:2012it,Abazov:2014cca}. While the two results are in modest agreement for 
$m_{t\bar{t}}<650\;\mbox{GeV/c}^2$, there is a significant deviation in the last bin (associated with large uncertainties) and which 
causes fitted slopes that are discordant at the $1.8\sigma$ level~\cite{Aguilar-Saavedra:2014kpa}. Please note, that the discrepancy 
of the corrected $A_{\mathrm{FB}}^{t\bar{t}}$ for events with $m_{t\bar{t}}\ge 450\;\mbox{GeV/c}^2$ compared to the NLO+EW prediction 
from~\cite{Kuhn:2011ri,Bernreuther:2012sx} of $3.0\sigma$ ($3.4\sigma$ if compared to MCFM) observed previously by the CDF 
collaboration~\cite{Aaltonen:2011kc} is reduced to $2.5\sigma$ in the final CDF measurement.\\ 
While the recent NNLO calculation~\cite{Czakon:2014xsa} shows a similar dependence of $A_{\mathrm{FB}}^{t\bar{t}}$ on $m_{t\bar{t}}$ and 
$|\Delta y|$ as the NLO (QCD+EW) predictions~\cite{Kuhn:2011ri,Bernreuther:2012sx} there is an offset in $A_{\mathrm{FB}}^{t\bar{t}}$ 
as a function of $p_{\mathrm{T}}^{t\bar{t}}$ for $p_{\mathrm{T}}^{t\bar{t}}>10\;\mbox{GeV/c}$ compared to the NLO (QCD+EW) 
predictions~\cite{Kuhn:2011ri,Bernreuther:2012sx}. Please note, that the predicted asymmetry values at 
background-subtracted level or in a visible phase space, rely on $A_{\mathrm{FB}}^{t\bar{t}}(p_{\mathrm{T}}^{t\bar{t}})$ from NLO+PS MCs 
and that the measured corrected asymmetries use for the correction of detector (minor for $A_{\mathrm{FB}}^{\ell}$) and acceptance 
effects NLO+PS MCs. Hence, if $\Delta y=y_t-y_{\bar{t}}$ as a function of $p_{\mathrm{T}}^{t\bar{t}}$ would not be modelled entirely 
correctly in NLO+PS MCs this could probably give rise to some measurement biases.\\

A very nice review about the measured asymmetries in top quark pair production at the Tevatron and LHC as well as about 
possible new physics models that could explain the asymmetry excess seen at the Tevatron is presented in 
Ref.~\cite{Aguilar-Saavedra:2014kpa} (status June 2014). The authors point out, that the explanation of the asymmetry 
excess with new physics faces two serious problems. First almost all successful models are rather ad hoc, secondly most 
new physics models (even with non generic choices of parameters) still predict a series of observable signals that have 
not been found. Possible new physics models that could comply with all the top quark measurements, including among 
others the asymmetries at Tevatron and LHC and differential $t\bar{t}$ distributions, are for example a $s$-channel 
colour-octet vector boson (with some non-trivial ingredients)~\cite{Ferrario:2008wm} or a light isodoublet exchanged in the 
$t$-channel~\cite{Nelson:2011us}. In the review of Ref.~\cite{Aguilar-Saavedra:2014kpa} the authors mention that there is still 
the possibility open that higher order QCD corrections are larger than expected and increase significantly the value of the 
predicted asymmetry at the Tevatron. However the authors see one issue of this possible explanation. As these corrections are 
isospin preserving, they will simultaneously raise the prediction of the charge asymmetry at the LHC, making it inconsistent 
with the present measurements.\\ 
The recent NNLO calculation~\cite{Czakon:2014xsa} predicts indeed an about 8\% larger inclusive asymmetry $A_{\mathrm{FB}}^{t\bar{t}}$ 
for the Tevatron than previous calculations at NLO~\cite{Kuhn:2011ri,Bernreuther:2012sx}. 
Therefore, it would be important that also the NNLO prediction for the inclusive asymmetry $A_{\mathrm{C}}^y$ and as a 
function of $|y_{t\bar{t}}|$, $m_{t\bar{t}}$, and $p_{\mathrm{T}}^{t\bar{t}}$ would become available to see whether the 
NNLO calculation is in agreement with the measurements from the LHC.\\
 
In Ref.~\cite{Westhoff:2015mfa} prospects to investigate the top-quark charge asymmetry at the LHC in and beyond the 
standard model are summarised. Here, particular attention is given to observables in $t\bar{t}$ production in association 
with a jet, a photon or a $W$ boson. In all these cases the absolute value of the predicted charge asymmetry would be 
increased in the SM (and in new physics models) compared to the predicted value in pure $t\bar{t}$ production.\\
An other way to proceed could be the measurement of collider independent asymmetries~\cite{AguilarSaavedra:2012rx}, which 
are as pointed out in Ref.~\cite{Aguilar-Saavedra:2014kpa} very demanding from the experimental side, but would offer a 
unique possibility of testing at the LHC the same quantities that are in the origin of the Tevatron asymmetry.





\section{Search for Resonant Top Quark Pair Production}
\label{sec:mttbar}

In this section a short motivation of searches for resonant $t\bar{t}$ production is presented. 
Then, the experimental studies and the achievements are described using as example analysis the first
search for resonant top quark pair production in the lepton+jets channel from CMS~\cite{Chatrchyan:2012cx}.
As the search reach of heavy new particles decaying to top quark pairs has been extended 
into the TeV region as soon as the LHC started running at $\sqrt{s}=7\,\mbox{TeV}$ in 2010, 
only an overview of results from the LHC is presented in the following.

\subsection[Motivation of the Search for Resonant $t\bar{t}$ Production]{Motivation of the Search for Resonant \boldmath{$t\bar{t}$} Production}

In the SM, the spectrum of the invariant mass of the top quark pair $m_{t\bar{t}}$ starts at 
the threshold of $~2m_t$, reaches its maximum just above the production threshold and it is 
steeply falling towards large $m_{t\bar{t}}$. The higher the centre-of-mass energy of the collider 
is, the larger is the tail in the $m_{t\bar{t}}$ spectrum. The differential $t\bar{t}$ cross section 
as a function of $m_{t\bar{t}}$ has been measured at the Tevatron~\cite{Aaltonen:2009iz,Abazov:2014vga} 
and at the LHC~\cite{Chatrchyan:2012saa,Khachatryan:2015oqa,Aad:2012hg,Aad:2014zka,Aad:2015eia}. 
At both accelerators, the SM predictions describe the $m_{t\bar{t}}$ spectrum within uncertainties 
reasonably well. Recent calculations in the non-relativistic QCD 
framework~\cite{Kiyo:2008bv,Hagiwara:2008df} predict for the LHC, that just below the 
$2m_t$ threshold an additional resonance occurs. The resonance is a remnant of a loosely bound 
$t\bar{t}$ state in a colour-singlet $S$-wave configuration in the gluon fusion channel. 
Resolving this close to threshold resonance is experimentally very challenging and so far no measurement 
on this exists.\\

Although the SM is amazingly successful in describing particle physics data, there are also very 
important open questions not addressed by the SM: neither the existence of dark matter or the 
huge matter antimatter asymmetry in the universe nor a unified treatment of all known forces, 
including gravity, are explained. It is expected that the SM is a low energy approximation of a 
more fundamental description of nature. With a mass of 
$m_t=173.34\pm 0.76\;\mbox{GeV/c}^2$~\cite{ATLAS:2014wva} the top quark is by far the heaviest 
known elementary particle. Due to its very large mass and hence its large coupling to the Higgs 
boson, it is widely believed that the top quark plays a key role in all theoretical models 
extending the Higgs mechanism of the SM and addressing one or more open questions not explained 
by the SM. Many of these theories predict the existence of heavy resonances decaying to top 
quark pairs resulting in an additional resonant component to the SM $t\bar{t}$ production, 
which would disturb the $m_{t\bar{t}}$ spectrum predicted in the SM~\cite{Frederix:2007gi}. 
There exist many examples of such resonances that decay preferentially into $t\bar{t}$. 
These include models with massive colour-singlet Z-like bosons in extended gauge 
theories~\cite{Rosner:1996eb,Lynch:2000md,Carena:2004xs}, 
colorons~\cite{Hill:1991at,Harris:2011ez,Hill:1993hs,Hill:1994hp} or 
axigluons~\cite{Frampton:1987dn,Choudhury:2007ux}, models in which a pseudoscalar Higgs 
boson may couple strongly to top quarks~\cite{Dicus:1994bm}, and models with extra dimensions, 
such as Kaluza-Klein (KK) excitations of gluons~\cite{Agashe:2006hk} or 
gravitons~\cite{Davoudiasl:1999jd} in various extensions of the 
Randall-Sundrum model~\cite{Randall:1999ee} or in the ADD extra-dimensional model~\cite{ArkaniHamed:1998rs}.\\

As discussed in chapter~\ref{sec:asymmetry} discrepancies compared to the SM prediction have been observed 
since 2008 in the charge asymmetry in top quark pair production measured at the Tevatron, which are still 
present but are less significant with the final Tevatron data set. Assuming this anomaly is due to new physics 
above the TeV scale, an enhancement of the $t\bar{t}$ rate at high invariant mass could be visible at the 
LHC~\cite{AguilarSaavedra:2011vw,Delaunay:2011gv}.\\

At the Tevatron searches for resonant top quark pair production have been performed by both experiments 
CDF~\cite{Aaltonen:2012af,Aaltonen:2007ag,Aaltonen:2007ak,Aaltonen:2009tx,Aaltonen:2011ts,Aaltonen:2011vi} 
and D0~\cite{Abazov:2011gv,Abazov:2008ny}. No evidence for resonant production has been found and upper limits 
on the production cross section of narrow resonances ($Z'$ with mass below $915\;\mbox{GeV/c}^2$ at 95\% 
confidence level (C.L.)~\cite{Aaltonen:2012af}) decaying into $t\bar{t}$ have been set.


\subsection[Search for Resonant $t\bar{t}$ Production in the Lepton+Jets Channel at the LHC]{Search for Resonant \boldmath{$t\bar{t}$} Production in the Lepton+Jets Channel at the LHC}
In the diploma thesis of Ref.~\cite{Ott:2009aka} finished in 2009 a Monte Carlo simulation based 
study for narrow heavy resonances decaying to $t\bar{t}$ in the mass range $m_{t\bar{t}}>1\;\mbox{TeV/c}^2$ 
has been performed using $t\bar{t}$ events in the muon+jets channel. 
In this study a scenario of MC data corresponding to an integrated luminosity of $L=200\;\mbox{pb}^{-1}$ 
collected with the CMS detector and at a LHC centre-of-mass energy of 10 TeV has been employed.\\ 

Top quarks emerging from heavy resonances have usually a large momentum (large ``boost'') resulting in 
top quark decay products that have only small angular separation and hence produce signals that 
overlap in the detector. So the charged lepton from the leptonic top quark decay $t_l$ 
($t\rightarrow b\ell\nu$) is less isolated for boosted top quarks and the probability that 
the charged lepton is merged with the $b$-jet originating from the same top quark rises 
with top quark momentum. Furthermore, the jets originating from the hadronic top quark decay 
$t_h$ ($t\rightarrow bq\bar{q}'$) merge first partially and finally fully with increasing 
top quark momentum. Due to this the ``standard'' $t\bar{t}$ event selection requiring one 
isolated charged lepton, at least four hadronic jets whereof one is identified as $b$-jet 
(as for example used in the charge asymmetry analyses from CDF and CMS, see chapter~\ref{sec:asymmetry}), 
fails to select efficiently boosted top quark pairs.\\ 

Therefore, an event selection optimised for the boosted $t\bar{t}$ topology in the muon+jets channel 
has been developed in Ref.~\cite{Ott:2009aka}. This includes the triggers, the number of reconstructed 
jets, the isolation criterion of charged leptons and additional cuts to suppress multi-jet background 
which gets largely enhanced by loosening the standard $t\bar{t}$ selection. Furthermore, also the 
reconstruction of the $t\bar{t}$ system has been adjusted in~\cite{Ott:2009aka} compared to the one 
used in the charge asymmetry analyses of Ref.~\cite{Peiffer2011,Aaltonen:2008hc} to cope with the boosted 
$t\bar{t}$ topology. Here, each jet is either assigned to the leptonic top quark decay $t_l$ or to the 
hadronic top quark decay $t_h$ and the event hypothesis that describes the boosted $t\bar{t}$ kinematic 
best is selected by a requirement based on differences in the $\eta$-$\phi$-plane of $t_l$ and its decay 
products and of $t_l$ and $t_h$.\\

The analysis strategy developed in~\cite{Ott:2009aka} has been applied to real CMS data at $\sqrt{s}=7\;\mbox{TeV}$ 
corresponding to an integrated luminosity of $36\;\mbox{pb}^{-1}$ in the PhD thesis of Ref.~\cite{Peiffer2011}. 
Again, only $t\bar{t}$ events in the muon+jets channel are considered. In this thesis~\cite{Peiffer2011} a method 
to estimate and model the remaining multi-jet background entirely from data has been developed and applied. 
A simultaneous binned likelihood fit in $m_{t\bar{t}}$ in the signal region and in $L_{\mathrm{T}}$, defined as the scalar 
sum of the charge lepton transverse energy and missing transverse energy, in a control region enriched with 
background events is performed.\\ 
In the absence of an observed narrow resonance, limits on the production cross section of the order of several 
pb at 95\% C.L. are set for masses above $1\,\mbox{TeV/c}^2$. As noted in~\cite{Peiffer2011}, the reached cross 
section limits
could be improved in this analysis by a factor of up to four for resonance masses of several $\mbox{TeV/c}^2$ 
compared to a similar analysis optimised for resonance searches close to the threshold.\\

Starting from the same analysis strategy as in the muon+jets analysis~\cite{Peiffer2011} a first study using 
the electron+jets channel has been performed in the diploma thesis of Ref.~\cite{Voigt2011}. The study showed 
that the multi-jet background is more severe in the electron+jets channel than in the muon+jets channel and 
that an additional handling is needed.\\

Using CMS data at $\sqrt{s}=7\;\mbox{TeV}$ corresponding to an integrated luminosity 
of $5.0\;\mbox{fb}^{-1}$ the CMS collaboration published at the end of 2012 a search for 
resonant top quark pair production in the $e/\mu$+jets decay channel~\cite{Chatrchyan:2012cx}. 
The range of $(0.5-3.0)\,\mbox{TeV/c}^2$ in $m_{t\bar{t}}$ is covered by the combination of 
two dedicated searches: one optimised for resonances for masses smaller than $1\,\mbox{TeV/c}^2$ 
(threshold region), and a second one optimised for masses larger than 1 $\mbox{TeV/c}^2$ (boosted region). 
The PhD thesis of Ref.~\cite{Ott:2013cut} is the basis of the search in the boosted region performed 
in the muon+jets channel and contains several improvements compared to the previous 
studies~\cite{Ott:2009aka,Peiffer2011,Voigt2011}.\\

All analyses of the publication~\cite{Chatrchyan:2012cx} split the data set in categories depending 
on the number of reconstructed jets and the number of jets identified as $b$-jets and perform a 
template-based statistical evaluation of the reconstructed $m_{t\bar{t}}$ distribution in all categories. 
In all analyses jets are reconstructed using an anti-$k_T$ algorithm with a distance parameter of 
$R=0.5$~\cite{Cacciari:2008gp} and corrections to account for the $\eta$ and $p_{\mathrm{T}}$ dependent detector 
response to jets~\cite{Chatrchyan:2011ds} and the effect of pileup are applied. Jets associated to $b$ 
quarks are identified using the Combined Secondary Vertex (CSV) algorithm~\cite{CMS:2009gxa} that reconstructs 
the secondary vertex corresponding to the decay of a $B$ hadron. When no secondary vertex is found, the 
significance of the impact parameter with respect to the primary vertex of the second most displaced track 
is used as a discriminator to distinguish decay products of a $B$ hadron from prompt tracks~\cite{CMS:2009gxa}.\\ 

The event selection in the boosted region has been refined in ~\cite{Chatrchyan:2012cx,Ott:2013cut} 
compared to the initial studies described above. In the updated analysis, events with one muon (electron) 
without isolation requirement but with $p_{\mathrm{T}}>42\,\mbox{GeV/c}$ and $|\eta|<2.1$ ($p_{\mathrm{T}}>70\,\mbox{GeV/c}$ and 
$|\eta|<2.5$), at least two reconstructed jets with $p_{\mathrm{T}}>50\,\mbox{GeV/c}$ and $|\eta|<2.4$, whereof the 
leading jet has to fulfil $p_{\mathrm{T}}>250\,\mbox{GeV/c}$ ($p_{\mathrm{T}}>150\,\mbox{GeV/c}$), and with $L_{\mathrm{T}}>150\;\mbox{GeV}$ 
are selected. Additional cuts to suppress multi-jet background are applied, whereat stronger and more 
sophisticated cuts are applied in the electron+jets channel. The two boosted analyses select 1200 candidate 
events with a $t\bar{t}$ purity of almost 80\%.\\
Resonant top quark production is modelled using PYTHIA 8.1, $t\bar{t}$ and $W/Z$+jets production are modelled 
using MADGRAPH 5.1 interfaced to PYTHIA 6.4 for the parton shower simulation, and single top quark production 
is modelled using POWHEG interfaced to PYTHIA for the parton showering. All top quark MC samples are generated 
with a top quark mass of $m_t=172.5\,\mbox{GeV/c}^2$.\\
The reconstruction method firstly developed in~\cite{Ott:2009aka} has been refined in~\cite{Ott:2013cut}. Here, 
a two-term $\chi^2$ to select the best event hypothesis is constructed from the sum of the normalised squared 
deviations of the leptonic top quark mass and the hadronic top quark mass. The event hypothesis with the smallest 
$\chi^2$ value ($\chi^2_{min}$) is selected. To reduce background further, both boosted analyses require $\chi^2_{min}<8$.\\ 

\begin{figure}[t]
\begin{center}
\subfigure[]{
\includegraphics[width=0.45\textwidth]{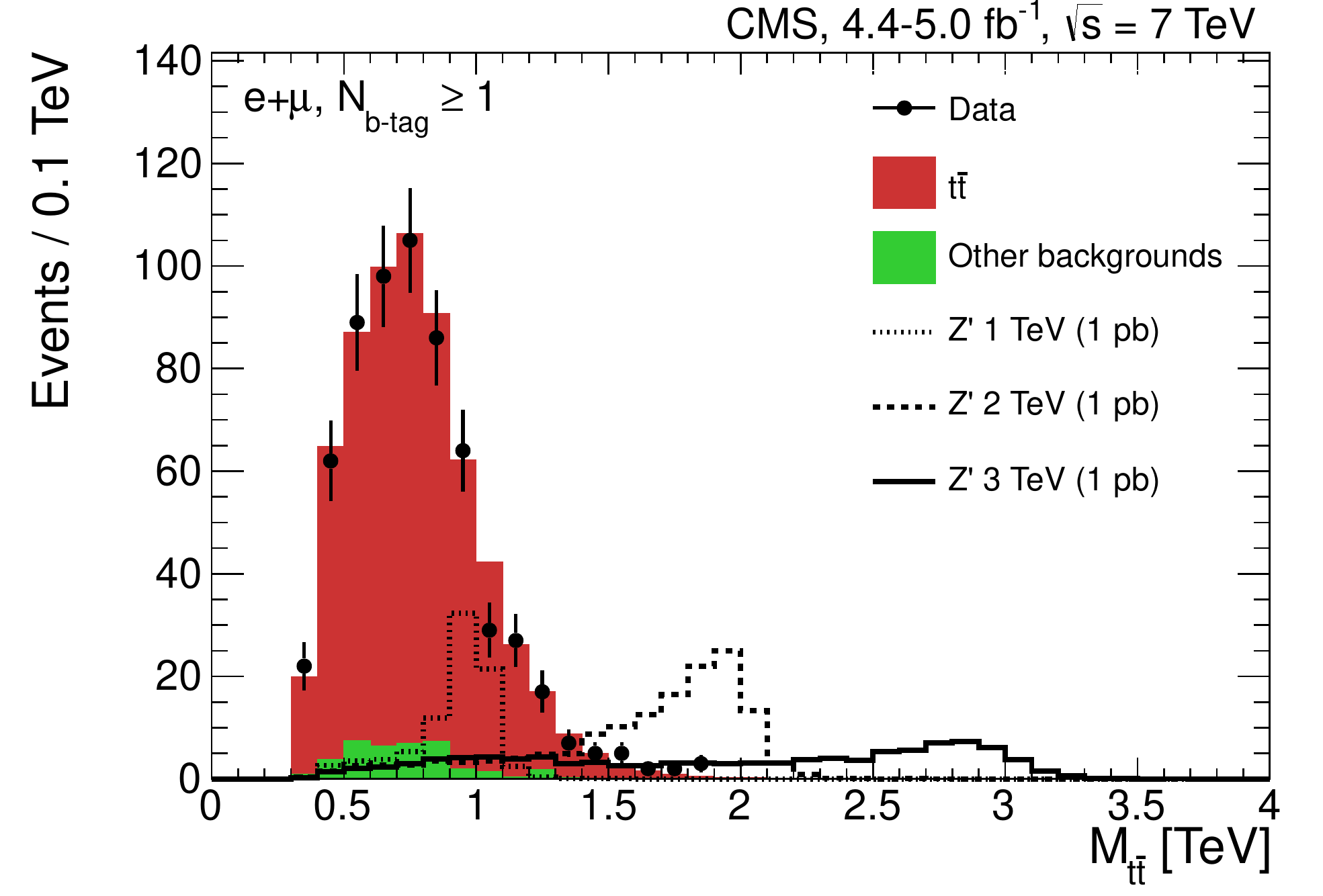}
}
\hspace*{0.3cm}
\subfigure[]{
\includegraphics[width=0.45\textwidth]{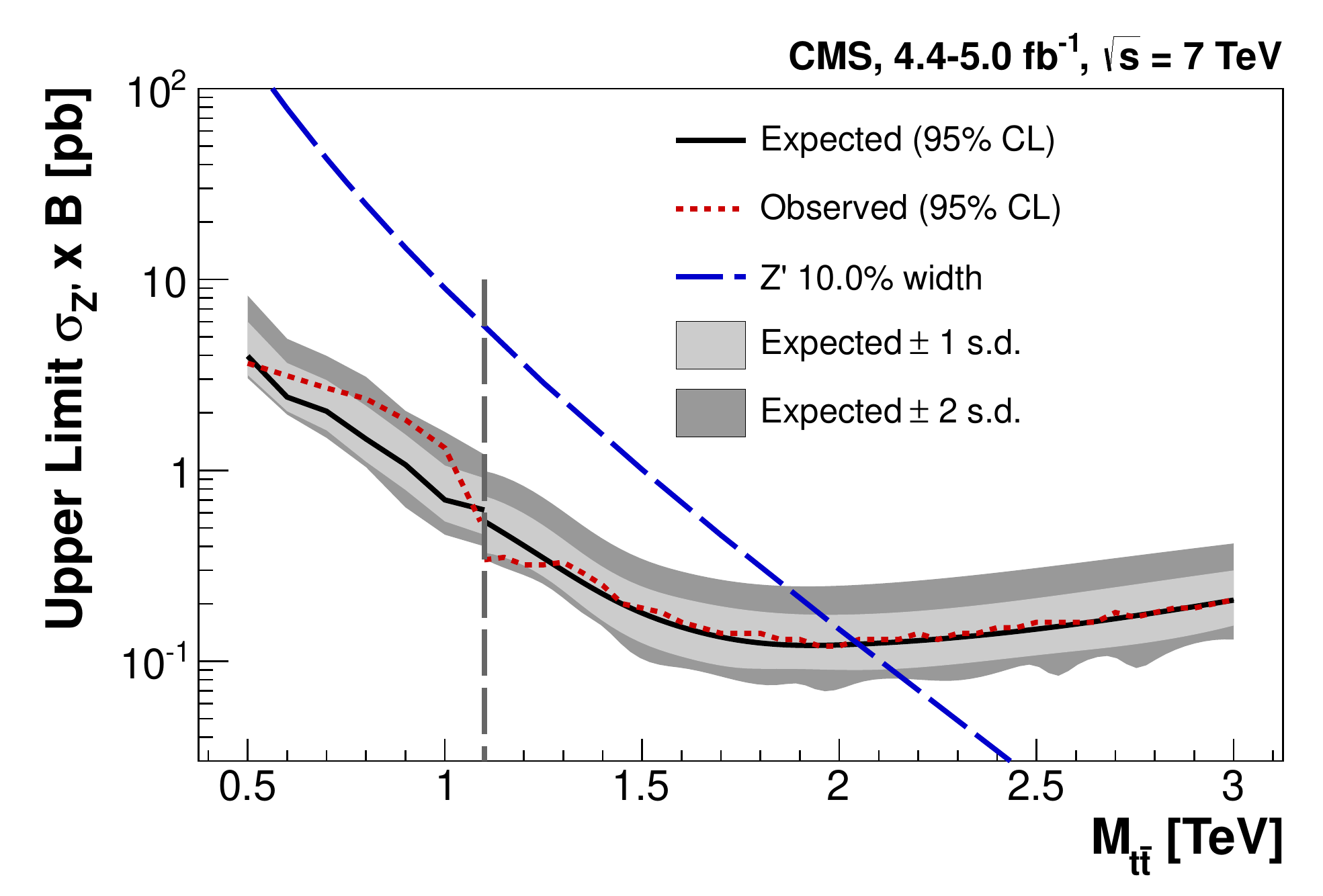}
}
\end{center}
\vspace*{-0.5cm}
\caption{Search for resonant $t\bar{t}$ production in events with $e/\mu$+jets signature
as performed by the CMS collaboration using data at $\sqrt{s}=7\;\mbox{TeV}$ corresponding 
to an integrated luminosity of $5.0\;\mbox{fb}^{-1}$~\cite{Chatrchyan:2012cx}. 
(a) Comparison of the reconstructed $M_{t\bar{t}}=m_{t\bar{t}}$ spectrum in data and SM predictions 
for the boosted analysis
with $\ge 1$ b-tagged jets. Expected signal contributions for narrow-width topcolor $Z'$ 
models~\cite{Harris:2011ez} ($\Gamma_{Z'}/m_{Z'} = 1.2\%$) at 
different masses are also shown. A cross section times branching fraction of 1.0 pb is used for 
the normalisation of the $Z'$ samples.
(b) The 95\% C.L. upper limits on the product of the production cross section of a wide $Z'$
($\Gamma_{Z'}/m_{Z'} = 10\%$)~\cite{Harris:2011ez}, $\sigma_{Z'}$, and the branching fraction 
B of hypothesised resonances that decay into $t\bar{t}$ as a function of the invariant mass 
of the resonance. The $\pm 1$ and $\pm 2$ standard deviation excursions from the expected 
limits are also shown. The vertical dashed line indicates the transition between the 
threshold and the boosted analyses, chosen based on the sensitivity of the expected limit.
}
\label{fig:Mtt_CMS}
\end{figure}
The statistical inference of all analyses has been performed using the package 
THE\-TA~\cite{JOTT2012}, which has been developed in the process of the PhD thesis of 
Ref.~\cite{Ott:2013cut} and has been widely used within the CMS collaboration.
As no excess of events over the expected yield in the reconstructed $m_{t\bar{t}}$ 
distribution is observed (see Figure~\ref{fig:Mtt_CMS} (a)) limits at 95\% C.L. are 
set on the production of heavy non-SM particles: topcolor Z' bosons~\cite{Harris:2011ez} 
with narrow and wide (10\%) width, and Kaluza-Klein excitations of a gluon~\cite{Agashe:2006hk}. 
The most stringent upper limit from the Tevatron accelerator of 
$915\,\mbox{GeV/c}^2$~\cite{Aaltonen:2012af} at 95\% C.L. for a narrow $Z'$ 
resonance has been improved to $1.49\,\mbox{TeV/c}^2$ at 95\% C.L. and wide $Z'$ 
resonances have been excluded up to a mass of $2.04\,\mbox{TeV/c}^2$ at 95\% C.L. as shown 
in Figure~\ref{fig:Mtt_CMS} (b).\\

The CMS collaboration published with the same data set also searches for resonant $t\bar{t}$ production in the dilepton channel~\cite{Chatrchyan:2012yca} ($\ell = e,\mu$) and in the all-hadronic channel~\cite{Chatrchyan:2012ku}. The CMS search in the dilepton channel is optimised for the threshold region as two isolated charged leptons are required. In contrast, the CMS analysis in the all-hadronic channel is designed for the boosted region, where the hadronic top quark decay products are partially or fully merged into a single jet.
Here, new methods to analyse the jet substructure of fully merged jets from the hadronic top quark decay (top-tagging)~\cite{Kaplan:2008ie,CMS:2009lxa} and of merged jets from the hadronic $W$-boson decay ($W$-tagging)~\cite{Ellis:2009su,Ellis:2009me,Butterworth:2008iy} are employed and suppress the non-top multi-jet background.\\ 

The limits set in the $e/\mu$+jets channel~\cite{Aaltonen:2012af} represent the most stringent limits in the $(0.5-2.0)\,\mbox{TeV/c}^2$ mass range, while for masses above $2\,\mbox{TeV/c}^2$ the limits obtained in the all-hadronic channel~\cite{Chatrchyan:2012ku} are slightly more stringent. It has been studied in Ref.~\cite{Ott:2009aka} that top-tagging algorithms~\cite{Kaplan:2008ie,CMS:2009lxa} to identify fully merged jets of hadronic top quark decays could also be useful for future analyses in the $e/\mu$+jets channel, in particular for analyses of Run II where the reachable $m_{t\bar{t}}$ range is extended largely due to the increased centre-of-mass energy of the LHC of (13-14) TeV.\\

The ATLAS collaboration published in 2012 searches for resonances decaying to $t\bar{t}$ using data collected at $\sqrt{s}=7\;\mbox{TeV}$ and corresponding to an integrated luminosity of $2\;\mbox{fb}^{-1}$. A search optimised for the threshold region has been performed in the $e/\mu$+jets channel and in the dilepton channel~\cite{Aad:2012wm}, whereat the dilepton channel is considered mainly as an independent cross check because it yields less stringent limits. A second search performed in the $e/\mu$+jets channel is optimised for the boosted region~\cite{Aad:2012dpa}. Here, events are selected with one isolated charged lepton and with one reconstructed fat jet identified as jet originating from a top quark by the top-tagging method proposed by Seymour~\cite{Seymour:1993mx}. Compared to the search optimised in the threshold region~\cite{Aad:2012wm} significantly better limits in the $(1-2)\,\mbox{TeV/c}^2$ region are obtained in the search optimised for the boosted topology~\cite{Aad:2012dpa}. Due to using less than the half of the 2011 data set the ATLAS limits in the boosted $e/\mu$+jets channel presented in Ref.~\cite{Aad:2012dpa} are substantially less stringent than those published by the CMS collaboration about three months later but exploiting the entire 7 TeV data set~\cite{Aaltonen:2012af}.\\

Beginning of 2013 the ATLAS collaboration extended the search for resonant $t\bar{t}$ production towards the all-hadronic channel~\cite{Aad:2012raa} using 7 TeV data corresponding to an integrated luminosity of $4.7\;\mbox{fb}^{-1}$. Here, two analyses optimised for the medium ($p_{\mathrm{T}}>200\,\mbox{GeV/c}$) and high ($p_{\mathrm{T}}>450\,\mbox{GeV/c}$) top-quark jet transverse momentum region, respectively, are combined. The analysis in the medium boosted region utilises the HEPTopTagger method~\cite{Plehn:2009rk,Plehn:2010st}, while the analysis in the highly boosted region employs the Top Template Tagger method~\cite{Almeida:2010pa,Almeida:2011aa}. Compared to the search from the CMS collaboration performed in the all-hadronic channel~\cite{Chatrchyan:2012ku} the $m_{t\bar{t}}$ region studied could be extended from $1\,\mbox{TeV/c}^2$ down to $0.7\,\mbox{TeV/c}^2$. In the $m_{t\bar{t}}$ region covered by both experiments similar limits on heavy resonances could be set.\\

In summer 2013 the ATLAS collaboration has updated and improved the searches for resonant $t\bar{t}$ production in the $e/\mu$+jets channel~\cite{Aad:2013nca} using 7 TeV data corresponding to an integrated luminosity of $4.7\;\mbox{fb}^{-1}$. This updated analysis is a combination of two ana\-ly\-ses, one optimised for the threshold region and the other for the boosted region. In order to cope with the boosted topology a special single fat-jet trigger has been utilised, a new isolation variable, named mini-isolation and suggested in Ref.~\cite{Rehermann:2010vq} yielding high efficiency in both search regions, has been employed, and again top-tagging~\cite{Ellis:1993tq,Seymour:1993mx} is applied to identify fat jets originating from the merged decay products of the hadronic top quark decay. The combination of the two analyses results in the most stringent limits using 7 TeV LHC data, excluding a narrow $Z'$ resonance up to a mass of $1.74\,\mbox{TeV/c}^2$ at 95\% C.L.\\

End of 2013 the CMS collaboration updated the search for heavy resonances decaying to top quark pairs using data with $\sqrt{s}= 8\;\mbox{TeV}$ and corresponding to an integrated luminosity of $19.7\;\mbox{fb}^{-1}$~\cite{Chatrchyan:2013lca}. The results are based on a combination of two analyses in the $e/\mu$+jets channel optimised for the threshold and boosted region, respectively, and of an analysis in the all-hadronic channel. The analyses follow the techniques explored in the CMS searches using 7 TeV data~\cite{Chatrchyan:2012cx,Chatrchyan:2012ku}. As no excess of events over the expected yield in the reconstructed $m_{t\bar{t}}$ distribution is observed limits are set on the production of heavy non-SM particles, excluding a narrow $Z'$, a $Z'$ with a width of 10\%, and a Kaluza-Klein excitation of a gluon up to a mass of $2.1\,\mbox{TeV/c}^2$, $2.7\,\mbox{TeV/c}^2$, and $2.5\,\mbox{TeV/c}^2$ at 95\% C.L., respectively. These limits are the most stringent limits of heavy resonances decaying to $t\bar{t}$ today (status May 2015).\\ 

In May 2015 the ATLAS collaboration published a search for $t\bar{t}$ resonances in events with $e/\mu$+jets signature using data with $\sqrt{s}= 8\;\mbox{TeV}$ and corresponding to an integrated luminosity of $20.3\;\mbox{fb}^{-1}$~\cite{Aad:2015fna}. As before, two analyses, one optimised for the threshold region and the other for the boosted region, have been performed using similar techniques as in Ref.~\cite{Aad:2013nca}. The combination of the two analyses results in a similar expected search sensitivity and in observed limits on heavy $t\bar{t}$ resonances that are only slightly less stringent than those obtained by the CMS collaboration employing the $e/\mu$+jets and all-hadronic channel~\cite{Chatrchyan:2013lca}.\\

With the start of the LHC Run II in 2015 with $\sqrt{s}=(13-14)\,\mbox{TeV}$, a new, and so far unexplored region in the $m_{t\bar{t}}$ spectrum will become accessible which will boost the searches for heavy resonances. In this newly accessible region top quarks will be highly boosted and future analyses will greatly benefit from the full potential of top-tagging algorithms.


\section{Experimental Studies of \boldmath{$t$}-Channel Single Top Quark Production}
\label{sec:singletop}

In this section a short motivation of single top quark production studies
is presented followed by a short overview of the measurements performed at the Tevatron, 
where single top quark production has been observed firstly in 2009~\cite{Aaltonen:2009jj,Abazov:2009ii}.
Then, the experimental studies performed at the LHC and the achievements are described 
using as example analyses the first measurement of the $t$-channel single top quark
production cross section at CMS~\cite{Chatrchyan:2011vp} and the $t$-channel cross section
measurement with the smallest relative uncertainty at CMS in Run I~\cite{Chatrchyan:2012ep}.

\subsection{Introduction to Single Top Quark Production}
\label{sec:singletop_intro}

The electroweak production of single top quarks via the charged current provides a powerful test of the electroweak interaction 
and to some extent to the strong interaction (QCD correction and $b$-PDF) predicted within the SM as well as sensitivity to new 
physics beyond the SM. As the cross section for single top quark production is proportional to the square of the 
Cabibbo-Kobayashi-Maskawa (CKM) matrix~\cite{Kobayashi:1973fv} element $|V_{tb}|$ the measurement of the single top 
quark cross section allows a direct determination of $|V_{tb}|$.
The CKM matrix element $|V_{tb}|$ is sensitive to contributions from additional vector-like quarks or a chiral fourth generation of quarks~\cite{Chanowitz:2009mz,Alwall:2006bx}, as well as to other new phenomena~\cite{Tait:2000sh} leading to a measured strength of $|V_{tb}|$ different from the SM prediction. Please note, that the simple extension of the SM by an additional chiral generation of fermions (SM4) is ruled out by recent experimental Higgs data~\cite{Lenz:2013iha}.\\
New physics could alter the production cross sections of the three different single top quark channels differently~\cite{Tait:2000sh,Cao:2007ea}, making it a powerful probe of BSM models and allowing it to distinguish between different BSM scenarios.
For example $s$-channel production is highly sensitive to new particles like a heavy $W'$ boson~\cite{Simmons:1996ws,Datta:2000gm,Sullivan:2002jt} or a charged Higgs boson~\cite{Li:1996bh,Tait:2000sh} but will also be affected by new physics involving flavour changing neutral currents~\cite{Tait:2000sh}. On the other hand, $t$-channel single top quark production is in particular sensitive to anomalous $Wtb$ couplings~\cite{Heinson:1996zm,Espriu:2001vj} or flavour changing neutral currents~\cite{Tait:1996dv,Luke:1993cy,Han:1998tp} (FCNC) between the top quark and any other quark. The associated $Wt$ production has in turn a high sensitivity to anomalous $Wtb$ couplings~\cite{Tait:2000sh,Cao:2007ea}, but is relatively insensitive to scenarios that affect the other single-top-quark production channels.\\
	
The electroweak single top quark production is sub-dominant at hadron colliders but it still has a cross section which is only a factor 2-3 smaller compared to the one of the dominant $t\bar{t}$ production. It is due to the huge amount of background events that lead to a similar event signature as singly produced top quarks, that it took 14 years longer to observe single top quark production than $t\bar{t}$ production, which was observed already in 1995 by the two Tevatron experiments~\cite{Abe:1995hr,Abachi:1995iq}.\\ 
At the Tevatron the dominant background to single top quark production after event selection is $W$-boson production in association with jets originating from heavy quarks or light quarks or gluons. As the amount of single top quark events is even after event selection substantially smaller than the uncertainty on the backgrounds it was necessary to use sophisticated multivariate methods and to combine them to find, metaphorically spoken, the needle in the haystack.\\ 

After first searches for single top quark production~\cite{Acosta:2004bs,Abazov:2005zz,Abazov:2006uq}, evidence has been 
reported by the D0~\cite{Abazov:2006gd,Abazov:2008kt} collaboration in 2007 and later also by the CDF 
collaboration~\cite{Aaltonen:2008sy} in 2008. Finally the observation of single top quark production in the sum of 
$s$- and $t$-channels was reported by the CDF~\cite{Aaltonen:2009jj,Aaltonen:2010fs,Aaltonen:2010jr} and D0~\cite{Abazov:2009ii,Abazov:2011pt} 
collaborations in March 2009.\\ 
In the next years the analyses on single top quark production have been extended and updated by both 
collaborations~\cite{Aaltonen:2014qja,Aaltonen:2014xta,Aaltonen:2014ura,Aaltonen:2014mza,Abazov:2009pa,Abazov:2009nu, Abazov:2011pm,Abazov:2011rz,Abazov:2013qka,CDF:2014uma,Aaltonen:2015cra}. In 2011 the D0 collaboration observed the single top quark $t$-channel process~\cite{Abazov:2011rz}. Evidence for single top quark $s$-channel production~\cite{Abazov:2013qka} has been reported firstly by the D0 collaboration in 2013 and has been confirmed by the CDF collaboration in 2014~\cite{Aaltonen:2014qja}. The combination of the $s$-channel cross section analyses performed by the CDF~\cite{Aaltonen:2014qja,Aaltonen:2014xta} and D0~\cite{Abazov:2013qka} collaborations resulted in the first observation of the single top quark $s$-channel process~\cite{CDF:2014uma}. The Tevatron combination for single top quark production in the sum of $s$- and $t$-channels and for $t$-channel production, separately, has been recently reported in Ref.~\cite{Aaltonen:2015cra}, using the final Tevatron data set. All measurements performed at the Tevatron are consistent with the SM approximate NNLO (based on threshold resummation at NNLL accuracy) theoretical predictions~\cite{Kidonakis:2010tc,Kidonakis:2011wy}. The relative uncertainty of single measurements of the $s$-channel cross section using the final data set are at the level of 30\% or slightly below~\cite{Aaltonen:2014xta,Abazov:2013qka}, while the relative uncertainty of single measurements of the $t$-channel cross section using the final data set is at the 20\% level~\cite{Aaltonen:2014mza,Abazov:2013qka}.

\subsection[Measurement of the $t$-Channel Single Top Quark Cross Section at the LHC]{Measurement of the \boldmath{$t$}-Channel Single Top Quark Cross Section at the LHC}
At the LHC with $\sqrt{s}=7\;\mbox{TeV}$ the cross section for $t$-channel single top quark 
production is increased by a factor 30 compared to the one at the Tevatron. Due to this and 
due to the fact that the cross section for $W$-boson production in association with jets ($W$+jets) 
does not increase that strongly as it is induced via $q\bar{q}$, the conditions for studying 
electroweak $t$-channel single top quark production are strongly improved at the LHC compared to 
those at the Tevatron.\\ 

In the PhD thesis of Ref.~\cite{Bauer:2010ssa} finished in 2010 a Monte Carlo simulation 
based study for the measurement of the $t$-channel single top quark production cross section has 
been performed using events in the muon+jets channel. Single top quark $t$-channel events are 
characterised by a spectator jet originating from a scattered light quark and produced in forward 
direction, a jet induced from the second $b$-quark (referred to as spectator $b$-quark or as 
2nd $b$-quark) from the initial gluon splitting, and the decay products of the top quark. In the study 
of Ref.~\cite{Bauer:2010ssa} a scenario of MC data corresponding to an integrated luminosity of 
$L=200\;\mbox{pb}^{-1}$ collected with the CMS detector and at a LHC centre-of-mass energy of 10 TeV 
has been employed. The analysis was designed to be applied to the first LHC data and it was kept simple 
to minimise data-MC comparison efforts right at the beginning of the LHC start.\\ 

In the process of this thesis~\cite{Bauer:2010ssa} detailed MC studies to model the $t$-channel process have been 
performed. In particular 
a dedicated matching procedure based on the kinematics of the 2nd $b$-quark~\cite{Boos:2006af,Lueck:2006hz,Lueck:2009zza} 
and applied to the MADGRAPH simulations of the $t$-channel process to account for crucial NLO contributions has been 
implemented for the usuage in CMS and validated with other MC generators. These MADGRAPH samples 
have been used to model the $t$-channel single top quark process in the first paper about 
single top quark production from the CMS collaboration~\cite{Chatrchyan:2011vp}.\\ 

Based on the knowledge from single top quark measurements at the Tevatron (see section~\ref{sec:singletop_intro}), 
the event selection, the reconstruction of the $t$-channel kinematics, 
and a method to model multi-jet background have been developed for the LHC (CMS) in the thesis of Ref.~\cite{Bauer:2010ssa}, 
and two variables to extract the $t$-channel single top quark content have been investigated in the presence of systematic uncertainties.\\
One variable is the reconstructed top quark mass $m_{\ell\nu b}$, the other is the cosine of the angle $\theta^{*}$ between 
the direction of the outgoing charged lepton and the spin axis of the top (anti-top) quark, approximated by the light jet 
recoiling against the top quark (spectator jet basis), in the top (anti-top) quark rest frame~\cite{Mahlon:1996pn,Motylinski:2009kt,Mahlon:1999gz}. 
This observable has a distinct slope in $t$-channel signal events, coming from the fact, that top quarks produced singly via the $t$-channel 
are about 100\% polarised along the direction of the $d$-type quark due to the V-A structure of the electroweak interaction~\cite{Mahlon:1999gz}. 
It has been found in~\cite{Bauer:2010ssa} that $m_{l\nu b}$ has the larger discrimination power against the different background 
processes compared to $\cos\theta^{*}$, but that $\cos\theta^{*}$ is much less affected by theoretical and instrumental sources 
of systematic uncertainties, resulting in the same expected performance of the two variables if systematic uncertainties are considered.\\ 
Furthermore, it has been demonstrated in Ref.~\cite{Bauer:2010ssa} that selecting events with large absolute pseudorapitidy 
value of the light recoil jet $|\eta_{j'}|$ could result in high-purity single top quark data samples.\\ 

Using LHC data with $\sqrt{s}=7\;\mbox{TeV}$ and corresponding to an integrated luminosity of 
$36\;\mbox{pb}^{-1}$ the CMS collaboration found evidence for $t$-channel single top quark 
production at the LHC~\cite{Chatrchyan:2011vp}. This measurement combines two complementary 
analyses with similar event selection. The first analysis, referred to as 2D analysis, 
is based on the MC studies described in Ref.~\cite{Bauer:2010ssa} and here a 2D fit in the 
variables $\cos\theta^{*}$ and $|\eta_{j'}|$ is performed to extract the $t$-channel signal 
content. It is the first single top quark analysis that does not depend on multivariate techniques. 
Furthermore, it uses data to model the most difficult backgrounds (multi-jet and $W$+jets). 
The second analysis exploits the fact that already in the first year of the LHC data taking a 
comprehensive data understanding was achieved and employs a multivariate analysis technique 
with boosted decision trees (BDTs)~\cite{cart84,adaboost}. The BDT analysis probes the overall 
compatibility of the signal event candidates with the event topology of $t$-channel single top 
quark production and relies for the modelling of $W$+jets background on MC simulation.\\
 
The event selections applied in the two analyses are inspired by the one developed in 
Ref. \cite{Bauer:2010ssa}, but have been optimised for 7 TeV conditions and extended 
to the electron+jets channel. In the paper~\cite{Chatrchyan:2011vp} events with one 
isolated muon (electron) with $p_{\mathrm{T}}>20\,(30)\;\mbox{GeV/c}$ and $|\eta|<2.1\,(2.5)$, 
exactly two jets reconstructed using the anti-$k_T$ algorithm~\cite{Cacciari:2008gp} 
with distance parameter $R=0.5$ and corrected for $\eta$ and $p_{\mathrm{T}}$-dependent detector 
response~\cite{CMS:2010ata} and with $p_{\mathrm{T}}>30\;\mbox{GeV/c}$ and $|\eta|<5.0$ are selected. 
Exactly one jet has to be identified as $b$-jet by passing the tight quality criteria of 
an algorithm that orders the tracks in impact parameter significance and discriminates using 
the track with the third highest significance~\cite{CMS:2010hua}. To reduce multi-jet 
background a cut on the transverse mass of the $W$-boson of $M_{\mathrm{T}}>40\,(50)\;\mbox{GeV/c}^2$ is 
applied for the $\mu$+jets ($e$+jets) channel. The 2D analysis rejects events if the jet 
failing the tight threshold of the $b$-tagging algorithm passes a loose threshold on the 
impact parameter significance of the second track. On the other hand, the BDT analysis 
rejects events where the jets are back-to-back, which are found to be poorly reproduced 
by the $W$+jets MC simulation. The 2D (BDT) analysis selects 184 (221) candidate events with 
a signal purity of about 18\% in both selections.\\

The $t$-channel single top quark events have been generated with the 
MADGRAPH 4.4~\cite{Maltoni:2002qb,Alwall:2007st} event generator. 
In order to obtain a reasonable approximation of the signal kinematic 
at full NLO, the most crucial NLO $qg\rightarrow q'\,t\bar{b}$ contribution 
is combined with the LO ($qb\rightarrow q't$) diagram by a matching procedure 
based on~\cite{Boos:2006af} and implemented in a time-saving way for the MADGRAPH 
generator during the PhD thesis of Ref.~\cite{Bauer:2010ssa}.\\ 
Main backgrounds are $t\bar{t}$, $W$+jets, single top $Wt$ and multi-jet production. 
The $t\bar{t}$, single-top s-channel and $Wt$ production, and $W/Z$+jets are also modelled 
using MADGRAPH. Diboson production ($WW,WZ,ZZ$) is generated using 
PYTHIA 6.4~\cite{Sjostrand:2006za}. The top quark mass is set to $m_t=172.5\;\mbox{GeV/c}^2$ 
in all MC samples containing top quarks.\\ 
In both analyses, multi-jet events are entirely modelled from data based on the method 
developed in Ref.~\cite{Bauer:2010ssa} and using events which pass looser isolation 
requirements. In the 2D analysis also the $W$+jets background is modelled using two different sidebands.\\

\begin{figure}[t]
\begin{center}
\subfigure[]{
\includegraphics[width=0.45\textwidth]{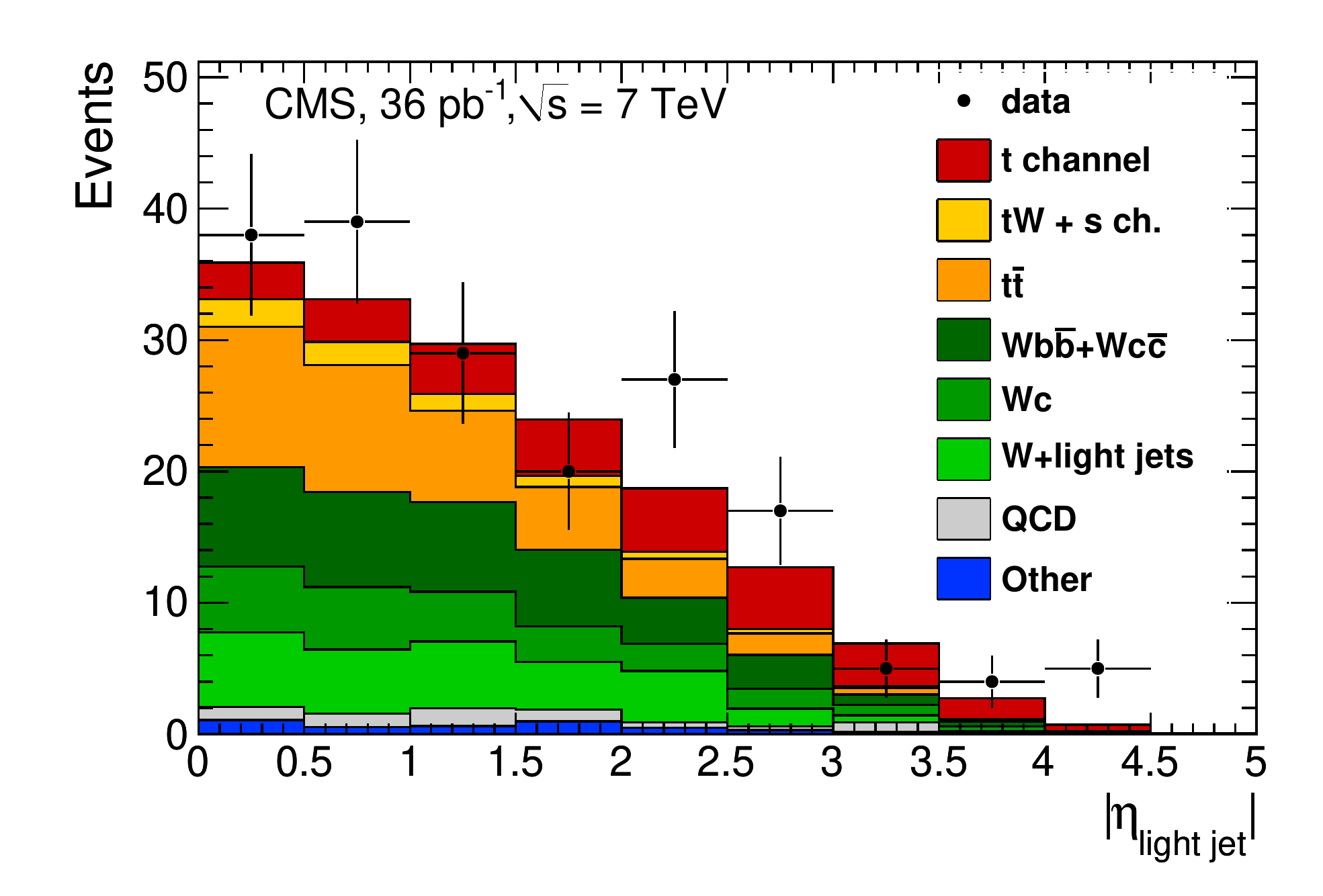}
}
\hspace*{0.3cm}
\subfigure[]{
\includegraphics[width=0.45\textwidth]{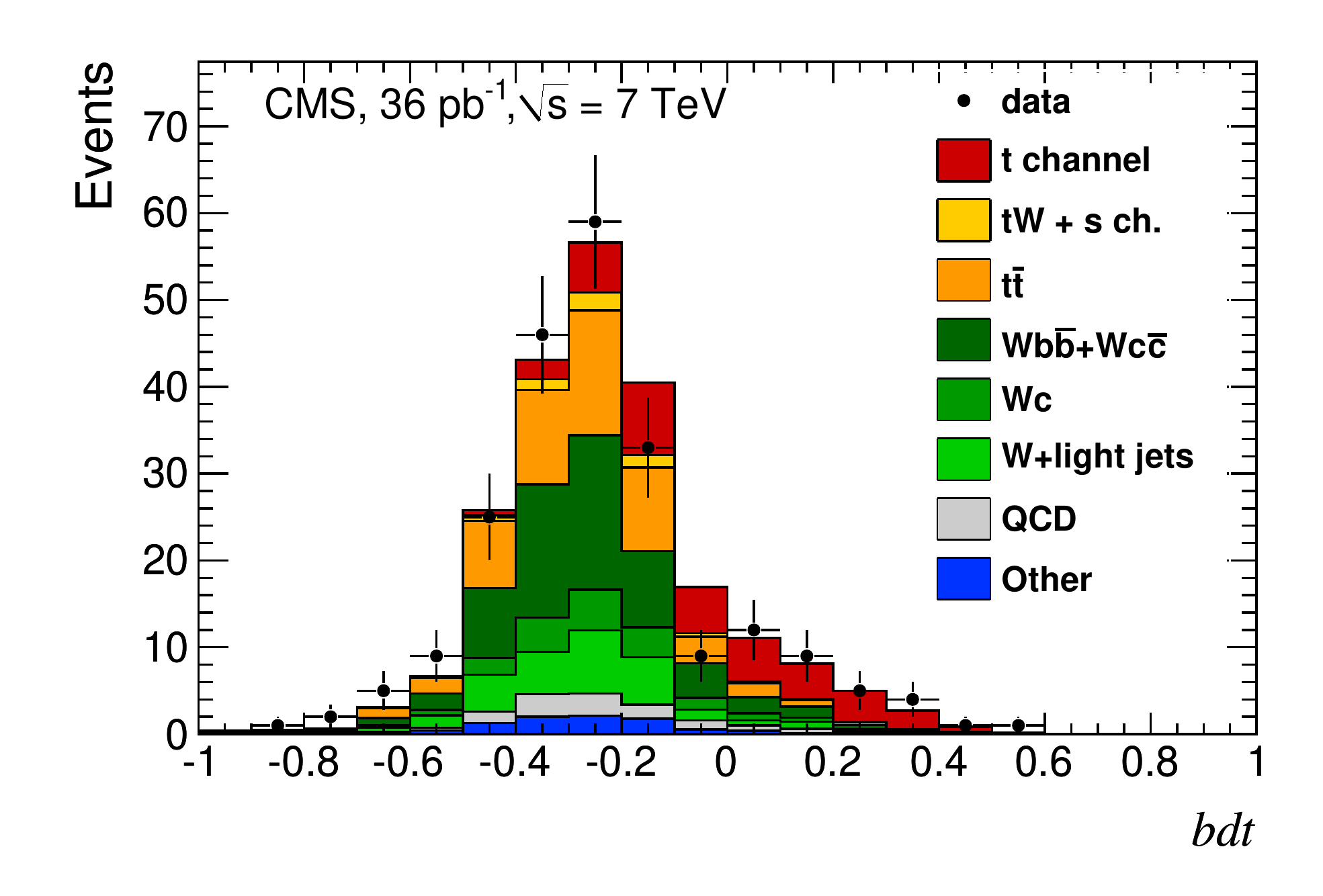}
}
\end{center}
\vspace*{-0.5cm}
\caption{Single top $t$-channel production cross section using events
with $e/\mu$+jets signature as measured by the CMS collaboration using 
data at $\sqrt{s}=7\;\mbox{TeV}$ corresponding to an integrated luminosity of 
$36\;\mbox{pb}^{-1}$~\cite{Chatrchyan:2011vp}. 
(a) Pseudorapidity of the untagged jet, $\eta_j'=\eta_{\mbox{\tiny{light jet}}}$, after 
applying the 2D selection. The normalisation of multi-jet and $W+$light-parton events
is determined from data, $t\bar{t}$ and $W/Z+$heavy-parton events are normalised to the result
of a dedicated $t\bar{t}$ cross section study, all other processes are normalised to the 
theoretical expectations.
(b) Boosted decision tree discriminant (bdt)
after applying the complete BDT selection, with the signal
scaled to the measured cross section and all systematic uncertainties 
and backgrounds scaled to the medians of their posterior
distributions.
}
\label{fig:t-channel_CMS_first}
\end{figure}
In both analyses the kinematic of $t$-channel single top quark production is reconstructed 
using the method developed in Ref.~\cite{Bauer:2010ssa}. The $t$-channel single top quark content 
is extracted from a fit to the 2D plane $\cos\theta^{*}$-$|\eta_{j'}|$ and to the BDT discriminant, 
respectively. In Figure~\ref{fig:t-channel_CMS_first} (a) $|\eta_{j'}|$ is presented after the 2D event 
selection and in Figure~\ref{fig:t-channel_CMS_first} (b) the boosted decision discriminant
(bdt) is shown after applying the complete BDT selection.
The results of the two analyses are compatible and the statistical correlation has been 
determined to be 51\% using pseudo experiments.\\ 

The results of the 2D and BDT analysis have been combined using the Best Linear Unbiased 
Estimator (BLUE) technique~\cite{Lyons:1988rp} and taking into account the statistical 
correlation and treating all systematic uncertainties as fully correlated except those coming from estimates 
based on data. The combination yields a measured $t$-channel cross section of 
$\sigma=83.6\pm 29.8\,\mbox{(stat.+syst.)}\pm 3.3\,\mbox{(lumi)}\;\mbox{pb}$ which is consistent with 
the approximate NNLO (based on threshold resummation at NNLL accuracy) SM prediction~\cite{Kidonakis:2011wy}. 
As the BDT analysis contributes with a weight of almost 90\%, the significance of the combination is well 
approximated by the significance of the BDT analysis of $3.5\sigma$. Assuming that $|V_{tb}|\gg |V_{td}|, |V_{ts}|$ 
an effective value of $|V_{tb}|=1.14\pm 0.22\,\mbox{(exp.)}\pm 0.02\,\mbox{(theory)}$ is determined.\\

In the diploma thesis of Ref.~\cite{Roecker:2011} finished in March 2011 it has been demonstrated that the 
sensitivity of the simple analysis strategy developed in~\cite{Bauer:2010ssa} could be strongly improved by 
combining many variables into a neural network using the NEUROBAYES package~\cite{Feindt:2004wla,Feindt:2006pm}, 
which confirms the improvement seen in the expected significance when comparing the BDT and 2D analyses of 
Ref.~\cite{Chatrchyan:2011vp}. In the process of this thesis single top quark MC samples have been generated 
for the first time at CMS using the NLO+PS generator POWHEG~\cite{Frixione:2007vw,Alioli:2010xd,Alioli:2009je,Re:2010bp} 
interfaced to PYTHIA 6.4~\cite{Sjostrand:2006za} and have been validated in dedicated studies.\\ 

In the diploma thesis of Ref.~\cite{Descroix:2011} $b$-tagging has been studied in the context of top quark physics. 
It has been demonstrated that the tight working point of the robust $b$-tagging algorithm used in 
the CMS paper~\cite{Chatrchyan:2011vp} in the general $t$-channel single top quark event selection to find the 
$b$-jet from the top quark decay is very close to optimal and that there is no need for choosing a different 
working point of this robust $b$-tagging algorithm. Furthermore, it has been shown that employing the Combined 
Secondary Vertex (CSV) algorithm (see section~\ref{sec:mttbar}) could improve future single top quark analyses 
given that the CSV tagger is fully commissioned.\\

Using 7 TeV data collected with the CMS detector and corresponding to an integrated luminosity of 
$0.92\;\mbox{fb}^{-1}$ $t$-channel single top quark production employing the electron+jets channel 
has been studied in the diploma thesis of Ref.~\cite{Hansen:2012kia}.\\ 
Only in the first part of the 2011 data taking single electron triggers with reasonably small $p_{\mathrm{T}}$ 
threshold were employed. Later, triggers requiring one electron and at least one central jet, that has 
been identified as $b$-jet based on the impact parameter significance of the tracks within the jet~\cite{CMS:2012gik}, 
are employed. In the diploma thesis of Ref.~\cite{Hansen:2012kia} these triggers have been used for the first time at 
CMS and trigger turn-on curves for the hadronic legs of the trigger have been determined and applied to simulate in 
MC the efficiency of those triggers. Furthermore, the method to model multi-jet background used in~\cite{Chatrchyan:2011vp} had 
to be adjusted for electron+jets events to ensure reasonable modelling of the data with the larger data set.\\ 
Based on the work in Ref.~\cite{Roecker:2011} a neural network is used to optimise the discrimination power between 
$t$-channel signal and the different background processes. An analysis improvement established in Ref.~\cite{Hansen:2012kia} 
is the extraction of the signal content by performing a simultaneous fit to the neural network discriminants in several 
categories. These categories split the data according to the number of selected jets and selected $b$-jets, denoted 
as $n\mbox{-jets}\;m\mbox{-btags}$, and consist of very different relative signal and individual background contributions. 
This allows to constrain in-situ individual background components and other sources of systematic uncertainties like 
$b$-tagging efficiency directly from the categories enriched with these backgrounds and from the interplay between 
the different categories.\\

Using CMS data collected for the $\mu$+jets and $e$+jets final states at $\sqrt{s}=7\;\mbox{TeV}$ and corresponding 
to an integrated luminosity of $1.17\;\mbox{fb}^{-1}$ and $1.56\;\mbox{fb}^{-1}$, respectively, the $t$-channel 
single top quark cross section has been measured end of 2012 by the CMS collaboration~\cite{Chatrchyan:2012ep} 
with a relative uncertainty slightly below 10\%. This measurement is a combination of three analyses where two 
different approaches have been followed. The first approach exploits the distributions of the pseudorapidity of 
the light recoil jet $|\eta_{j'}|$ and of the reconstructed top quark mass and uses data to model backgrounds and 
is referred to as $|\eta_{j'}|$-analysis. The second approach is based on multivariate methods that probe the 
overall compatibility of the signal event candidates with the event topology of $t$-channel single top quark 
production and that aim for a precise measurement by optimising the discrimination between signal and background. 
Due to the high complexity of the second approach, two independent multivariate analyses have been conducted that 
cross-check each other. One is based on Neural Networks (referred to as NN-analysis) and the other on Boosted Decision 
Trees (referred to as BDT-analysis). After validating that the results of all three analyses are consistent with each 
other, the final result is obtained by combining the three analyses using BLUE~\cite{Lyons:1988rp}. The NN-analysis 
of the CMS publication~\cite{Chatrchyan:2012ep} is based on the PhD analysis of Ref.~\cite{Martschei:2012vxa}.\\
 
Events with a muon are triggered with an isolated single muon trigger with $p_{\mathrm{T}}>17\;\mbox{GeV/c}$. For the electron+jets 
channel only in the initial data taking period, corresponding to an integrated luminosity of $216\;\mbox{pb}^{-1}$, an 
isolated lepton trigger was used with $p_{\mathrm{T}}>27\;\mbox{GeV/c}$. In the remaining data taking period, this single electron 
trigger was heavily prescaled and a trigger selecting at least one isolated electron with $p_{\mathrm{T}}>25\;\mbox{GeV/c}$ and a 
jet identified as $b$-jet based on the impact parameter significance~\cite{CMS:2012gik}, was used. 2D trigger turn-on 
curves as a function of the transverse momentum of the jet and the $b$-tagging discriminant have been applied to 
simulate in MC the efficiencies of the electron+$b$-jet trigger. Here, the method developed in Ref.~\cite{Hansen:2012kia} 
has been extended in Ref.~\cite{Martschei:2012vxa} to the two dimensional plane to account for the observed correlation between 
the jet trigger element and the trigger element that identifies the $b$-jet which are required at two different steps in 
the trigger chain.\\
Events with an isolated muon (electron) with $p_{\mathrm{T}}>20\,(30)\;\mbox{GeV/c}$ and $|\eta|<2.1\,(2.5)$, at least two jets 
with $p_{\mathrm{T}}>30\;\mbox{GeV/c}$ and $|\eta|<4.5$ and with a transverse mass of the $W$-boson of $M_{\mathrm{T}}>40\;\mbox{GeV/c}^2$ 
(muon) and with a missing transverse energy of $E_{\mathrm{T}}^{mis}>35\;\mbox{GeV}$ (electron), respectively, are selected. 
As studied in Ref.~\cite{Hansen:2012kia} the multi-jet contamination is larger in the electron+jets channel and a cut on $E_{\mathrm{T}}^{mis}$ 
instead of $M_{\mathrm{T}}$ is found to be more efficient in reducing multi-jet background events.\\ 

\begin{figure}[t]
\begin{center}
\subfigure[]{
\includegraphics[width=0.45\textwidth]{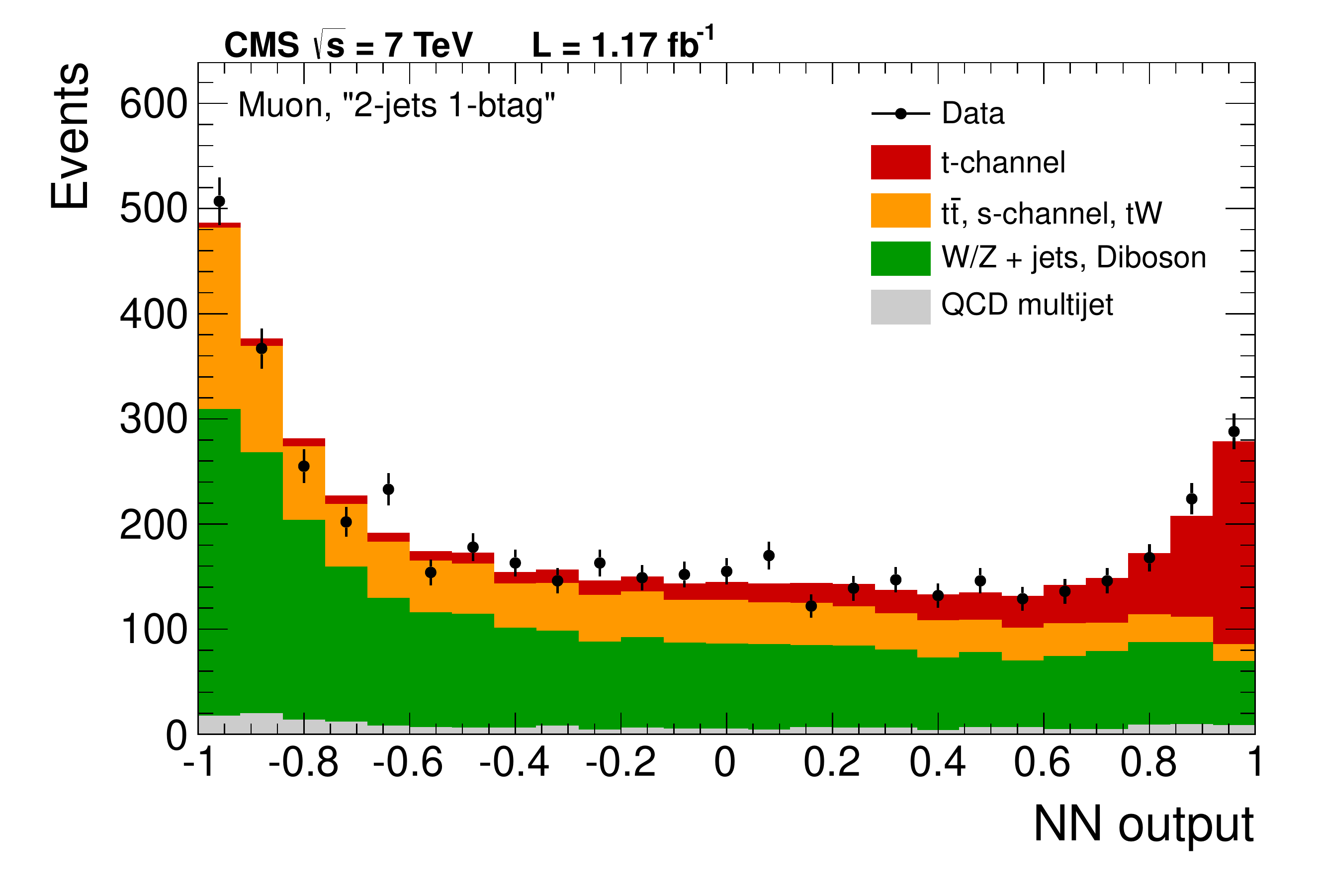}
}
\hspace*{0.3cm}
\subfigure[]{
\includegraphics[width=0.45\textwidth]{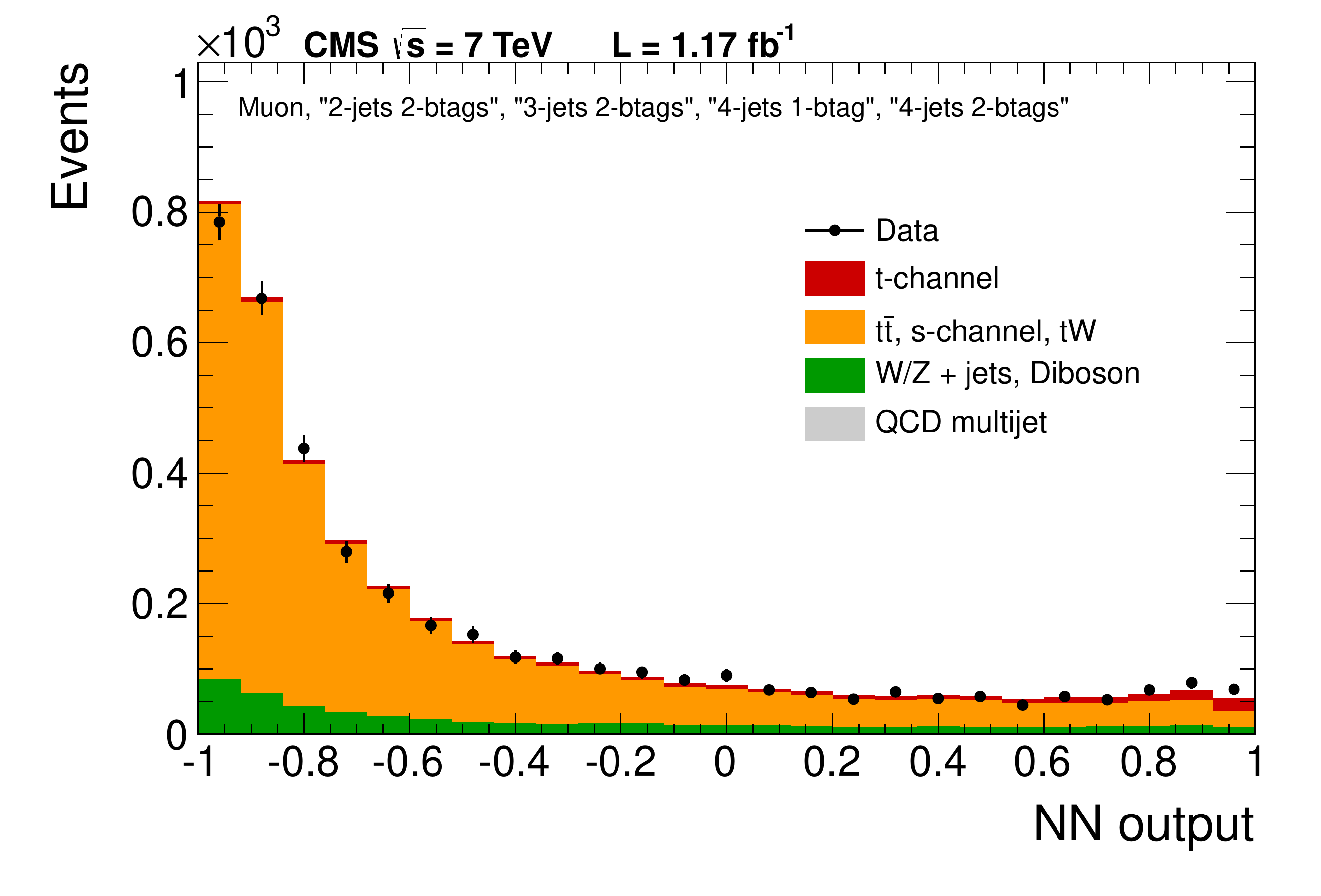}
}
\end{center}
\vspace*{-0.5cm}
\caption{Single top $t$-channel production cross section using events
with $e$+jets and $\mu$+jets signature as measured by the CMS collaboration using 
data at $\sqrt{s}=7\;\mbox{TeV}$ corresponding to an integrated luminosity of 
$1.56\;\mbox{fb}^{-1}$ and $1.17\;\mbox{fb}^{-1}$, respectively~\cite{Chatrchyan:2012ep}. 
(a) Distribution of the NN discriminator output in the muon channel for the “2-jets 1-btag” 
category. Simulated signal and background contributions
are scaled to the best fit results.
(b) Distribution of the NN discriminator output in the background dominated region.
All events from the signal depleted categories “2-jets 2-btags”, “3-jets 2-btags”, “4-jets 1-btag”,
and “4-jets 2-btags” are combined for the muon channel. Simulated signal and background 
contributions are scaled to the best fit results.
}
\label{fig:t-channel_CMS_second}
\end{figure}
Signal and control samples are defined through different event categories defined by the 
number of selected jets and selected $b$-jets ($n\mbox{-jets}\;m\mbox{-btags}$). 
These categories have been introduced for the NN-analysis in Ref.~\cite{Hansen:2012kia} and 
the same algorithm to identify $b$-jets as in the first CMS publication~\cite{Chatrchyan:2011vp} 
about single top quark production is used, which employs a tight working point of the impact 
parameter significance based algorithm~\cite{CMS:2012gik}. Although the usage of the CSV 
$b$-tagging algorithm could improve the single top quark analysis as shown 
in Ref.~\cite{Descroix:2011} this $b$-tagger has not been used in the paper as in the electron+$b$-jet 
trigger $b$-jets were identified based on the impact parameter significance. 
The $t$-channel single top quark events are primarily contained in the categories 
``2-jets 1-btag'' and ``3-jets 1-btag'' (2nd $b$-jet is mostly out of acceptance, 
in particular out of the central tracking detector acceptance). The ``2/3-jets 0-btags'' 
categories are enriched in events with $W$-bosons produced in association with light 
partons (u,d,s,g), while the ``3-jets 2-btags'' and ``4-jets 0/1/2-btags'' categories 
are enriched in top quark pair events.\\ 
The NN and BDT analyses use the six categories ``2/3/4-jets 1/2-btags'' to extract 
the signal content, whereof the background enriched categories are used to constrain 
sources of systematic uncertainties like the $b$-tagging efficiency and background 
normalisations in-situ. The three categories ``2/3/4-jets 0-btags'' are used to check 
the modelling of input variables. In Figure~\ref{fig:t-channel_CMS_second} the distribution
of the NN discriminator output is shown for the signal category ``2jets 1-tag'' in (a)
and for the signal depleted categories ``2/3-jets 2-btags'' and 
``4-jets 1/2-btags'' in (b).\\
The $|\eta_j'|$ analysis extracts the signal content 
from the ``2jets 1-tag'' category with an additional cut on the reconstructed top 
quark mass $130\,\mbox{GeV/c}^2<m_{\ell\nu b}<220\,\mbox{GeV/c}^{2}$, but 
uses the ``2-jets 0-btags'' and ``3-jets 2-btags'' categories to check the 
modelling of $W$+jets and $t\bar{t}$ background, respectively. The $|\eta_j'|$ analysis 
selects 4664 candidate events with a signal purity of about 20\%.\\

Single top quark production in the $t$ and $s$-channel as well as single top quark production in association 
with a $W$-boson ($Wt$) are simulated using the NLO+PS generator POWHEG~\cite{Frixione:2007vw,Alioli:2010xd,Alioli:2009je,Re:2010bp} 
interfaced to PYTHIA 6.4~\cite{Sjostrand:2006za} for the parton shower simulation.\\ 
Multi-jet background is modelled using data. For events with muons the same method as in the the previous 
CMS publication~\cite{Chatrchyan:2011vp} on $t$-channel single top quark production is employed, while for 
events with electrons the method developed in Ref.~\cite{Hansen:2012kia} is used. The $|\eta_j'|$ analysis 
models also the $W$+jets background using data from the sideband in the $m_{\ell\nu b}$ distribution.\\
The $t$-channel single top quark kinematic is reconstructed in the ``2-jets 1-btag'' category as in 
Ref.~\cite{Chatrchyan:2011vp,Bauer:2010ssa} but in the other categories the detailed assignment of the $b$-jet 
from the top quark decay and of the light recoil jet $j'$ has been optimised for the purpose of each analysis 
and differs among them.\\

The $|\eta_j'|$ analysis performs in the ``2-jets 1-btag'' category a maximum-likelihood fit 
to the observed distribution of $|\eta_j'|$ to extract the signal content.\\ 
The NN and BDT analyses employ a Bayesian approach~\cite{bayesian} to measure the $t$-channel 
single top quark production cross section and to determine the cross section simultaneously 
from data distributions of the multivariate discriminant in the six categories. 
The posterior distribution for the signal strength is obtained by integrating 
(marginalising) the posterior in all nuisance parameters using the Markov Chain MC 
method~\cite{Metropolis:1953,Hastings:1970} as implemented in the package 
THETA~\cite{JOTT2012}. The central value of this integrated posterior gives the $t$-channel 
cross section and the central 68\% quantile represents the total marginalised uncertainty 
of the measurement including the statistical uncertainties and the uncertainties 
included as parameters in the marginalisation.\\ 
Modelling systematic uncertainties as nuisance parameters requires in particular an 
assumption for the shape dependence of the discriminator distributions on a certain 
source of systematic for signal and background processes. An assumed dependence that 
does not cover all possible variations can lead to an over-constrain on the observed 
residual uncertainty~\cite{Barlow:2002yb}. In case of theoretical uncertainties this 
dependence is often unknown. Therefore, theoretical uncertainties are not included as 
additional parameters in the marginalisation procedure but their effect on the cross 
section measurement is estimated by performing pseudo-experiments. The estimated 
uncertainty is then added in quadrature to the total marginalised uncertainty.\\
Special care has been taken to ensure that all possible variations of the shape dependence 
of the discriminator distributions on experimental systematic sources are covered. 
In particular, the complete parametrisation of the jet energy scale as in 
Ref.~\cite{Chatrchyan:2011ds} is considered, possible variations of the $b$-tagging 
scale factor as a function of $p_{\mathrm{T}}$ and $|\eta|$ using Chebyshev polynomials are considered, 
and in case of the $W$+jets background normalisation the various sub-processes are treated 
uncorrelated as well as the different jet categories.\\

The $t$-channel cross section measurements of all three analyses are consistent with each other. 
The NN and BDT analyses achieve relative uncertainties on the $t$-channel cross section of about 
10\% each, while the $|\eta_j'|$ analysis yields a relative uncertainty of about 14\%.\\  
The three analyses have been combined using BLUE~\cite{Lyons:1988rp} taking into account the 
statistical correlation between each pair of measurements determined from pseudo-experiments and 
ranging between 60\% and 74\%. Most of the systematic uncertainties are treated as fully correlated 
among the three analyses. However, some uncertainties are only loosely correlated between the $|\eta_j'|$
analysis and the analyses exploiting multivariate techniques (BDT and NN) as different methods are 
followed in these two complementary approaches. In these cases the assumed correlation has been varied 
largely but no appreciable variation in the uncertainty has been found.\\
The combination of the three analyses yields as final result a measured $t$-channel cross section of 
$\sigma=(67.2\pm 6.1)\;\mbox{pb}=67.2\pm 3.7\,\mbox{(stat.)}\pm 3.0\,\mbox{(syst.)}\pm 3.5\,\mbox{(theo.)}\pm 1.5\,\mbox{(lumi)}\;\mbox{pb}$ 
which is well consistent with the SM approximate NNLO prediction~\cite{Kidonakis:2011wy}. Assuming that 
$|V_{tb}|\gg |V_{td}|, |V_{ts}|$ and taking into account the possible presence of an anomalous form factor 
$f_{L_V}$~\cite{AguilarSaavedra:2008zc,Kane:1991bg,Rizzo:1995uv} of the $Wtb$ vertex, 
$|f_{L_V}V_{tb}|=1.020\pm 0.046\,\mbox{(exp.)}\pm 0.017\,\mbox{(theo.)}$ is determined. From this result 
the confidence interval $0.92<|V_{tb}|\le 1$ is determined using the unified approach of Feldman and 
Cousins~\cite{Feldman:1997qc} and assuming the constraint $|V_{tb}|\le 1$ and $f_{L_V}=1$ as predicted in the SM.\\  

The ATLAS collaboration published in September 2012 a measurement of the $t$-channel single top quark cross section using data at $\sqrt{s}=7\;\mbox{TeV}$ and corresponding to an integrated luminosity of $1.04\;\mbox{fb}^{-1}$. Events with $e/\mu$+jets signature are considered. To discriminate signal from backgrounds a neural network~\cite{Feindt:2004wla,Feindt:2006pm} has been used and the signal content is extracted from a simultaneous maximum-likelihood fit to the data distributions of the neural network discriminant of events with two and three jets. The observed significance of $t$-channel production is $7.2\sigma$ and the measured cross section of $\sigma=83\pm 4\;\mbox{(stat.)}\;^{+ 20}_{-19}\;\mbox{(syst.)}\;\mbox{pb}$ is consistent with the SM approximate NNLO prediction~\cite{Kidonakis:2011wy} and with the first measurement from the CMS collaboration~\cite{Chatrchyan:2011vp} published in 2011 and the second measurement~\cite{Chatrchyan:2012ep}, which appeared only few months later than the measurement from the ATLAS collaboration. The relative uncertainty of this $t$-channel cross section measurement is 24\%. Assuming that $|V_{tb}|\gg |V_{td}|, |V_{ts}|$ $|V_{tb}|=1.13^{+0.14}_{-0.13}$ is measured.\\
In addition the $t$-channel cross section has been measured using a simple cut-based analysis. The cross section obtained in this analysis is consistent with the result using the neural network but less accurate. Using the cut-based analysis the cross sections for top and anti-top quark production have been measured, separately, and are found to be consistent with the SM predictions of Ref.~\cite{Kidonakis:2011wy}.\\

End of 2014 the ATLAS collaboration published an updated and largely extended study on $t$-channel single top quark production~\cite{Aad:2014fwa} using data at $\sqrt{s}=7\;\mbox{TeV}$ and corresponding to an integrated luminosity of $4.59\;\mbox{fb}^{-1}$. Again events with $e/\mu$+jets signature are exploited and a multivariate analysis based on neural networks~\cite{Feindt:2004wla,Feindt:2006pm} is employed. The inclusive $t$-channel cross section is measured to be $\sigma=(68\pm 8)\;\mbox{pb}$ which is well consistent with the SM approximate NNLO prediction~\cite{Kidonakis:2011wy} and the second measurement from the CMS collaboration~\cite{Chatrchyan:2012ep}. With a relative uncertainty of about 12\% this measurement is only slightly less accurate than the most accurate single measurements from the CMS collaboration (BDT and NN analysis)~\cite{Chatrchyan:2012ep}. The ATLAS collaboration determines from this inclusive measurement $|V_{tb}|=1.02\pm 0.07$ assuming $|V_{tb}|\gg |V_{td}|, |V_{ts}|$.\\ 
Furthermore, the ATLAS collaboration measured in Ref.~\cite{Aad:2014fwa} the cross section for top ($\sigma_t$) and anti-top ($\sigma_{\bar{t}}$) quarks, separately, as well as the ratio $R=\sigma_t/\sigma_{\bar{t}}$. These measurements are consistent with the NLO predictions from~\cite{Campbell:2009ss,Kant:2014oha}. The measured ratio $R$ shows already with this data set some sensitivity to favour/disfavour certain NLO parton distribution functions (PDFs) used in the NLO calculation (5F scheme)~\cite{Campbell:2009ss,Kant:2014oha}.\\ 
By applying a cut on the neural network discriminant a sample enriched in $t$-channel single top quark events is obtained. From this enriched data sample differential and normalised differential cross sections for top and anti-top quarks as a function of the transverse momentum and the rapidity of the top (anti-top) quark, respectively, are extracted using an iterative Bayesian method~\cite{DAgostini:1994zf} for unfolding. These differential measurements represent the very first measurements of this kind in single top quark production. Overall, good agreement is observed between the measured differential cross sections and the NLO QCD predictions from MCFM~\cite{Campbell:2010ff}.\\

In June 2014 the CMS collaboration measured the $t$-channel single top quark production cross section at 
$\sqrt{s}=8\;\mbox{TeV}$~\cite{Khachatryan:2014iya} using data corresponding to an integrated luminosity of 
$19.7\;\mbox{fb}^{-1}$. As in previous measurements at 7 TeV~\cite{Chatrchyan:2011vp,Chatrchyan:2012ep}, events 
with $e/\mu$+jets signature are employed. In this analysis the method used in the previous $|\eta_j'|$ analysis 
described in Ref.~\cite{Chatrchyan:2012ep} is utilised. One of the differences is, that not only multi-jet and 
$W$+jets background are modelled from data but that also $t\bar{t}$ background is modelled at least partially 
from data because of the larger $t\bar{t}$ contamination at 8 TeV than at 7 TeV.\\ 
The measured $t$-channel cross section of $\sigma=83.6\pm 2.3\;\mbox{(stat.)}\pm 7.4\;\mbox{(syst.)}\;\mbox{pb}$ 
is consistent with the SM approximate NNLO prediction~\cite{Kidonakis:2011wy} and with the NLO QCD prediction in 
the 5F scheme from~\cite{Campbell:2009gj} and is used to extract 
$|f_{L_V}V_{tb}|=0.979  \pm 0.045\;\mbox{(exp.)}\pm  0.016\;\mbox{(theo.)}$ assuming $|V_{tb}|\gg |V_{td}|, |V_{ts}|$ 
and taking into account the possible presence of an anomalous form factor $f_{L_V}$~\cite{AguilarSaavedra:2008zc,Kane:1991bg,Rizzo:1995uv} 
of the $Wtb$ vertex. This result has been combined with the previous CMS measurement at 7 TeV~\cite{Chatrchyan:2012ep}, 
yielding the most precise measurement of its kind up to date: $|f_{L_V}V_{tb}| = 0.998 \pm 0.038\;\mbox{(exp.)} \pm  0.016\;\mbox{(theo.)}$. 
Furthermore, the cross sections for top and anti-top quark production and the ratio $R$ has been measured and no deviation from 
NLO SM expectation is found.\\

In January 2015 the CMS collaboration measured for the first time the $W$-boson helicity fractions in events with single top quark signature~\cite{Khachatryan:2014vma} using data with $\sqrt{s} = 8 \;\mbox{TeV}$ and corresponding to an integrated luminosity of $19.7\;\mbox{fb}^{-1}$. The measured helicity fractions are in agreement with the standard model NNLO predictions~\cite{Czarnecki:2010gb} and have similar precision to those based on $t\bar{t}$ events.\\ 

In June 2012 the ATLAS collaboration published a search for single top-quarks via flavour changing neutral currents~\cite{Aad:2012gd} ($qg\rightarrow t$) using data at $\sqrt{s}=7\;\mbox{TeV}$ and corresponding to an integrated luminosity of $2.05\;\mbox{fb}^{-1}$. In the absence of an excess of such events upper limits on the production cross-section and anomalous $ugt$ and $ctg$ couplings are set, which are the most stringent to date (status May 2015) on FCNC single top-quark production processes for $qg \rightarrow t$.\\

In August 2012 the ATLAS collaboration reported first evidence ($3.3\sigma$) of $Wt$ 
production~\cite{Aad:2012xca} using data at $\sqrt{s}=7\;\mbox{TeV}$ and corresponding 
to an integrated luminosity of $2.05\;\mbox{fb}^{-1}$. Using events with two charged 
leptons ($\ell=e,\mu$) a multivariate analysis with BDTs has been performed to increase 
the discrimination power between $Wt$ production and $t\bar{t}$ production. Beginning 
of 2013 the CMS collaboration confirmed evidence for single top $Wt$ production 
($4.0\sigma$) \cite{Chatrchyan:2012zca} using data at $\sqrt{s}=7\;\mbox{TeV}$ 
and corresponding to an integrated luminosity of $4.9\;\mbox{fb}^{-1}$. 
Also here, events with two charged leptons ($\ell=e,\mu$) are used and the measurement 
is performed by employing a multivariate technique (BDT) to separate the $t\bar{t}$ 
background from the signal. In June 2014 the CMS collaboration reported observation 
($6.1\sigma$) of $Wt$ production~\cite{Chatrchyan:2014tua} using data at 
$\sqrt{s}=8\;\mbox{TeV}$ and corresponding to an integrated luminosity of 
$12.2\;\mbox{fb}^{-1}$. The measured cross section of $\sigma=12.3\pm 5.4\;\mbox{pb}$ 
is in agreement with the SM approximate NNLO prediction of Ref.~\cite{Kidonakis:2012rm}.\\ 

Beginning of 2015 the ATLAS collaboration published a search for $s$-channel single top quark production~\cite{Aad:2014aia} using data at $\sqrt{s}=8\;\mbox{TeV}$ and corresponding to an integrated luminosity of $20.3\;\mbox{fb}^{-1}$. The analysis leads to an upper limit on the $s$-channel single top-quark production cross-section of $14.6\;\mbox{pb}$ at 95\% C.L. (expected: $15.7\;\mbox{pb}$ at 95\% C.L. for signal-plus-background and $9.4\;\mbox{pb}$ at 95\% C.L. for background-only hypothesis).\\

Within the LHC Run I $t$-channel single top quark production has been well established and measurements of so far unprecedented precision could be conducted due to the much higher centre-of-mass energy compared to the Tevatron and due to the better conditions in terms of background processes. In Run II there is great potential for precise measurements of the inclusive and differential cross sections in the $t$-channel, but also properties like the $W$-boson helicity, the top quark mass or the polarisation of top quarks, will be measurable in the $t$-channel with reasonable precision. Furthermore, searches for new physics (e.g. anomalous $Wtb$ couplings or FCNC) will profit from the larger centre-of-mass energy. 

%


\section{Summary}
\label{sec:summary}

In this review article three promising aspects of top quark production
have been discussed in more detail: the charge asymmetry in top quark pair production, 
the search for resonant top quark pair production and electroweak single top quark production.
The general analysis strategy of each topic was explained using selected analyses from 
the CDF and CMS collaborations and put into the context of the global status at the beginning of LHC
Run II and progress in this field.\\

Concerning the charge asymmetry in top quark pair production, it seems that the puzzle about the large
asymmetries measured at the Tevatron is solved, and we learned that the full NNLO QCD corrections and
the full electroweak NLO corrections are important for the modelling of top quark pair production.
In summary, there is no indication for new physics from the final Tevatron analyses on this topic and
from the LHC Run I measurements.\\

In case of the search for resonant top quark pair production, the analysis strategies for top quark pair
events with boosted event topology have been set up and established during LHC Run I. So far, there is 
no indication for resonant top quark pair production. Due to the higher centre-of mass energy 
the search sensitivity will be significantly increased in the LHC Run II data taking period.\\

As third topic the $t$-channel single top quark production has been discussed in more detail. 
The measured cross sections are in good agreement with the SM prediction and at a centre-of-mass energy
of 7 TeV and 8 TeV, the $t$-channel single top quark production cross section has been measured
with a relative uncertainty of below 9.5\%. In Run II the $t$-channel will certainly be the working
horse for all the top quark property measurements performed for electroweak interaction.\\

In general there are many analyses involving top quarks that will strongly profit from the
higher centre-of mass energy in Run II resulting in many promising possibilities to test
the SM and to search for new physics with top quarks.






\bibliographystyle{bibliostylefile}
\bibliography{TopProduction_JWK}




\end{document}